\documentclass[onecollarge,final]{svjour2}       
\smartqed  
\usepackage{graphicx}
\usepackage{mathptmx}      
\usepackage{amsmath,citesort}
\newcommand{\rd}{\mathrm{d}}
\newcommand{\e}{\mathrm{e}}
\newcommand{\expct}[1]{\langle #1 \rangle}
\newcommand{\Expct}[1]{\left\langle #1 \right\rangle}
\newcommand{\cum}[1]{{\langle #1 \rangle}_{\rm c}}

\newcommand{\diff}[2]{\frac{\mathrm{d} #1}{\mathrm{d} #2}}
\newcommand{\prt}[2]{\frac{\partial #1}{\partial #2}}
\renewcommand{\(}{\left(}
\renewcommand{\)}{\right)}
\renewcommand{\[}{\left[}
\renewcommand{\]}{\right]}
\newcommand{\im}{\mathrm{Im}}
\newcommand{\re}{\mathrm{Re}}
\newcommand{\Std}{\mathrm{Std}}
\newcommand{\cdf}{\mathrm{cdf}}
\newcommand{\unit}[1]{~\mathrm{#1}}
\begin{document}

\title{Evidence for geometry-dependent universal fluctuations of the Kardar-Parisi-Zhang interfaces in liquid-crystal turbulence}

\titlerunning{KPZ universal fluctuations in liquid-crystal turbulence}        

\author{Kazumasa A. Takeuchi         \and
        Masaki Sano 
}


\institute{K. A. Takeuchi \and M. Sano \at
              Department of Physics, the University of Tokyo, 7-3-1 Hongo, Bunkyo-ku, Tokyo 113-0033, Japan. \\
              \email{kat@kaztake.org}\\
              Tel. and Fax: +81-3-5841-4183
}

\date{Received: date / Accepted: date}

\maketitle

\begin{abstract}
We provide a comprehensive report on scale-invariant fluctuations
 of growing interfaces in liquid-crystal turbulence,
 for which we recently found evidence that
 they belong to the Kardar-Parisi-Zhang (KPZ) universality class
 for $1+1$ dimensions
 [Phys. Rev. Lett. \textbf{104}, 230601 (2010);
 Sci. Rep. \textbf{1}, 34 (2011)].
Here we investigate both circular and flat interfaces
 and report their statistics in detail.
First we demonstrate that their fluctuations show
 not only the KPZ scaling exponents but beyond:
 they asymptotically share
 even the precise forms of the distribution function
 and the spatial correlation function
 in common with solvable models of the KPZ class,
 demonstrating also an intimate relation to random matrix theory.
We then determine other statistical properties
 for which no exact theoretical predictions were made,
 in particular the temporal correlation function
 and the persistence probabilities.
Experimental results on finite-time effects
 and extreme-value statistics are also presented.
Throughout the paper, emphasis is put on
 how the universal statistical properties depend
 on the global geometry of the interfaces,
 i.e., whether the interfaces are circular or flat.
We thereby corroborate the powerful yet geometry-dependent universality
 of the KPZ class, which governs growing interfaces driven out of equilibrium.
\end{abstract}

\tableofcontents

\section{Introduction}

Discovery and understanding of universality due to scale invariance,
 i.e., absence of characteristic length and time scales,
 marked a watershed in statistical physics.
Classical examples are critical phenomena at thermal equilibrium,
 for which universality of critical exponents has been confirmed
 in a wide variety of experiments and deeply understood
 with theoretical frameworks such as renormalization group theory
 \cite{Stanley-Book1987,Henkel-Book1999}.
For two-dimensional systems at criticality,
 conformal field theory yields a classification of universality classes
 and unveils universality in far more detailed quantities
 such as the distribution function and the correlation function
 \cite{Henkel-Book1999}.
Moreover, interest in scale-invariant phenomena is not restricted
 to fundamental areas of physics, as exemplified by
 vast applications of scaling laws of Brownian motion
 in various fields of science.

Scale invariance is also important for systems driven out of equilibrium,
 as the way universality arises therefrom does not \textit{a priori} require
 thermal equilibrium.
Studies on fully developed turbulence \cite{Frisch-Book1995}
 and non-equilibrium critical phenomena \cite{Hinrichsen-AP2000},
 for example,
 have indeed underpinned emergence of universality out of equilibrium.
It is often manifested as scaling laws characterized by universal exponents
 akin to those for equilibrium critical phenomena,
 but more detailed statistical properties
 remain largely inaccessible in this context.
Here, focusing on scale-invariant growth processes
 \cite{Barabasi.Stanley-Book1995,Meakin-PR1993,HalpinHealy.Zhang-PR1995,Krug-AP1997},
 we present an experimental case study on such non-equilibrium universality,
 which allows us to investigate
 detailed statistical properties beyond the scaling exponents.

Growth phenomena can be roughly grouped into two categories:
 those driven by local interactions,
 such as spreading of fires and penetration of water into porous media,
 and those due to nonlocal interactions
 like formation of snowflakes and metallic dendrites.
Here we concentrate on the local growth processes,
 for which one may expect that detailed characteristics of interactions
 are scaled out at macroscopic levels
 and thus certain universality may be anticipated.
It is known both from experiments and numerical models
 \cite{Barabasi.Stanley-Book1995,Meakin-PR1993,HalpinHealy.Zhang-PR1995,Krug-AP1997}
 that such processes typically produce compact clusters
 with rough, self-affine shapes of interfaces.
To quantify this self-affinity,
 one often measures the local height $h(x,t)$ of the interfaces
 growing, e.g., on a flat substrate,
 and compute the interface width $w(l,t)$ defined as the standard deviation
 of $h(x,t)$ over a length $l$:
\begin{equation}
 w(l,t) \equiv \expct{\Std[h(x,t)]_l},
 \hspace{20pt}
 \Std[h(x,t)]_l \equiv \sqrt{\Expct{\[h(x,t) - \expct{h(x,t)}_l\]^2}_l},  \label{eq:WidthDef}
\end{equation}
 where the average $\expct{\cdots}_l$ is taken
 within a segment of length $l$ around the given position $x$
 and $\expct{\cdots}$ along each interface and then over all the samples.
The self-affinity of the interfaces is then
 testified by the following power law called the Family-Vicsek scaling
 \cite{Family.Vicsek-JPA1985}:
\begin{equation}
 w(l,t) \sim t^\beta F_w(lt^{-1/z}) \sim \begin{cases} l^\alpha & \text{for $l \ll l_*$}, \\ t^\beta & \text{for $l \gg l_*$}, \end{cases}  \label{eq:FamilyVicsekWidth}
\end{equation}
 with a scaling function $F_w$,
 two characteristic exponents $\alpha$ and $\beta$,
 the dynamic exponent $z \equiv \alpha/\beta$,
 and a crossover length scale $l_* \sim t^{1/z}$.
One may similarly use the height-difference correlation function
\begin{equation}
 C_{\rm h}(l,t) \equiv \expct{[h(x+l,t) - h(x,t)]^2},  \label{eq:CorrHeightDiffDef}
\end{equation}
 to measure a typical change in the height over a length $l$,
 which should also exhibit the Family-Vicsek scaling
\begin{equation}
 C_{\rm h}(l,t)^{1/2} \sim t^\beta F_h(lt^{-1/z}) \sim \begin{cases} l^\alpha & \text{for $l \ll l_*$}, \\ t^\beta & \text{for $l \gg l_*$}. \end{cases}  \label{eq:FamilyVicsekCorrHeightDiff}
\end{equation}
Self-affine growth of interfaces is often characterized
 by these two scaling exponents $\alpha$ and $\beta$.

The simplest macroscopic theory for such local growth processes is given
 on the basis of the continuum equation
 called the Kardar-Parisi-Zhang (KPZ) equation
 \cite{Kardar.etal-PRL1986}:
\begin{equation}
 \prt{}{t}h(x,t) = v_0 + \nu \nabla^2 h + \frac{\lambda}{2}(\nabla h)^2 + \xi(x,t)  \label{eq:KPZEqDef}
\end{equation}
 with white Gaussian noise
\begin{equation}
\begin{array}{l}
 \expct{\xi(x,t)} = 0, \\
 \expct{\xi(x,t)\xi(x',t')} = D\delta(x-x')\delta(t-t').
\end{array} \label{eq:XiDef}
\end{equation}
Note that the constant driving force $v_0$ can be absorbed
 by the transformation $h \to h + v_0t$
 and thus is often set to be zero in the literature.
For $1+1$ dimensions,
 i.e., for one-dimensional interfaces growing in two-dimensional space,
 the values of the scaling exponents are known exactly
 by symmetry arguments as well as by renormalization group technique
 \cite{Kardar.etal-PRL1986,Forster.etal-PRA1977,Barabasi.Stanley-Book1995,HalpinHealy.Zhang-PR1995}
 to be $\alpha^{\rm KPZ} = 1/2$ and $\beta^{\rm KPZ} = 1/3$.
In particular, $\alpha^{\rm KPZ} = 1/2$ stems from the fact that
 the spatial profile of the stationary KPZ interface is equivalent
 to the locus of the one-dimensional Brownian motion,
 with space and time coordinates being $h$ and $x$
 \cite{Barabasi.Stanley-Book1995,HalpinHealy.Zhang-PR1995};
 hence, because of the square-root growth of its displacement,
 $w \sim l^{1/2}$.
Since these scaling exponents characterize macroscopic dynamics,
 their values are expected to be universal regardless
 of microscopic details of growth processes.
It has been indeed repeatedly confirmed in a wide variety of numerical models
 and theoretical situations
 \cite{Barabasi.Stanley-Book1995,Meakin-PR1993,HalpinHealy.Zhang-PR1995,Krug-AP1997},
 constituting the KPZ universality class.
Furthermore, for $1+1$ dimensions,
 a series of remarkable theoretical achievements
 have been made in the last decade
 \cite{Kriecherbauer.Krug-JPA2010,Sasamoto.Spohn-JSM2010,Corwin-RMTA2012},
 where, for some solvable models,
 the asymptotic distribution function of the interface fluctuations
 has even been calculated.
Surprisingly, the derived distribution function
 depends on whether the interfaces are initially flat or curved
\cite{Prahofer.Spohn-PRL2000,Prahofer.Spohn-PA2000,Kriecherbauer.Krug-JPA2010,Sasamoto.Spohn-JSM2010,Corwin-RMTA2012}
 but is nevertheless expected to be universal.
The $(1+1)$-dimensional KPZ class would therefore serve as a touchstone
 for such unprecedented universality beyond the scaling exponents
 in scale-invariant systems far from equilibrium.

In this regard, experimental investigations of growing interfaces
 are of major importance, in particular to assess the scope and the robustness
 of the KPZ universality class.
A substantial number of experiments have been performed and reported
 that rough, self-affine interfaces indeed arise
 in various kinds of local growth processes,
 such as fluid flow in porous media, paper wetting,
 colony of proliferating bacteria, and molecular deposition,
 to name but a few
 \cite{Barabasi.Stanley-Book1995,Meakin-PR1993,HalpinHealy.Zhang-PR1995,Krug-AP1997}.
Although in most cases the measured values of the exponents are significantly
 different from those of the KPZ class
 \cite{Barabasi.Stanley-Book1995,Meakin-PR1993,HalpinHealy.Zhang-PR1995,Krug-AP1997},
 typically because of quenched disorder
 and/or effectively long-range interactions,
 a few studies have reported direct evidence of the KPZ-class exponents
 in real experiments:
 colony growth of mutant \textit{Bacillus subtilis} (bacteria)
 \cite{Wakita.etal-JPSJ1997}%
\footnote{
Note that similar experiments
 with wild-type bacteria have shown non-KPZ exponents
 \cite{Vicsek.etal-PA1990,Wakita.etal-JPSJ1997,Barabasi.Stanley-Book1995}.
}
 as well as of Vero cells (eukaryote)
 \cite{Huergo.etal-PRE2010,Huergo.etal-PRE2011}
 and slow combustion of paper
 \cite{Maunuksela.etal-PRL1997,Myllys.etal-PRE2001,Myllys.etal-PRL2000},
 to the knowledge of the authors.
These experiments have clearly shown the KPZ scaling exponents,
 but an unavoidable difficulty in them is that for each realization
 one needs to prepare a sample with due care and then
 observe it for long time, one to a few days for the colony experiments.
This restricts the amount of available data,
 and hence detailed statistical properties of keen theoretical interest,
 such as the distribution function, have not been studied experimentally.
The only exception is Miettinen \textit{et al.}'s encouraging result
 in the paper combustion experiment \cite{Miettinen.etal-EPJB2005},
 which claimed agreement with theoretical predictions
 on the distribution function,
 but their analysis does not seem to be sufficient
 to draw a significant conclusion on it \cite{Takeuchi-c2012}.

\begin{figure}[t]
 \begin{center}
  \includegraphics[clip]{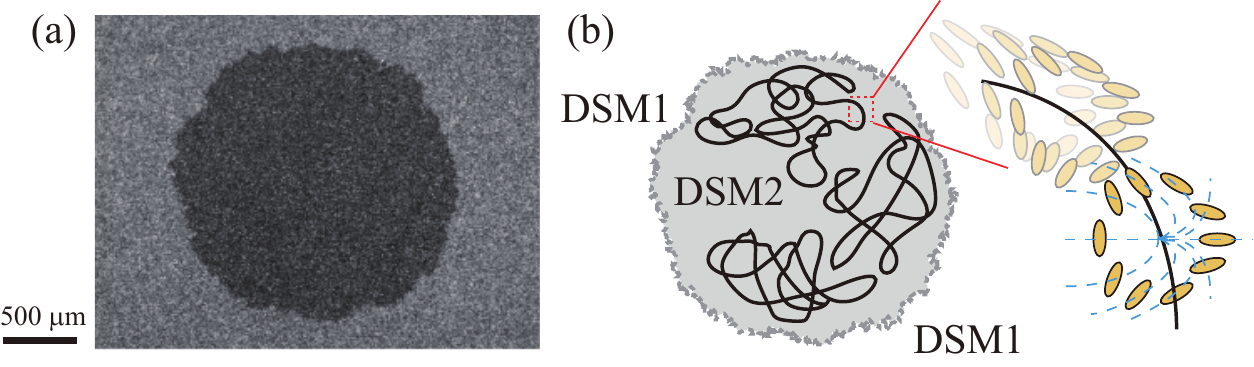}
  \caption{DSM1 and DSM2. (a) Real image of a DSM2 cluster (black) growing outward amidst the metastable DSM1 phase (gray), obtained by the method detailed in Sect.~\ref{sec:ExpSystem}. The same image is used for the third snapshot of Fig.~\ref{fig:Snapshots}(a). (b) Sketch of DSM1 and DSM2. The DSM2 state is composed of a high density of disclinations, i.e., loops of singularity in the liquid-crystal orientation, whereas there are hardly any topological defects in the DSM1 state. The black solid loops in the sketch stand for disclinations, yellow ovals for liquid-crystal molecules, and blue dashed curves for lines of equal molecular orientation. Note that, because of the high density of disclinations, DSM2 scatters incident light more strongly and hence appears darker when observed by transmitted light.}
  \label{fig:DSM}
 \end{center}
\end{figure}

To overcome all these difficulties,
 we have focused on the electroconvection of nematic liquid crystals
 \cite{deGennes.Prost-Book1995,Kai.Zimmermann-PTPS1989}
 and investigated growing clusters of topological-defect turbulence
 \cite{Takeuchi.Sano-PRL2010,Takeuchi.etal-SR2011}.
The electroconvection, driven by an ac electric field
 applied to a thin container of liquid crystal,
 exhibits two distinct turbulent states called the dynamic scattering modes
 1 and 2 (DSM1 and DSM2) for large amplitudes of the applied voltage
 \cite{deGennes.Prost-Book1995,Kai.Zimmermann-PTPS1989}.
These are spatiotemporal chaos,
 in which the velocity and director fields of the liquid crystal
 as well as the density field of electric charges strongly fluctuate
 in space and time with short-range correlations%
\footnote{
Therefore, strictly, the DSMs are unlike the fully developed turbulence
 in isotropic fluid, which is scale-invariant
 and characterized by various scaling laws \cite{Frisch-Book1995}.
In the present paper, however, we use the term ``turbulence,''
 following the convention of the electroconvection community.
}.
In addition, the DSM2 state is composed of a high density
 of topological defects called the disclinations \cite{Kai.etal-PRL1990}
 (Fig.~\ref{fig:DSM}),
 which are constantly elongated, split, and transported
 by fluctuating turbulent flow around.
Upon applying a large voltage $V$,
 we first observe the DSM1 state with practically no disclinations.
It lasts until a disclination is finally created
 by strong shear of the turbulence or by external perturbation,
 which immediately multiplies and forms an expanding DSM2 region
 for large enough voltages [Fig.~\ref{fig:DSM}(a)].
In other words, 
 the DSM1 state is metastable and
 eventually replaced by the stable DSM2 state
 for such large applied voltages.
Therefore, once a DSM2 nucleus is created amidst the DSM1 state,
 spontaneously or externally,
 it forms a growing compact cluster bordered by a moving interface.
This has turned out to roughen in the course of time
 \cite{Takeuchi.Sano-PRL2010,Takeuchi.etal-SR2011}
 as one may expect for such a local growth process.

This DSM2 growth provides an ideal experimental situation
 to study local growth processes for a number of reasons:
(i) Local interactions. 
The interface is the front of disclinations,
 proliferated and transported by turbulent flow with short-range correlation.
This constitutes the local and stochastic growth of the interface.
(ii) Effectively no quenched disorder.
The stochasticity of the process is due to intrinsic turbulent flow,
 which overwhelms cell heterogeneity.
(iii) Easily repeatable in a highly controlled condition.
The DSM2 growth can be triggered by laser
 and can also be reset to the initial fully-DSM1 state
 by switching the applied voltage off and on.
(iv) Control of the initial cluster shape.
It is simply determined by the form of the laser beam
 used for the DSM2 nucleation.
As reported in the preceding letters
 \cite{Takeuchi.Sano-PRL2010,Takeuchi.etal-SR2011},
 these advantages indeed allowed us to perform series of experiments
 for both globally circular and flat interfaces
 and to obtain accurate data with good statistics.
We then found evidence of the KPZ scaling laws
 and, in particular, of the predicted distribution functions
 dependent on the global cluster shape.

In the present paper, we provide a detailed report
 on these results and beyond, showing also the results
 of comprehensive statistical analyses of the DSM2 interface fluctuations.
They include experimental tests of various theoretical predictions
 made for solvable models,
 e.g., those for the spatial correlation function
 and on extreme-value statistics,
 but equal emphasis is put on quantities without any rigorous predictions,
 such as the time correlation function
 and the spatial and temporal persistence probabilities.
Results from numerical simulations
 made by one of the authors \cite{Takeuchi-JSM2012}
 are also occasionally referred to,
 in order to argue universality of these unsolved quantities.
All in all, the present work is intended to provide a comprehensive report
 on experimentally determined statistical properties
 of the DSM2 growing interfaces, which are presumably shared
 in the $(1+1)$-dimensional KPZ class as universal properties.

\section{Experimental system}  \label{sec:ExpSystem}

\begin{figure}[t]
 \begin{center}
  \includegraphics[clip]{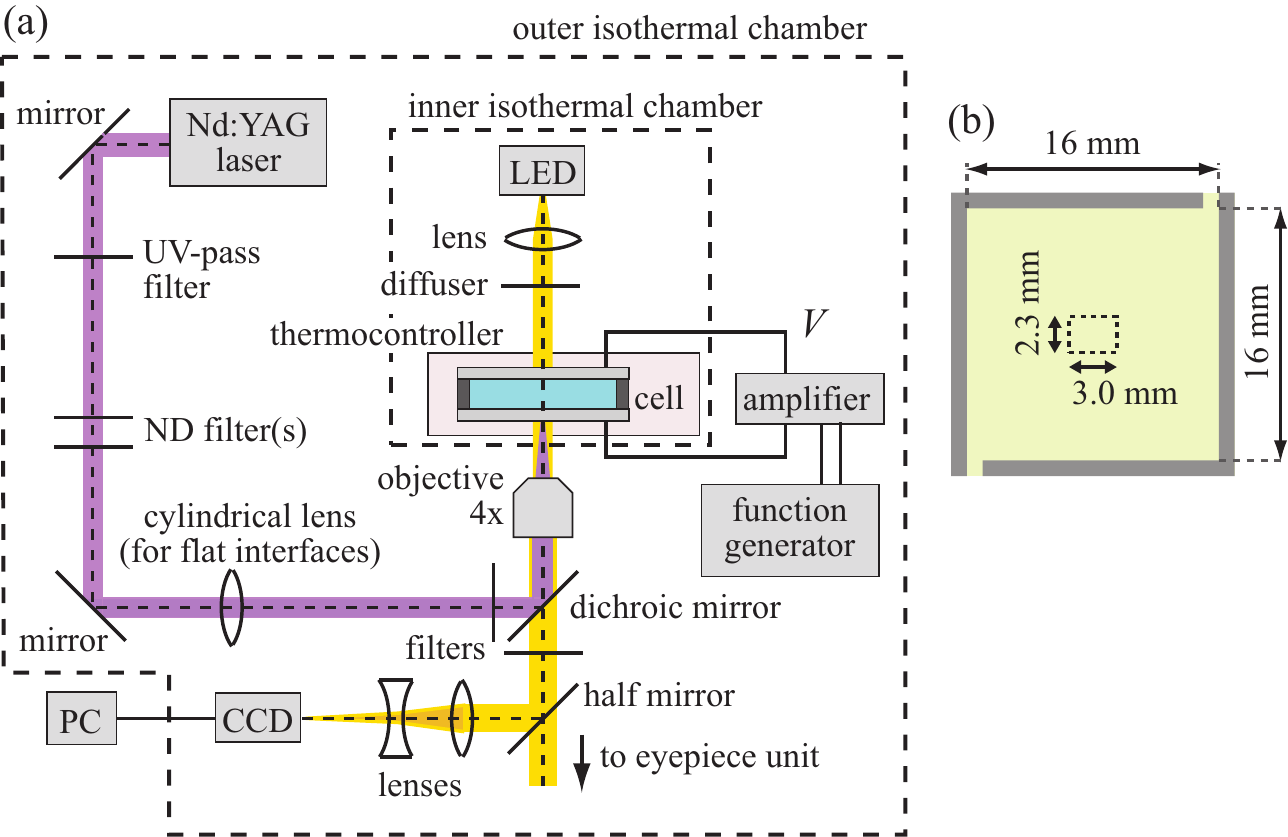}
  \caption{Sketch of the experimental setup. (a) The whole system. CCD: charge-coupled device camera, LED: light-emitting diode, ND: neutral density, PC: computer, UV: ultraviolet. The third harmonic of the Nd:YAG laser at $355\unit{nm}$ is used in the experiment. See text for details. (b) Top view of the central part of the cell. The yellow region is filled with liquid crystal, while the gray region is a side wall made of polyester films. The dashed rectangle indicates the observed region, which is set near the center of the cell.}
  \label{fig:ExpSetup}
 \end{center}
\end{figure}%

\begin{figure}[t]
 \begin{center}
  \includegraphics[clip]{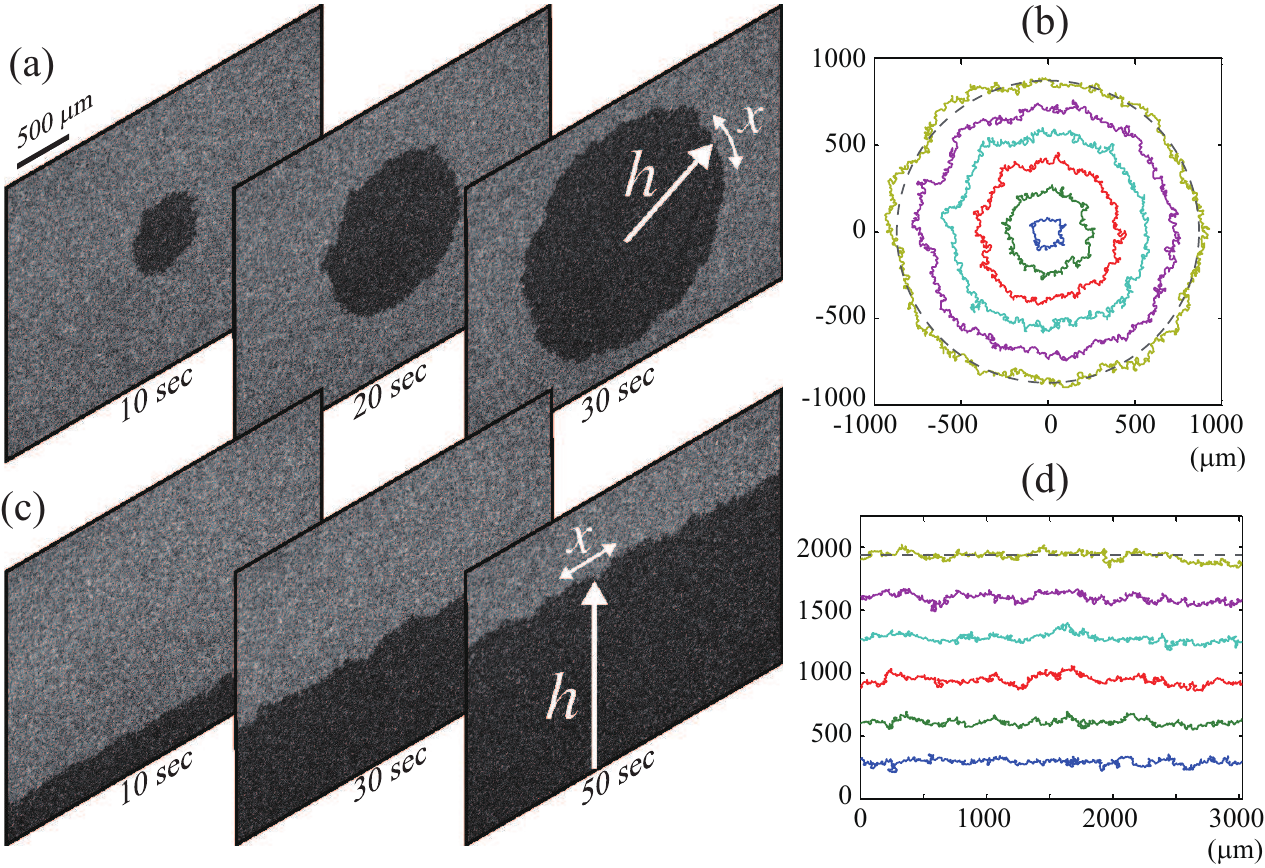}
  \caption{Growing DSM2 cluster with a circular (a,b) and flat (c,d) interface. (a,c) Raw images. Indicated below each image is the elapsed time after the emission of laser pulses. (b,d) Snapshots of the interfaces at $t = 3, 8, \dots, 28\unit{s}$ for the circular case (b) and $t=10, 20, \dots, 60\unit{s}$ for the flat case (d). The gray dashed lines indicate the mean radius (height) of all the circular (flat) interfaces recorded at $t=28\unit{s}$ ($t=60\unit{s}$). See also Supplementary Movies of Ref.~\cite{Takeuchi.etal-SR2011}.}
  \label{fig:Snapshots}
 \end{center}
\end{figure}

Our experimental setup,
 which has also been used for our preceding results
 \cite{Takeuchi.Sano-PRL2010,Takeuchi.etal-SR2011},
 is outlined in Fig.~\ref{fig:ExpSetup}.
At the heart of the system is a quasi-two-dimensional convection cell
 of size $16\unit{mm} \times 16\unit{mm} \times 12\unit{\mu{}m}$,
 filled with a nematic liquid crystal sample detailed below.
The cell surfaces are treated so that the liquid crystal molecules are aligned
 perpendicularly to the cell (homeotropic alignment)
 when no voltage is applied,
 in order to work with isotropic growth in the horizontal plane.
The cell temperature is kept constant at $25 \unit{^\circ{}C}$
 with typical fluctuations less than $1\unit{mK}$ throughout the experiment.
For each realization of a growing interface,
 we apply a voltage of $26\unit{V}$ at $250\unit{Hz}$
 (with typical fluctuations less than $1\unit{mV}$)
 to set the system first in the DSM1 state.
We then wait for a few seconds 
 and shoot a couple of ultraviolet laser pulses to nucleate a DSM2 seed.
It forms a growing circular interface when the laser beam is focused
 on a point in the cell [Fig.~\ref{fig:Snapshots}(a,b)],
 whereas a flat interface can also be generated
 by shooting line-shaped pulses [Fig.~\ref{fig:Snapshots}(c,d)].
The chosen voltage of $26\unit{V}$ is sufficiently larger than
 that for the onset of the DSM1-DSM2 spatiotemporal intermittency regime
 \cite{Takeuchi.etal-PRL2007,Takeuchi.etal-PRE2009},
 which is $V_{\rm c} = 22.2\unit{V}$ in our sample,
 but not too large in order that
 spontaneous nucleation of DSM2 may occur only very rarely.
As a result, the created clusters are compact
 and constantly grow in the outward direction (Fig.~\ref{fig:Snapshots})
 without being influenced by spontaneously generated DSM2 regions, if any.
We record in total 955 isolated circular interfaces and 1128 flat ones
 by observing transmitted light, by which DSM1 and DSM2 are clearly contrasted
 because of stronger light scattering of the DSM2 state (Fig.~\ref{fig:DSM}).

More specifically, the convection cell is made of two parallel glass plates,
 spaced by polyester films of thickness $12\unit{\mu{}m}$ which enclose
 the convection region [Fig.~\ref{fig:ExpSetup}(b)].
The inner surfaces of the glass are covered with indium-tin oxide
 which serves as transparent electrodes.
On top of them we made a uniform coat of
 $N$,$N$-dimethyl-$N$-octadecyl-3-aminopropyltrimethoxysilyl chloride
 using a spin coater to realize the homeotropic alignment.
After assembly, the cell was filled with nematic liquid crystal
 $N$-(4-methoxybenzylidene)-4-butylaniline
 (MBBA) (purity $> 99.5\%$, Tokyo Chemical Industry),
 doped with 0.01 wt.\% of tetra-$n$-butylammonium bromide
 to increase the conductivity of the sample.
This sets the cutoff frequency between the conductive and dielectric regimes
 of the electroconvection
 \cite{deGennes.Prost-Book1995,Kai.Zimmermann-PTPS1989}
 to be $850 \pm 50\unit{Hz}$.
In order to avoid fluctuations of material parameters during the experiment,
 which affect for instance the growth speed of the interfaces,
 we need to maintain a precisely constant temperature of the cell.
It is achieved by enclosing the cell in a thermocontroller
 made of heating wires and Peltier elements
 (a more detailed description is given
 in Ref.~\cite{Takeuchi.etal-PRE2009})
 as well as by nested isothermal chambers
 composed of thermally insulated walls,
 which pump out inner heat through constant-temperature water circulators
 [Fig.~\ref{fig:ExpSetup}(a)].
As a result, temperature fluctuations of the cell
 were kept typically less than $1\unit{mK}$
 during the whole series of the experiments.

The nucleation of a DSM2 cluster is realized
 by shooting two successive Nd:YAG laser pulses
 of length $4$-$6\unit{ns}$ each
 (MiniLase II $20\unit{Hz}$, New Wave Research)
 \cite{Takeuchi.etal-PRE2009}.
A bandpass filter is placed on the beam line
 to extract their third harmonic at $355\unit{nm}$,
 neutral density filters to reduce their energy,
 and then, for the flat interfaces,
 a cylindrical lens to expand the beam
 in a transversal direction.
After a few more filters, the laser pulses are finally
 focused by a 4X objective lens
 and reach the sample with energy $6\unit{nJ}$ for the circular interfaces
 and $0.04\unit{nJ/\mu{}m}$ for the flat ones.
This creates a growing interface without any observable damage to the sample.
We then observe the interface by recording transmitted light
 with a charge-coupled device camera,
 until it expands beyond the camera field.
The resolution of the captured images is $4.74\unit{\mu{}m}$ per pixel.
Finally, we turn off the applied voltage to end the process,
 and after a few seconds, switch it on again
 and repeat the steps described here.
Excluding a few realizations in which
 an uncontrolled, spontaneous nucleation of DSM2
 occurred within the view field,
 we accumulated in this way 955 and 1128 records
 of circular and flat interfaces, respectively, as stated above.

To analyze the data,
 we first binarize the recorded images using the difference in the intensity
 between DSM1 and DSM2
 and locate the positions of the interfaces.
We define the local height $h(x,t)$
 as the distance from the initial interface position,
 i.e., the point or the line at which the laser pulses are shot
 [Fig.~\ref{fig:Snapshots}(a,c)].
The height is measured along the global moving direction of the interfaces
 and therefore is a function of the lateral coordinate $x$ and time $t$.
The observed interfaces have a number of tiny overhangs
 [Fig.~\ref{fig:Snapshots}(b,d)];
 although a recent study showed that overhangs are irrelevant
 for the scaling of the interfaces \cite{Rodriguez-Laguna.etal-JSM2011}, here,
 for the sake of simplicity and direct comparison to theoretical predictions,
 we take the mean of all the detected heights at a given coordinate $x$
 to define a single-valued function $h(x,t)$ for each interface.
The spatial profile $h(x,t)$ is statistically equivalent at any point $x$
 because of the isotropic and homogeneous growth of the interfaces,
 which, together with the large numbers of the realizations,
 provides accurate statistics for the interface fluctuations analyzed below.

Before presenting the results of the analysis,
 it is worth noting different characters of the ``system size'' $L$,
 or the total lateral length, of the circular and flat interfaces in general.
While the system size of the flat interfaces is chosen \textit{a priori}
 and fixed during the evolution, that of the circular interfaces
 is the circumference which
 grows linearly with time and is therefore not independent of dynamics.
This matters most when one takes an average of a stochastic variable,
 e.g., the interface height.
For the flat interfaces, the spatial and ensemble averages are equivalent
 provided that the system size $L$ is much larger than the correlation length
 $l_* \sim t^{1/z}$.
In contrast, for the circular interfaces,
 the two averages make a significant difference,
 because the system size is inevitably finite
 and the influence of finite-size effects varies in time.
To avoid this complication, we take below
 the ensemble average denoted by $\expct{\cdots}$ unless otherwise stipulated,
 which turns out to be the right choice
 when one measures characteristic quantities
 such as the growth exponent $\beta$.

\section{Experimental results}

\subsection{Scaling exponents}

\begin{figure}[t]
 \begin{center}
  \includegraphics[clip]{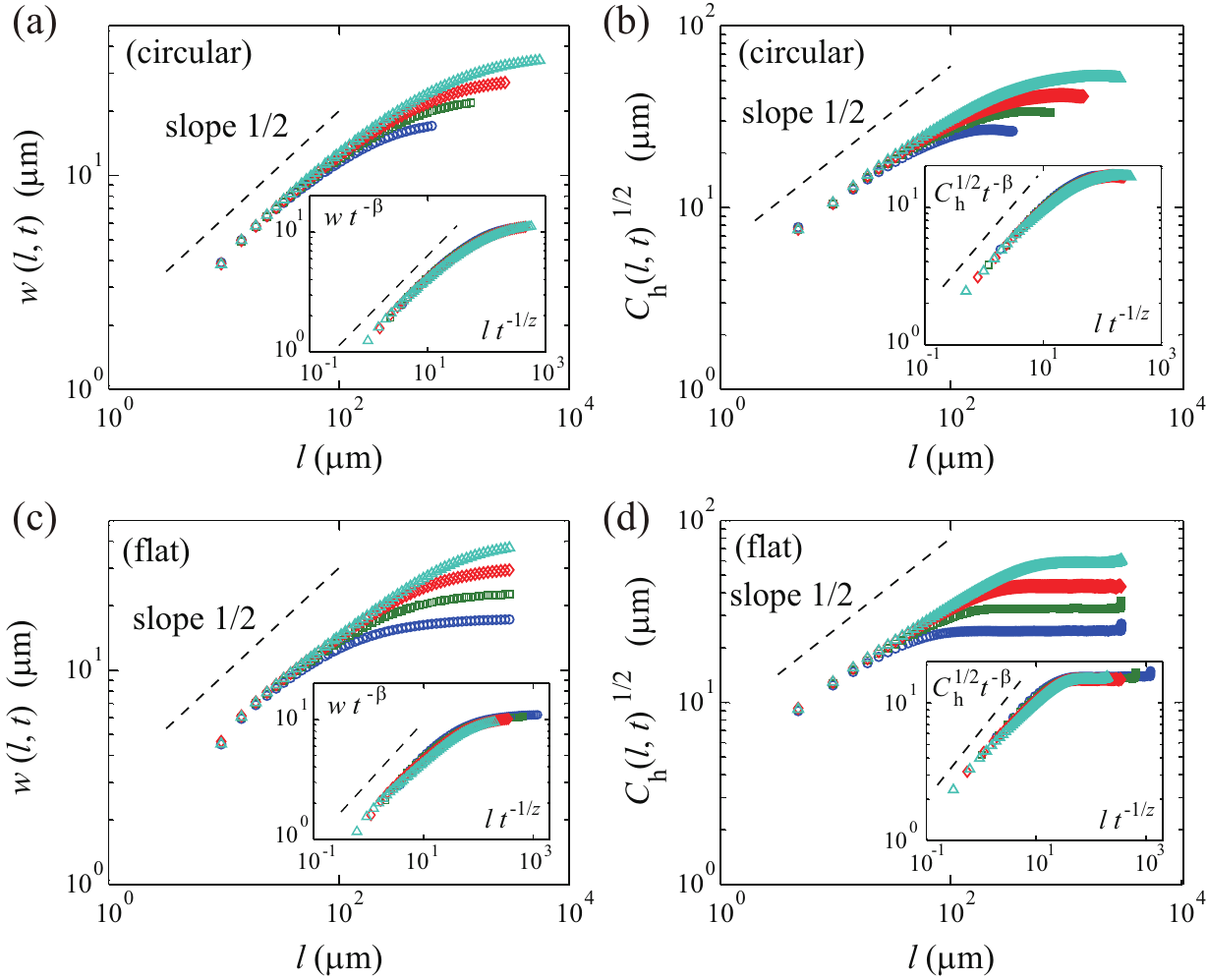}
  \caption{Family-Vicsek scaling. The interface width $w(l,t)$ (a,c) and the square root of the height-difference correlation function $C_{\rm h}(l,t)^{1/2}$ (b,d) are shown for different times $t$ for the circular (a,b) and flat (c,d) interfaces. The four data sets correspond to $t = 2\unit{s}, 4\unit{s}, 12\unit{s}, 30\unit{s}$ (a,b) and to $t = 4\unit{s}, 10\unit{s}, 25\unit{s}, 60\unit{s}$ (c,d)  from bottom to top. The insets show the same data with the rescaled axes, $wt^{-\beta}$ or $C_{\rm h}^{1/2}t^{-\beta}$ against $lt^{-1/z}$, with the KPZ exponents $\beta=1/3$ and $z=3/2$. The dashed lines are guides for the eyes indicating the slope for the KPZ exponent $\alpha^{\rm KPZ} = 1/2$.}
  \label{fig:FamilyVicsek}
 \end{center}
\end{figure}

First we test the Family-Vicsek scaling
 \eqref{eq:FamilyVicsekWidth} and \eqref{eq:FamilyVicsekCorrHeightDiff}
 and measure the roughness exponent $\alpha$ and the growth exponent $\beta$.
Figure \ref{fig:FamilyVicsek} shows the interface width $w(l,t)$
 and the square root
 of the height-difference correlation function $C_{\rm h}(l,t)^{1/2}$
 measured at different times $t$, for both circular and flat interfaces
 [Fig.~\ref{fig:FamilyVicsek}(a,b) and (c,d), respectively].
They grow algebraically for short lengths $l \ll l_*$ and
 converge to time-dependent constants for large $l$,
 in agreement with the Family-Vicsek scaling
 \eqref{eq:FamilyVicsekWidth} and \eqref{eq:FamilyVicsekCorrHeightDiff}.
Fitting $w \sim l^\alpha$ and $C_{\rm h}^{1/2} \sim l^\alpha$
 in the power-law regime
 of the data at the latest time in Fig.~\ref{fig:FamilyVicsek},
 we estimate $\alpha = 0.48(5)$ and $0.43(6)$
 for the circular and flat interfaces, respectively.
Here, the numbers in the parentheses indicate
 ranges of error in the last digit, estimated
 both from uncertainty in a single fit
 and from the dependence on the fitting range.
The estimated values of $\alpha$ for the two geometries
 are therefore
 consistent with the KPZ-class exponent $\alpha^{\rm KPZ}=1/2$,
 albeit barely for the flat case%
\footnote{
Since our experiment is aimed at accumulating detailed statistics
 on large-scale fluctuations, the chosen spatial resolution is unfortunately
 not optimal to measure the power laws $w \sim l^\alpha$
 and $C_{\rm h}^{1/2} \sim l^\alpha$ governing short lengths $l$.
We therefore consider that our estimates may admit larger uncertainties
 and in particular that the apparent slight discrepancy
 between $\alpha^{\rm KPZ}=1/2$ and $\alpha = 0.43(6)$ for the flat interfaces
 is not significant.
}.
The validity of $\alpha = 1/2$ in our experimental system will also be
 confirmed directly by data collapse for the Family-Vicsek scaling
 (Fig.~\ref{fig:FamilyVicsek} insets), as well as
 from various quantities and aspects presented throughout the paper.

\begin{figure}[t]
 \begin{center}
  \includegraphics[clip]{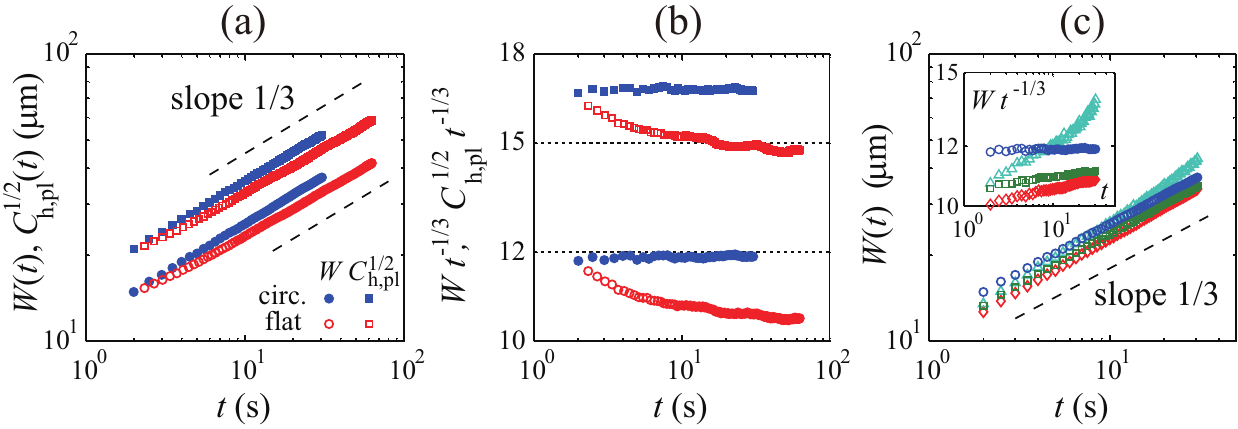}
  \caption{Temporal growth of the roughness. (a) Overall width $W(t)$ (circles) and plateau value of the square root of the height-difference correlation function, $C_{\rm h,pl}^{1/2}(t)$ (squares), for the circular (blue solid symbols) and flat (red open symbols) interfaces. (b) The same data as (a) but multiplied by $t^{-1/3}$ to enhance the visibility of the asymptotic power laws. (c) The overall width $W(t)$ of the circular interfaces measured with various definitions. Blue circles: the appropriate definition \eqref{eq:OverallWidthDef}, $W(t) = \sqrt{\expct{[h(x,t)-\expct{h}]^2}}$. Green squares: the samplewise definition, $W(t) = \expct{\sqrt{\expct{[h(x,t)-\expct{h}_{\rm s}]^2}_{\rm s}}}$. Red diamonds and turquoise triangles: same as Eq.~\eqref{eq:OverallWidthDef} but the origin is set to be the center of mass of the cluster and the interface, respectively. The same data are multiplied by $t^{-1/3}$ in the inset. The dashed lines indicate the exponent $1/3$.}
  \label{fig:Beta}
 \end{center}
\end{figure}

For large enough lengths $l$,
 the roughness measures become insensitive to $l$ and grow with time $t$
 (Fig.~\ref{fig:FamilyVicsek}).
This temporal growth is quantified by measuring the overall width
\begin{equation}
 W(t) \equiv \sqrt{\expct{[h(x,t)-\expct{h}]^2}}  \label{eq:OverallWidthDef}
\end{equation}
 and the mean value of $C_{\rm h}(l,t)^{1/2}$ in the plateau region
 of Fig.~\ref{fig:FamilyVicsek}(b,d), denoted by $C_{\rm h,pl}(t)^{1/2}$.
These are shown in Fig.~\ref{fig:Beta}(a)
 for both circular and flat interfaces and evidence the expected power laws
 $W(t) \sim t^\beta$ and $C_{\rm h,pl}(t)^{1/2} \sim t^\beta$ at large $t$.
For the circular case (blue solid symbols), the power laws actually hold
 in the whole time span, providing remarkably accurate estimates of $\beta$:
 specifically, $\beta = 0.335(3)$ from $W(t)$
 and $\beta = 0.333(3)$ from $C_{\rm h,pl}(t)^{1/2}$,
 in excellent agreement with the KPZ value $\beta^{\rm KPZ} = 1/3$
 [see also Fig.~\ref{fig:Beta}(b)].
The agreement also holds for the flat interfaces,
 though finite-time effects are visible in this case
 for the early stage [red open symbols in Fig.~\ref{fig:Beta}(a,b)].
Fitting is therefore performed for large $t$,
 providing $\beta = 0.319(15)$ for $W(t)$
 and $\beta = 0.313(24)$ for $C_{\rm h,pl}(t)^{1/2}$,
 again in agreement with $\beta^{\rm KPZ} = 1/3$.
These finite-time effects lower the apparent exponent values
 for small $t$ [Fig.~\ref{fig:Beta}(a)].
This may appear to be a crossover from the Edwards-Wilkinson regime,
 which is characterized by $\beta = 1/4$
 and governs the early stage of the KPZ-class interfaces in general
 \cite{Krug-AP1997,Amar.Family-PRA1992},
 but in Sect.~\ref{sec:DistFunc} we shall find it more reasonable
 to consider that this is rather
 an algebraically decaying finite-time correction
 superimposed on the asymptotic power law $t^{1/3}$.

It is worth noting here that, for the circular interfaces,
 the measurement of the exponent $\beta$ has a few pitfalls
 that have not been necessarily noticed in earlier investigations.
First, as mentioned in the previous section,
 the system size of the circular interfaces is not independent of dynamics
 and thus the ensemble average in Eq.~\eqref{eq:OverallWidthDef}
 \textit{cannot} be replaced by the spatial, or samplewise, average.
Measuring and averaging the width of each interface,
 i.e., $\expct{\sqrt{\expct{[h(x,t)-\expct{h}_{\rm s}]^2}_{\rm s}}}$
 with the spatial average $\expct{\cdots}_{\rm s}$,
 yield $\beta = 0.351(3)$ [squares in Fig.~\ref{fig:Beta}(c)],
 which is slightly but significantly larger than $\beta = 0.335(3)$
 obtained from $W(t)$ with the ensemble average,
 Eq.~\eqref{eq:OverallWidthDef}.
This overestimation can be easily understood,
 because for earlier time the effective system size is smaller
 and hence the samplewise standard deviation
 is underestimated more strongly.
It implies that the limit $w(\infty,t)$ does \textit{not} replace
 the overall width $W(t)$
 and hence, strictly, the Family-Vicsek scaling
 in the form \eqref{eq:FamilyVicsekWidth}
 does not describe the correct time dependence $t^\beta$
 for the circular interfaces.
Another subtlety in the definition of the width (or any other quantities)
 concerns the origin of the cluster,
 which is used to define the local height $h(x,t)$.
Although the most natural and theoretically sound definition of the origin
 is the location of the cluster seed
 (in our experiment, the point shot by laser pulses),
 the center of mass of the cluster or of the interface has also been used
 occasionally in the literature, mainly because of technical difficulties
 in locating the true origin experimentally.
This, however, affects the value of $\beta$
 because of random movement of the center of mass akin to Brownian motion,
 as pointed out by earlier studies
 \cite{FerreiraJr.Alves-JSM2006,Paiva.FerreiraJr-JPA2007,Kuennen.Wang-JSM2008}.
In our system $\beta = 0.360(2)$ and $0.415(13)$ are obtained
 when the center of mass of the cluster and that of the interface,
 respectively, are used
 [diamonds and triangles, respectively, in Fig.~\ref{fig:Beta}(c)],
 which are significantly larger than
 the unbiased estimate $\beta = 0.335(3)$ obtained with the true, fixed origin.

The agreement of the characteristic exponents $\alpha$ and $\beta$
 with the KPZ class is crosschecked by data collapse
 of $w(l,t)$ and $C_{\rm h}(l,t)$.
The functional form of the Family-Vicsek scaling
 \eqref{eq:FamilyVicsekWidth} and \eqref{eq:FamilyVicsekCorrHeightDiff}
 implies that $w(l,t)$ and $C_{\rm h}(l,t)^{1/2}$ at different times
 (main panels of Fig.~\ref{fig:FamilyVicsek})
 should overlap onto a single curve
 when $w t^{-\beta}$ and $C_{\rm h}^{1/2}t^{-\beta}$ are plotted
 against $lt^{-1/z}$.
This is indeed confirmed in the insets of Fig.~\ref{fig:FamilyVicsek},
 in which we rescale the data in the main panels
 using the KPZ exponents $\beta = 1/3$ and $z = 3/2$.
For all these results, we conclude that
 the scale-invariant growth of the DSM2 interfaces is governed
 by the KPZ universality class,
 with the characteristic exponents $\alpha=1/2$ and $\beta=1/3$
 regardless of the cluster shape.

\subsection{Cumulants and amplitude ratios}

Now we investigate detailed statistical properties
 of the scale-invariant fluctuations
 in the DSM2 interface growth,
 mainly in view of the recent rigorous theoretical developments
 \cite{Kriecherbauer.Krug-JPA2010,Sasamoto.Spohn-JSM2010,Corwin-RMTA2012}.
The roughness growth with exponent $\beta = 1/3$ implies that
 the local time evolution of the interface height $h$ is composed
 of a deterministic linear growth term and a stochastic $t^{1/3}$ term
 as follows:
\begin{equation}
 h \simeq v_\infty t + (\Gamma t)^{1/3} \chi,  \label{eq:Height}
\end{equation}
 with two constant parameters $v_\infty$ and $\Gamma$
 and with a random variable $\chi$ that captures
 the fluctuations of the growing interfaces.
Note that Eq.~\eqref{eq:Height} is meant to describe
 local (one-point) statistical properties of the height $h$,
 while its correlation in space and time
 will be studied in subsequent sections.

\begin{figure}[t]
 \begin{center}
  \includegraphics[clip]{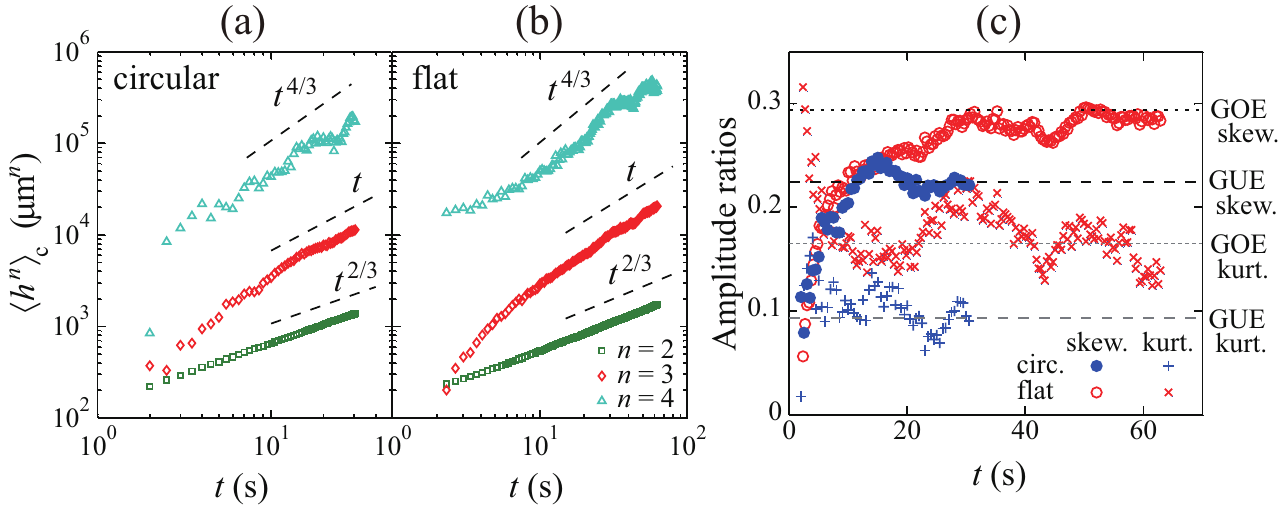}
  \caption{Cumulants and their amplitude ratios. (a,b) Cumulants $\expct{h^n}_{\rm c}$ against time $t$ for the circular (a) and flat (b) interfaces. The dashed lines indicate the exponents $n/3$ for the corresponding cumulants. (c) Skewness $\expct{h^3}_{\rm c} / \expct{h^2}_{\rm c}^{3/2}$ (solid and open circles) and kurtosis $\expct{h^4}_{\rm c} / \expct{h^2}_{\rm c}^2$ (pluses and crosses) against time $t$ for the circular (blue) and flat (red) interfaces. The horizontal lines indicate the values of the skewness (black) and the kurtosis (gray) for the GUE (dashed) and GOE (dotted) TW distributions \cite{Prahofer.Spohn-PRL2000}.}
  \label{fig:Cumulants}
 \end{center}
\end{figure}

To characterize the distribution, we first compute
 the second- to fourth-order cumulants
 $\expct{h^n}_{\rm c}$ of the local height,
 defined by $\expct{h^2}_{\rm c} = \expct{\delta{}h^2} = W(t)^2$,
 $\expct{h^3}_{\rm c} = \expct{\delta{}h^3}$
 and $\expct{h^4}_{\rm c} = \expct{\delta{}h^4} - 3\expct{\delta{}h^2}^2$
 with $\delta{}h \equiv h - \expct{h}$.
They naturally grow with time as $t^{n/3}$ [Fig.~\ref{fig:Cumulants}(a,b)].
These cumulants then determine
 the skewness $\expct{h^3}_{\rm c} / \expct{h^2}_{\rm c}^{3/2}$
 and the kurtosis $\expct{h^4}_{\rm c} / \expct{h^2}_{\rm c}^2$
 as their amplitude ratios,
 which are equal to those for $\chi$
 irrespective of the values of the two parameters.
The result is shown in Fig.~\ref{fig:Cumulants}(c)
 for both circular and flat interfaces (blue and red symbols, respectively).
First, we notice that both the skewness (solid and open circles)
 and the kurtosis (pluses and crosses)
 converge to some non-zero values,
 which indicate that the interface fluctuations are not Gaussian,
 and that they are significantly different
 between the circular and flat interfaces.

Remarkably, these asymptotic values coincide
 with the skewness and the kurtosis of the Tracy-Widom (TW) distributions
 \cite{Tracy.Widom-CMP1994,Tracy.Widom-CMP1996},
 which have been defined and developed in a completely different context
 of random matrix theory \cite{Mehta-Book2004}.
They are the distribution for the largest eigenvalue of large random matrices
 in certain kinds of the Gaussian ensemble,
 i.e., matrices whose elements are drawn from the Gaussian distribution.
The measured values of the skewness and the kurtosis
 for the circular interfaces [blue symbols in Fig.~\ref{fig:Cumulants}(c)]
 are found to be very close to those of the TW distribution
 for the Gaussian unitary ensemble (GUE) \cite{Tracy.Widom-CMP1994},
 or the largest-eigenvalue distribution
 of large complex Hermitian random matrices.
In contrast, those for the flat interfaces (red symbols)
 agree with the values of the TW distribution
 for the Gaussian orthogonal ensemble (GOE) \cite{Tracy.Widom-CMP1996},
 or for large real-valued symmetric random matrices.
These results suggest that the random variable $\chi$
 in Eq.~\eqref{eq:Height} asymptotically obeys
 the GUE and GOE TW distributions
 for the circular and flat interfaces, respectively.

It is important to note that the GUE TW distribution
 for the circular, or, more generally, curved growing interfaces
 was first identified rigorously by Johansson \cite{Johansson-CMP2000}
 on the basis of notable mathematical progress
 in related combinatorial problems \cite{Baik.etal-JAMS1999}.
Subsequent studies then derived the GUE TW distribution
 for a number of solvable models in the $(1+1)$-dimensional KPZ class,
 namely the totally and partially asymmetric simple exclusion processes
 (TASEP and PASEP) \cite{Johansson-CMP2000,Tracy.Widom-CMP2009},
 the polynuclear growth (PNG) model
 \cite{Prahofer.Spohn-PRL2000,Prahofer.Spohn-PA2000}%
\footnote{
The TASEP and the PNG model can actually be dealt with in the single framework
 of the directed polymer problem \cite{Sasamoto.Spohn-JSM2010}.
} and also the KPZ equation
 \cite{Sasamoto.Spohn-PRL2010,Sasamoto.Spohn-NPB2010,Amir.etal-CPAM2011,Calabrese.etal-EL2010,Dotsenko-EL2010}%
\footnote{
The derivations in Refs.~\cite{Calabrese.etal-EL2010,Dotsenko-EL2010} rely
 on the replica method \cite{Mezard.etal-Book1987},
 while those in
 Refs.~\cite{Sasamoto.Spohn-PRL2010,Sasamoto.Spohn-NPB2010,Amir.etal-CPAM2011}
 do not.
}.
Similarly, for the flat interfaces,
 since the work by Pr\"ahofer and Spohn
 \cite{Prahofer.Spohn-PRL2000,Prahofer.Spohn-PA2000}
 which clarified the connection to the corresponding combinatorial problem
 \cite{Baik.Rains-DMJ2001,Baik.Rains-MSRIP2001},
 the GOE TW distribution has been shown
 for the PNG model \cite{Prahofer.Spohn-PRL2000,Prahofer.Spohn-PA2000},
 the TASEP \cite{Borodin.etal-JSP2007}
 and the KPZ equation \cite{Calabrese.LeDoussal-PRL2011}.
Given these exact and nontrivial results,
 which have been proved however only for the few highly simplified models
 with more or less related analytical techniques,
 it is essential to test the emergence of the TW distributions
 in a real experiment.
This is what we reported briefly in the preceding letters
 \cite{Takeuchi.Sano-PRL2010,Takeuchi.etal-SR2011}
 and shall present in detail in the following two sections.

\subsection{Parameter estimation}

For the test, one needs to perform a direct comparison of
 $\chi$ in Eq.~\eqref{eq:Height}
 with random variables $\chi_{\rm GUE}$ and $\chi_{\rm GOE}$
 obeying the GUE and GOE TW distributions, 
 respectively.
Here, in view of the theoretical results for the solvable models
 \cite{Prahofer.Spohn-PRL2000,Prahofer.Spohn-PA2000,Borodin.etal-JSP2007,Calabrese.LeDoussal-PRL2011},
 we multiply the conventional definition for the GOE TW random variable
 by $2^{-2/3}$ to define our $\chi_{\rm GOE}$.
This choice allows us to use the single expression \eqref{eq:Height}
 for both circular and flat cases and thus to avoid unnecessary complication.
With this in mind, we first estimate the values of the two parameters,
 $v_\infty$ and $\Gamma$, which appear in Eq.~\eqref{eq:Height}.

\begin{figure}[t]
 \begin{center}
  \includegraphics[clip]{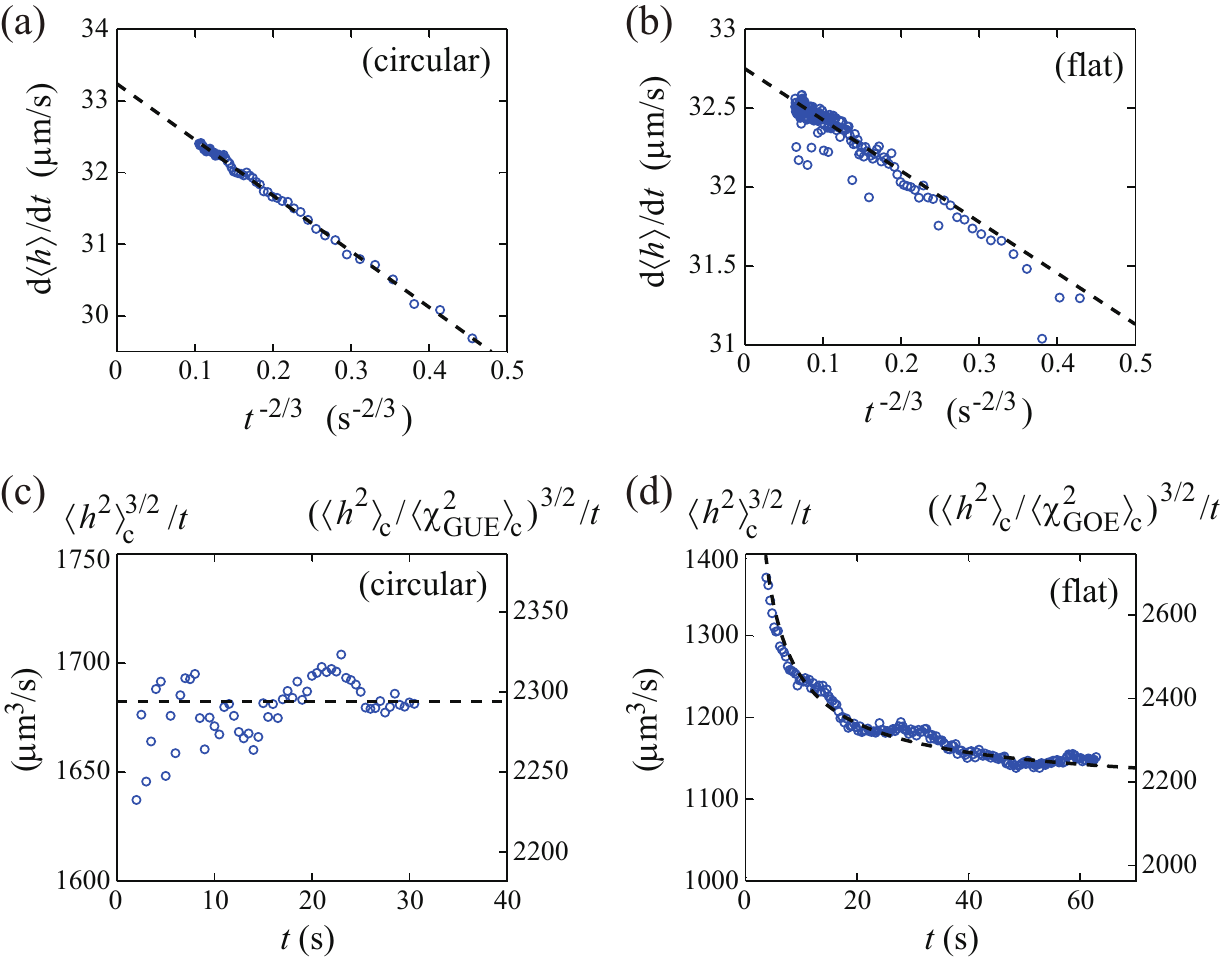}
  \caption{Parameter estimation. (a,b) Estimation of the linear growth rate $v_\infty$ for the circular (a) and flat (b) interfaces. The time derivative of the mean height $\rd\expct{h}/\rd t$, averaged here over $2.5\unit{s}$, is plotted against $t^{-2/3}$. The value of $v_\infty$ is estimated from the $y$-intercept of the linear regression (dashed line). (c,d) Estimation of the amplitude $\Gamma$ of the $t^{1/3}$-fluctuations for the circular (c) and flat (d) interfaces. The left axis indicates the raw amplitude of the second-order cumulant, $\cum{h^2}^{3/2}/t$, whereas the right axis displays the corresponding value for $\Gamma$. For the circular interfaces (c), the estimate of $\Gamma$ is given by averaging the data at late times $t \geq 10\unit{s}$ (dashed line). For the flat interfaces (d), in contrast, $\Gamma$ is given by fitting $at^{-2/3} + \Gamma$ (dashed line) to take into account the finite-time effect (see text).}
  \label{fig:Parameter}
 \end{center}
\end{figure}

The linear growth rate $v_\infty$ can be obtained
 as the asymptotic growth speed
 of the mean height \cite{Krug.etal-PRA1992},
 since from Eq.~\eqref{eq:Height}
\begin{equation}
 \diff{\expct{h}}{t} \simeq v_\infty + c_v t^{-2/3}  \label{eq:GrowthRate}
\end{equation}
 with a constant $c_v$.
Our experimental data are therefore plotted against $t^{-2/3}$
 in Fig.~\ref{fig:Parameter}(a,b)
 and indeed show this time dependence very clearly.
The linear regression then provides a precise estimate of $v_\infty$ at
\begin{equation}
 v_\infty = \begin{cases} 33.24(4)\unit{\mu{}m/s} & \text{(circular)}, \\ 32.75(3)\unit{\mu{}m/s} & \text{(flat)}. \end{cases}  \label{eq:EstimateVinf}
\end{equation}
Note that the estimates for the circular and flat interfaces are very close
 but significantly different.
We believe that this is because of possible tiny shift
 in the controlled temperature and of aging of our sample,
 which is a well-accepted property of MBBA
 \cite{deGennes.Prost-Book1995,Takeuchi.etal-PRE2009},
 during the days that separated the two series of the experiments.

The amplitude $\Gamma$ of the $t^{1/3}$-fluctuations is best estimated
 in our data from the amplitude of the second-order cumulant $\cum{h^2}$,
 or equivalently from the overall width $W(t)$,
 by using the relation $\cum{h^2} \simeq (\Gamma t)^{2/3} \cum{\chi^2}$.
In this method we need to set the variance of $\chi$ at an arbitrary value.
Given that this choice does not affect the following results
 as we confirmed in our preceding letter
 (Supplementary Information of Ref.~\cite{Takeuchi.etal-SR2011}),
 we choose such a value that we can compare $\chi$ directly
 with $\chi_{\rm GUE}$ and $\chi_{\rm GOE}$.
This is achieved by plotting $(\cum{h^2}/\cum{\chi_{\rm GUE,GOE}^2})^{3/2}/t$
 as a function of $t$ and reading its asymptotic value
 [Fig.~\ref{fig:Parameter}(c,d)].
For the circular interfaces [Fig.~\ref{fig:Parameter}(c)],
 the data are already stable from early times,
 so that the value of $\Gamma$ is obtained
 from the average taken over $t \geq 10\unit{s}$,
 which yields $\Gamma = 2.29(3) \times 10^3\unit{\mu{}m^3/s}$.
In contrast, finite-time effect is visible for the flat interfaces
 [Fig.~\ref{fig:Parameter}(d)]
 as we have already seen in their width [Fig.~\ref{fig:Beta}(a,b)].
To take it into consideration, we attempt a power-law fitting
 $a_1 t^{-\delta} + a_2$ to the experimental data
 and find that it works reasonably,
 with an estimate of the power at $\delta = 0.59(10)$.
It is noteworthy to remark that recent finite-time analyses
 of the TASEP, the PASEP and the PNG model
 \cite{Ferrari.Frings-JSP2011,Baik.Jenkins-a2011},
 as well as of the exact (curved) solution of the KPZ equation
 (see the next section for details),
 show that the finite-time effect
 in the second-order cumulant is at most $\mathcal{O}(t^{-2/3})$.
Our estimate of $\delta$ covers this power and in particular
 excludes any other multiples of $1/3$.
It is therefore reasonable to assume that the power $-2/3$
 also applies to our experimental data.
With this constraint on the power-law fitting
 and with a weight proportional to $(y-a_2)^{-2}$,
 where $y$ is the ordinate of the data and $a_2$ is the estimate
 from the preceding fit,
 we finally obtain $\Gamma = 2.15(10) \times 10^3\unit{\mu{}m^3/s}$
 for the flat interfaces.
Here the confidence interval covers such values of $\Gamma$ that
 $(y-\Gamma)$ as a function of $t$ does not apparently deviate
 from the power-law decay with exponent $-2/3$.
In summary, the parameter $\Gamma$ is estimated at
\begin{equation}
 \Gamma = \begin{cases} 2.29(3) \times 10^3\unit{\mu{}m^3/s} & \text{(circular)}, \\ 2.15(10) \times 10^3\unit{\mu{}m^3/s} & \text{(flat)}. \end{cases}  \label{eq:EstimateGamma}
\end{equation}

Note that the values of both $v_\infty$ and $\Gamma$
 are very close between the two cases.
Recalling the inevitable tiny change in the parameter values
 that may occur between the two sets of the experiments,
 we consider that the parameter values are, ideally,
 independent of the different cluster shapes.
This is also expected from the theoretical point of view,
 because the parameter $\Gamma$ in the present definition is given
 in terms of the parameters in the KPZ equation \eqref{eq:KPZEqDef}
 \cite{Prahofer.Spohn-PA2000,Kriecherbauer.Krug-JPA2010},
 as
\begin{equation}
 \Gamma = \frac{1}{2}A^2 |\lambda|,  \qquad \text{with} \quad A \equiv D/2\nu.  \label{eq:GammaDef}
\end{equation}
Concerning $v_\infty$,
 the parameter $\lambda$ of the KPZ equation \eqref{eq:KPZEqDef} is given by
 $\lambda = v_\infty''(0)$, where $v_\infty(u)$ is the asymptotic growth rate
 as a function of the average local inclination $u \equiv \expct{\prt{h}{x}}$
 \cite{Krug-JPA1989,Krug.etal-PRA1992}.
Since for the circular interfaces $v_\infty(u) = \sqrt{1+u^2}v_\infty$,
 it simply follows that
\begin{equation}
 v_\infty = \lambda.  \label{eq:VinfEqLambda}
\end{equation}
For the flat interfaces, one in principle needs to impose a tilt $u$
 to measure $v_\infty(u)$ \cite{Krug.etal-PRA1992},
 but supposing the rotational invariance of the system,
 one again finds $v_\infty(u) = \sqrt{1+u^2}v_\infty$ and thus
 Eq.~\eqref{eq:VinfEqLambda}.
This is justified by practically the same values of our estimates $v_\infty$
 in Eq.~\eqref{eq:EstimateVinf}.
In passing, the rotational invariance also implies $v_0 = 0$
 in the KPZ equation \eqref{eq:KPZEqDef},
 without the need to invoke the comoving frame.

\subsection{Distribution function} \label{sec:DistFunc}

\begin{figure}[t]
 \begin{center}
  \includegraphics[clip]{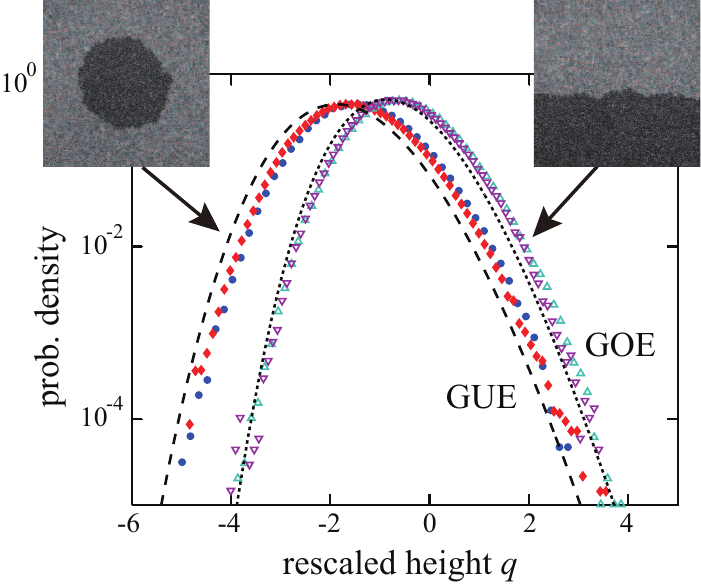}
  \caption{Histogram of the rescaled local height $q \equiv (h - v_\infty t)/(\Gamma t)^{1/3}$ for the circular (solid symbols) and flat (open symbols) interfaces. The blue circles and red diamonds display the histograms for the circular interfaces at $t = 10\unit{s}$ and $30\unit{s}$, respectively, while the turquoise up-triangles and purple down-triangles are for the flat interfaces at $t = 20\unit{s}$ and $60\unit{s}$, respectively. The dashed and dotted curves show the GUE and GOE TW distributions, respectively, defined by the random variables $\chi_{\rm GUE}$ and $\chi_{\rm GOE}$.}
  \label{fig:Dist}
 \end{center}
\end{figure}%

Using the two experimentally determined parameter values
 \eqref{eq:EstimateVinf} and \eqref{eq:EstimateGamma},
 we shall directly compare the interface fluctuations $\chi$
 with the mathematically defined random variables
 $\chi_{\rm GUE}$ and $\chi_{\rm GOE}$.
This is achieved by defining the rescaled local height
\begin{equation}
 q \equiv \frac{h - v_\infty t}{(\Gamma t)^{1/3}} \simeq \chi  \label{eq:RescaledHeightDef}
\end{equation}
 and producing its histogram for the circular and flat interfaces
 (solid and open symbols in Fig.~\ref{fig:Dist}, respectively).
The result clearly shows that
 the two cases exhibit distinct distributions,
 which are not centered nor symmetric, and hence clearly non-Gaussian.
Moreover, the histograms for the circular and flat interfaces are found,
 without any \textit{ad hoc} fitting,
 very close to the GUE (dashed curve) and GOE (dotted curve) TW distributions,
 respectively, as anticipated from their values
 of the skewness and the kurtosis
 as well as from the theoretical results for the solvable models.
The agreements are confirmed with resolution
 roughly down to $10^{-5}$ in the probability density.

\begin{figure}[t]
 \begin{center}
  \includegraphics[clip]{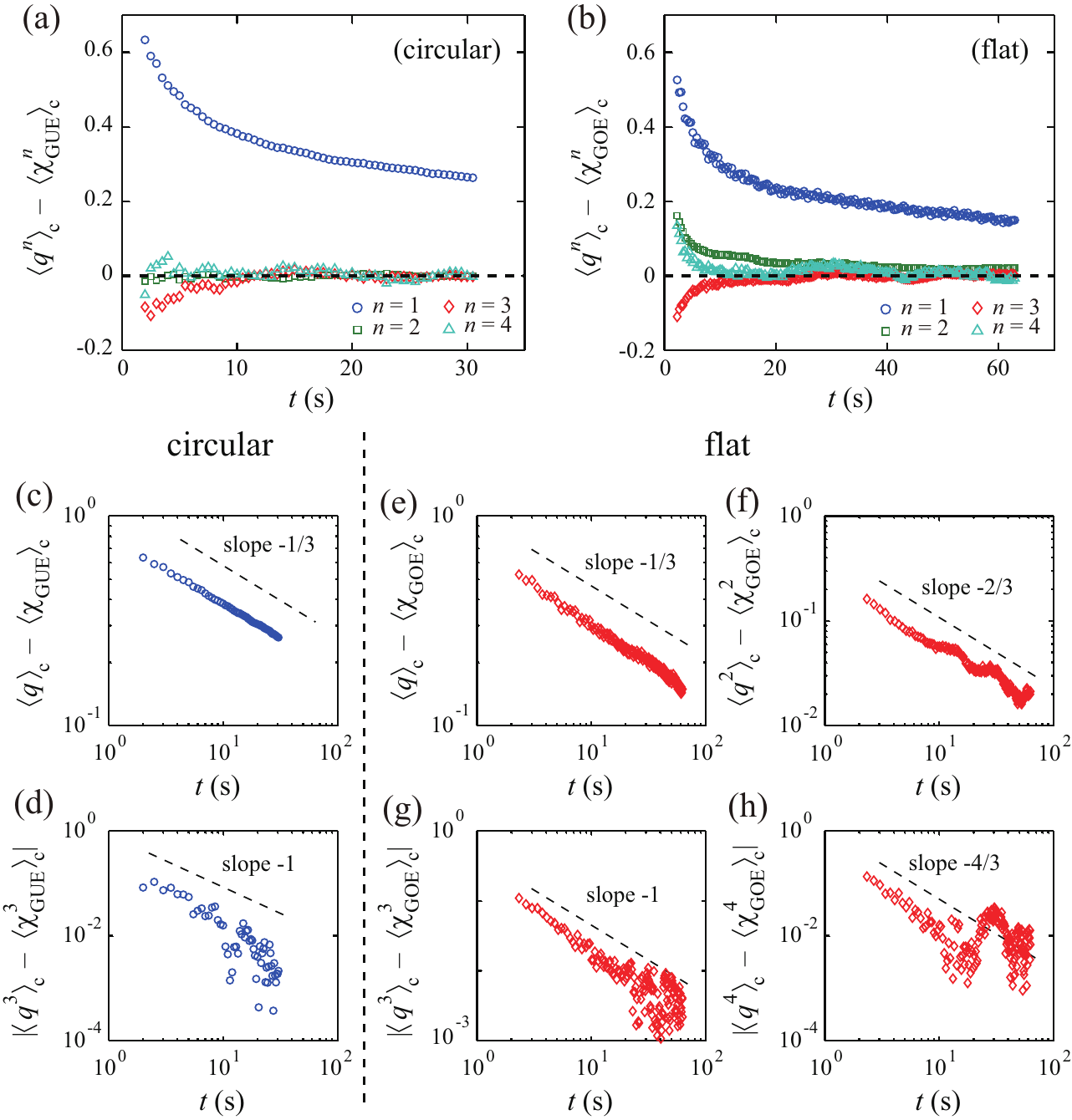}
  \caption{Finite-time effects in the $n$th-order cumulants $\cum{q^n}$. Displayed are the differences in the cumulants between the experimental data $\cum{q^n}$ and the corresponding TW distributions $\cum{\chi_{\rm GUE}^n}$ for the circular interfaces (a,c,d) and $\cum{\chi_{\rm GOE}^n}$ for the flat interfaces (b,e-h). The same data as in the panels (a,b) are shown in the panels (c-h) in the logarithmic scales with guides for the eyes showing the indicated exponent values. The absolute values of the differences are shown for the third- and fourth-order cumulants (d,g,h) in order to display both positive and negative differences, though they are essentially negative for the third-order cumulant in both cases (d,g) and positive for the fourth-order cumulant of the flat interfaces (h).}
  \label{fig:FiniteTime}
 \end{center}
\end{figure}

A closer inspection of the experimental data in Fig.~\ref{fig:Dist} reveals,
 however, a slight deviation from the theoretical curves,
 which is mostly a small horizontal translation that shrinks
 as time elapses.
To quantify this effect, we plot in Fig.~\ref{fig:FiniteTime}
 the time series of the difference between the $n$th-order cumulants
 of the measured rescaled height $q$ and those for the TW distributions.
We then find for both circular and flat interfaces
 [Fig.~\ref{fig:FiniteTime}(a,b)]
 that, indeed, the second- to fourth-order cumulants quickly converge
 to the values for the corresponding TW distributions,
 whereas the first-order cumulant $\expct{q}$, or the mean,
 shows a pronounced deviation as suggested in the histograms.
This, however, decreases in time, showing a clear power law
 proportional to $t^{-1/3}$ [Fig.~\ref{fig:FiniteTime}(c,e)],
 and thus vanishes in the asymptotic limit $t\to\infty$.
Similar finite-time corrections are in fact visible in other cumulants,
 though less clearly for higher-order ones.
The data for the flat interfaces [Fig.~\ref{fig:FiniteTime}(e-h)]
 indicate that all of the cumulants from $n=1$ to $4$ exhibit
 finite-time corrections proportional to $t^{-n/3}$.
This is also seen for the first- and third-order cumulants
 of the circular interfaces [Fig.~\ref{fig:FiniteTime}(c,d)],
 while the corrections in their second- and fourth-order cumulants
 are so small within the whole observation time
 [see Fig.~\ref{fig:FiniteTime}(a)] that we cannot recognize
 any systematic change in time,
 larger than experimental and statistical errors.
We are at present unable to explain why the finite-time corrections
 of the even-order cumulants apparently vanish for the circular interfaces.

It is intriguing to compare these experimental results with
 recent theoretical approaches on finite-time fluctuations
 in the solvable models.
Sasamoto and Spohn showed with their exact curved solution
 that the finite-time effects in the KPZ equation are essentially
 controlled by the Gumbel distribution, or more specifically,
 $q = \chi_{\rm GUE} - \gamma_t^{-1}\chi_{\rm G} + \mathcal{O}(t^{-4/3})$
 with $\gamma_t \equiv (\lambda^4 D^2 t/2^6\nu^5)^{1/3}$
 and $\chi_{\rm G}$ obeying the Gumbel probability density
 $\e^{-\chi_{\rm G}} \exp(-\e^{-\chi_{\rm G}})$
 \cite{Sasamoto.Spohn-PRL2010,Sasamoto.Spohn-NPB2010,Sasamoto-PC2012,Prolhac.Spohn-PRE2011}%
\footnote{
This can be easily shown from the finite-time expression
 of their exact solution reported
 in Refs.~\cite{Sasamoto.Spohn-PRL2010,Sasamoto.Spohn-NPB2010}.}.
This implies that the finite-time corrections in $\cum{q^n}$ are
 in the order of $t^{-n/3}$ up to $n=4$,
 which is consistent with our experimental results.
A difference, however, arises in the values of their coefficients.
In particular, for the first-order cumulant, 
 the finite-time effect in the KPZ equation is given by
 $\expct{q} - \expct{\chi_{\rm GUE}} \simeq -\gamma_t^{-1} \expct{\chi_{\rm G}} \approx -0.577 \gamma_t^{-1} \sim -t^{-1/3}$,
 which has the sign opposite to ours%
\footnote{
In Sasamoto and Spohn's solution
 \cite{Sasamoto.Spohn-PRL2010,Sasamoto.Spohn-NPB2010},
 one has, in addition to $-0.577$,
 another term that reads $2\log\alpha$ in their expression.
This however comes from their narrow wedge initial condition
 and cannot be independently chosen.
The sign of the first-order cumulant is therefore always negative
\cite{Sasamoto-PC2012}.
}
 (compare also the numerical evaluation
 in Ref.~\cite{Prolhac.Spohn-PRE2011}
 with our Fig.~\ref{fig:Dist}).
In contrast, Ferrari and Frings studied finite-time fluctuations
 of the TASEP, the PASEP and the PNG model, and showed that
 the leading correction is again $\mathcal{O}(t^{-1/3})$
 for the first-order cumulant \cite{Ferrari.Frings-JSP2011,Baik.Jenkins-a2011},
 with the positive sign for the TASEP like in our experiment.
Given that the TASEP and the KPZ equation correspond to the limit of
 strongly and weakly asymmetric growth, respectively
 \cite{Sasamoto.Spohn-PRL2010,Sasamoto.Spohn-NPB2010,Ferrari.Frings-JSP2011},
 our result on the sign of the correction in $\cum{q}$ implies
 that the DSM2 growth is also strongly asymmetric.
Concerning the higher-order cumulants,
 Ferrari and Frings showed that the correction is $\mathcal{O}(t^{-2/3})$
 for all the higher-order \textit{moments} $\expct{q^n}$
 when $q$ is appropriately shifted
 \cite{Ferrari.Frings-JSP2011,Baik.Jenkins-a2011},
 and hence \textit{at most} $\mathcal{O}(t^{-2/3})$
 for the corresponding cumulants $\cum{q^n}$.
On the numerical side, Oliveira and coworkers performed simulations
 of flat interfaces in some discrete models
 and discretized versions of the KPZ equation \cite{Oliveira.etal-PRE2012}.
They reported that the corrections in the \textit{cumulants} are 
 $\mathcal{O}(t^{-1/3})$ for the first order
 and $\mathcal{O}(t^{-2/3})$ for the second to fourth orders,
 in disagreement with the exponents we found
 for the third- and fourth-order cumulants [Fig.~\ref{fig:FiniteTime}(d,g,h)].
Although such finite-time effects are argued to be model-dependent,
 further study is clearly needed to elucidate
 these partial agreements and disagreements between our experimental results
 and the theoretical outcomes on the finite-time corrections.
An interesting conclusion that can be drawn from our finite-time analysis is
 that the random variable that should be added to Eq.~\eqref{eq:Height}
 as the leading finite-time correction term
 is statistically independent from the TW variable $\chi$,
 since otherwise the finite-time corrections
 for the second- and higher-order cumulants
 would be $\mathcal{O}(t^{-1/3})$ \cite{Ferrari.Frings-JSP2011}.

The identification of the different TW distributions
 for the two studied geometries
 implies that, at the level of the distribution function,
 the single KPZ universality class should be divided into at least two
 ``sub-universality classes'' corresponding to the curved and flat interfaces.
In the following sections, we shall argue that this splitting
 is not a particular feature for the distribution function, but
 on the contrary results in more striking differences
 in other statistical properties,
 especially in the correlation functions.

\subsection{Spatial correlation function}  \label{sec:SpaceCorrFunc}

The recent analytic developments for the solvable models are not restricted
 to the distribution function.
One of the other most studied quantities is the spatial correlation function
\begin{equation}
C_{\rm s}(l;t) \equiv \expct{h(x+l,t)h(x,t)} - \expct{h(x+l,t)}\expct{h(x,t)}.  \label{eq:SpatialCorrFuncDef}
\end{equation}
Theoretical studies have shown that in the asymptotic limit
 the spatial correlation of the interface fluctuations
 is given exactly by the \textit{time correlation} of the stochastic process
 called the Airy$_2$ process $\mathcal{A}_2(t')$ for the curved interfaces
 \cite{Prahofer.Spohn-JSP2002,Prolhac.Spohn-JSM2011a}
 and the Airy$_1$ process $\mathcal{A}_1(t')$ for the flat ones
 \cite{Sasamoto-JPA2005,Borodin.etal-CMP2008}.
The predictions in their general form read
\begin{equation}
 C_{\rm s}(l;t) \simeq (\Gamma t)^{2/3} g_i\( \frac{Al}{2}(\Gamma t)^{-2/3} \)  \label{eq:AiryCorrelation}
\end{equation}
 with $g_i(\zeta) \equiv \expct{\mathcal{A}_i(t'+\zeta)\mathcal{A}_i(t')}$,
 $A$ defined by Eq.~\eqref{eq:GammaDef}
 and $i=1$ for the flat interfaces and $2$ for the curved ones%
\footnote{
Different definitions of the Airy$_1$ process (by constant factors)
 are occasionally found in the literature.
Here, for the sake of simplicity, we adopt the definition that
 allows us to use the single mathematical expression \eqref{eq:AiryCorrelation}
 for both circular and flat interfaces.
In this definition, together with that for $\chi_{\rm GOE}$
 with the factor $2^{-2/3}$,
 we have in particular $\cum{\mathcal{A}_1^2} = \cum{\chi_{\rm GOE}^2}$.
\label{ft:Airy1Def}}.
Moreover, it is known \cite{Johansson-CMP2003}
 that the Airy$_2$ process coincides
 with the dynamics of the largest eigenvalue in Dyson's Brownian motion
 for GUE random matrices \cite{Mehta-Book2004}.
It implies that the spatial profile of a curved KPZ-class interface
 is equivalent to the locus of this largest-eigenvalue dynamics,
 to be compared with that of (usual) Brownian motion
 for the stationary interfaces
 \cite{Barabasi.Stanley-Book1995,HalpinHealy.Zhang-PR1995}.
In contrast, the Airy$_1$ process for the flat interfaces was recently found
 \textit{not} to be the largest-eigenvalue dynamics
 of Dyson's Brownian motion for GOE random matrices
 \cite{Bornemann.etal-JSP2008}.
This indicates that the statistics of the KPZ-class interfaces is not
 always connected to random matrix theory.

\begin{figure}[t]
 \begin{center}
  \includegraphics[clip]{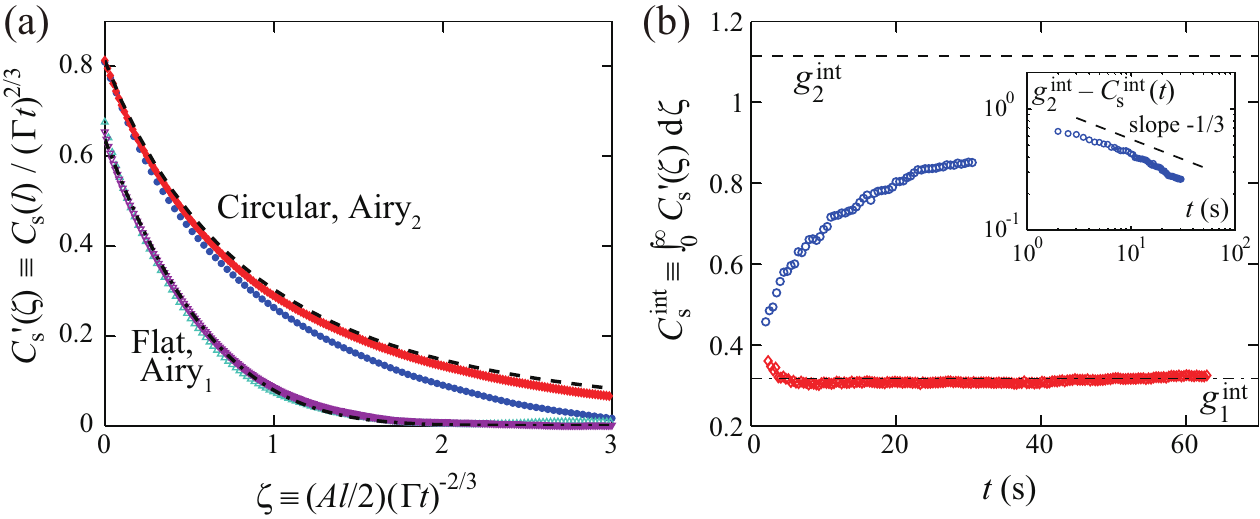}
  \caption{Spatial correlation function $C_{\rm s}(l;t)$. (a) Rescaled correlation function $C'_{\rm s}(\zeta;t) \equiv C_{\rm s}(l;t)/(\Gamma t)^{2/3}$ against rescaled length $\zeta \equiv (Al/2)(\Gamma t)^{-2/3}$. The symbols indicate the experimental data for the circular and flat interfaces (top and bottom pairs of symbols, respectively), obtained at $t = 10\unit{s}$ and $30\unit{s}$ for the former and $t=20\unit{s}$ and $60\unit{s}$ for the latter (from bottom to top). The dashed and dashed-dotted lines indicate the correlation function for the Airy$_2$ and Airy$_1$ processes, respectively, estimated numerically by Bornemann \textit{et al.} \cite{Bornemann-MC2010,Bornemann.etal-JSP2008}. (b) Integral of the rescaled correlation function $C_{\rm s}^{\rm int}(t) \equiv \int_0^\infty C'_{\rm s}(\zeta;t)\rd\zeta$ for the circular (blue circles) and flat (red diamonds) interfaces. The dashed and dashed-dotted lines indicate the values for the Airy$_2$ and Airy$_1$ processes, respectively, $g_i^{\rm int} \equiv \int_0^\infty g_i(\zeta)\rd\zeta$. The inset shows the difference $g_2^{\rm int} - C_{\rm s}^{\rm int}(t)$ for the circular interfaces, with a guide for the eyes indicating the slope $-1/3$.}
  \label{fig:SpaceCorr}
 \end{center}
\end{figure}%

We compute the spatial correlation function \eqref{eq:SpatialCorrFuncDef}
 from our experimental data and plot it in Fig.~\ref{fig:SpaceCorr}(a)
 within the rescaled axes
 $C'_{\rm s}(\zeta;t) \equiv C_{\rm s}(l;t)/(\Gamma t)^{2/3}$ against
 $\zeta \equiv (Al/2)(\Gamma t)^{-2/3}$.
Comparing the experimental results
 for the circular interfaces (top symbols), obtained at different times,
 with the Airy$_2$ correlation (dashed line),
 and the flat interfaces (bottom symbols) with the Airy$_1$ correlation
 (dashed-dotted line),
 we find both pairs in agreement at large times,
 with considerable finite-time effect for the circular interfaces.
To quantify it, we calculate the integral of the correlation function
\begin{equation}
 C_{\rm s}^{\rm int}(t) \equiv \int_0^\infty C'_{\rm s}(\zeta;t)\rd\zeta,  \label{eq:SpatialCorrIntDef}
\end{equation}
 which is in practice estimated within the range
 where $C'_{\rm s}(\zeta;t)$ is positive and decreasing,
 to get rid of statistical errors at large $\zeta$.
The result is shown in Fig.~\ref{fig:SpaceCorr}(b).
This indicates that $C_{\rm s}^{\rm int}(t)$ for the flat interfaces
 (red diamonds) reaches and stays close to the corresponding value
 $g_1^{\rm int}$ for the Airy$_1$ correlation (dashed-dotted line),
 while $C_{\rm s}^{\rm int}(t)$ for the circular interfaces (blue circles)
 is still approaching the value $g_2^{\rm int}$
 for the Airy$_2$ correlation (dashed line).
The difference however decreases in time as $t^{-1/3}$
 [inset of Fig.~\ref{fig:SpaceCorr}(b)]
 and therefore vanishes in the limit $t\to\infty$.

In short, we find that
 the spatial correlation of the circular and flat interfaces is indeed given,
 in the asymptotic limit, by the Airy$_2$ and Airy$_1$ correlations,
 respectively, confirming the universal correlation functions
 of the KPZ-class interfaces.
We note that this actually implies \textit{qualitative} difference
 between the circular and flat cases;
 it is theoretically known that the Airy$_2$ correlation
 for the circular interfaces decreases as $g_2(\zeta) \sim \zeta^{-2}$
 for large $\zeta$,
 while the Airy$_1$ correlation $g_1(\zeta)$ for the flat interfaces decays
 faster than exponentially \cite{Bornemann.etal-JSP2008}.

\subsection{Temporal correlation function}  \label{sec:TimeCorrFunc}

In contrast to the spatial correlation,
 correlation in time axis is a statistical property that
 has not been solved yet by analytical means.
It is characterized by the temporal correlation function
\begin{equation}
C_{\rm t}(t,t_0) \equiv \expct{h(x,t)h(x,t_0)} - \expct{h(x,t)}\expct{h(x,t_0)}.  \label{eq:TemporalCorrFuncDef}
\end{equation}
The temporal correlation should be measured along the directions in which
 fluctuations propagate in space-time, called the characteristic lines
 \cite{Ferrari-JSM2008,Corwin.etal-AIHPBPS2012}.
In our experiment, these are simply perpendicular to the mean spatial profile
 of the interfaces, namely the upward and radial directions
 for the flat and circular interfaces, respecitvely,
 which are represented in both cases
 by the fixed $x$ in Eq.~\eqref{eq:TemporalCorrFuncDef}.

\begin{figure}[t]
 \begin{center}
  \includegraphics[clip]{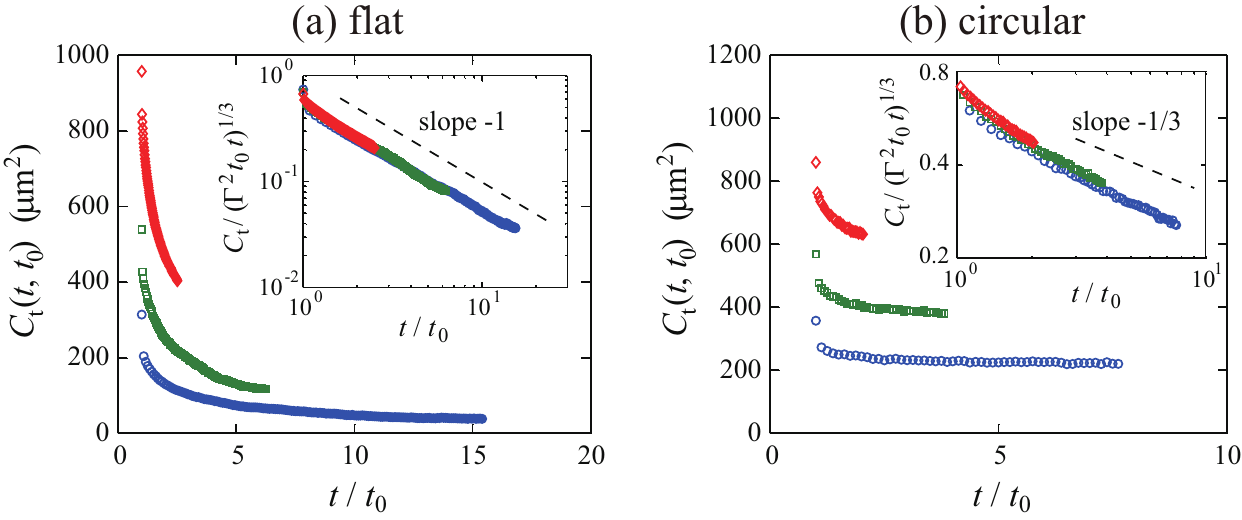}
  \caption{Temporal correlation function $C_{\rm t}(t,t_0)$ for the flat (a) and circular (b) interfaces, measured with different $t_0$. The reference time $t_0$ is $t_0 = 4\unit{s}, 10\unit{s}, 25\unit{s}$ for the flat case (a) and $t_0 = 4\unit{s}, 8\unit{s}, 15\unit{s}$ for the circular case (b) from bottom to top. The insets test the scaling ansatz \eqref{eq:TemporalCorrFuncAnsatz}. The dashed lines are guides for the eyes showing the indicated slopes.}
  \label{fig:TimeCorr1}
 \end{center}
\end{figure}%

Figure \ref{fig:TimeCorr1} displays the experimental results
 for $C_{\rm t}(t,t_0)$, obtained with different $t_0$
 for the flat (a) and circular (b) interfaces.
Here, again, we find different functional forms for the two cases;
 with increasing $t$,
 the temporal correlation function $C_{\rm t}(t,t_0)$ decays
 toward zero for the flat interfaces [Fig.~\ref{fig:TimeCorr1}(a)],
 while that for the circular ones seems to remain strictly positive
 indefinitely [Fig.~\ref{fig:TimeCorr1}(b)].
Recalling that the interface fluctuations grow as $(\Gamma t)^{1/3}$,
 one may argue that a natural scaling form for $C_{\rm t}(t,t_0)$
 would be
\begin{equation}
 C_{\rm t}(t,t_0) \simeq (\Gamma^2 t_0 t)^{1/3} F_{\rm t}(t/t_0),  \label{eq:TemporalCorrFuncAnsatz}
\end{equation}
 with a scaling function $F_{\rm t}$.
This indeed works for the flat interfaces,
 showing a long-time behavior $F_{\rm t}(t/t_0) \sim (t/t_0)^{-\bar\lambda}$
 with $\bar\lambda = 1$ [inset of Fig.~\ref{fig:TimeCorr1}(a)]
 in agreement with past numerical studies
 \cite{Kallabis.Krug-EL1999,Henkel.etal-PRE2012}.
For the circular interfaces, in contrast,
 the natural scaling \eqref{eq:TemporalCorrFuncAnsatz}
 does not seem to hold as well within our time window,
 as the data with different $t_0$ do not overlap on a single curve
 [inset of Fig.~\ref{fig:TimeCorr1}(b)].
The autocorrelation exponent $\bar\lambda$ would be formally $1/3$
 in this case, but this only reflects the observation that
 the unscaled $C_{\rm t}(t,t_0)$ converges to a non-zero value at $t\to\infty$.
These results support Kallabis and Krug's conjecture
 \cite{Kallabis.Krug-EL1999} that the following scaling relations
 for the linear growth equations \cite{Kallabis.Krug-EL1999,Singha-JSM2005}
 also hold for nonlinear growth processes, especially in the KPZ class:
\begin{equation}
 \bar\lambda = \begin{cases} \beta + d/z & \text{(flat)}, \\ \beta & \text{(circular)}, \end{cases}  \label{eq:BarLambda}
\end{equation}
 where $d$ is the spatial dimension.
Given the close relation between the temporal correlation function
 and the response function,
 as well as the recent interest in their aging dynamics
 \cite{Henkel.etal-PRE2012},
 the geometry dependence of the response function
 would be an interesting property that should be addressed in future studies.

\begin{figure}[t]
 \begin{center}
  \includegraphics[clip]{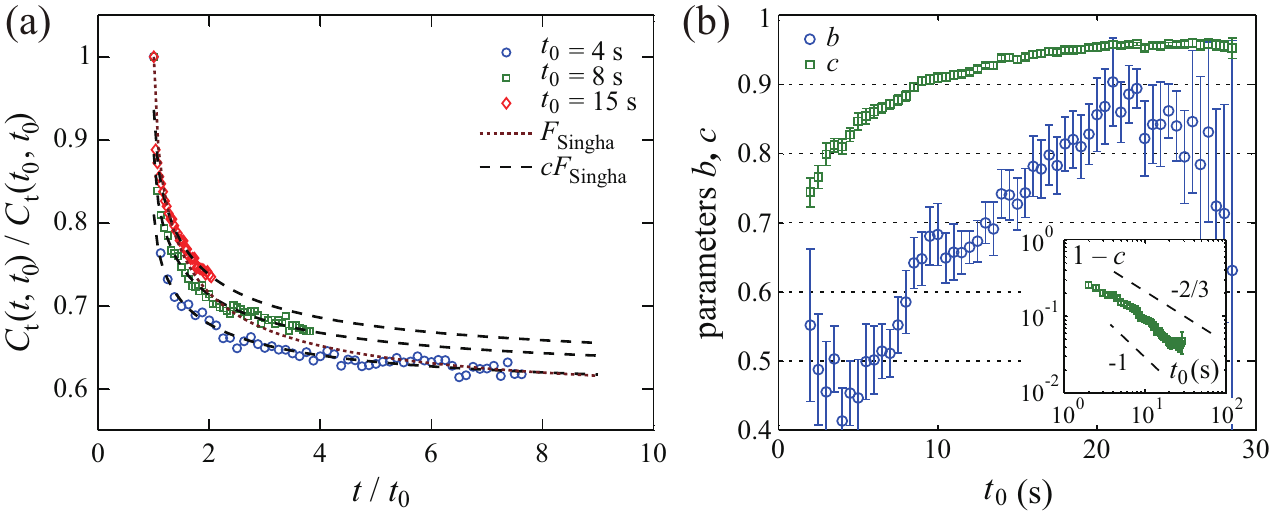}
  \caption{Functional form of the temporal correlation function $C_{\rm t}(t,t_0)$ for the circular interfaces. (a) $C_{\rm t}(t,t_0) / C_{\rm t}(t_0,t_0)$ against $t/t_0$ with different $t_0$ (as shown in the legend). The brown dotted curve indicates the result of the best fitting obtained with Singha's original equation \eqref{eq:Singha1}, while the black dashed ones display those for the modified equation \eqref{eq:Singha2} with the leftmost data point excluded. Notice that the correlation function remains far above zero. (b) The values of the fitting parameters $b$ and $c$ in Eq.~\eqref{eq:Singha2} as functions of $t_0$. The inset shows $1-c$ against $t_0$ in the logarithmic scales. The two dashed lines are guides for the eyes showing the slope $-2/3$ and $-1$.}
  \label{fig:TimeCorr2}
 \end{center}
\end{figure}%

Concerning the temporal correlation $C_{\rm t}(t,t_0)$
 for the circular interfaces,
 Singha performed an approximative theoretical calculation
 for the KPZ class, which yields \cite{Singha-JSM2005}
\begin{align}
 &\frac{C_{\rm t}(t,t_0)}{C_{\rm t}(t_0,t_0)} \approx F_{\rm Singha}(t/t_0; b), \label{eq:Singha1} \\
 &F_{\rm Singha}(x; b) \equiv \frac{\e^{b(1-1/\sqrt{x})} \Gamma(2/3, b(1-1/\sqrt{x}))}{\Gamma(2/3)}, \label{eq:FSinghaDef}
\end{align}
 with a single unknown parameter $b$,
 the upper incomplete Gamma function
 $\Gamma(s,x) \equiv \int_x^\infty y^{s-1} \e^{-y} \rd y$
 and the Gamma function $\Gamma(s) = \Gamma(s,0)$.
This however does not fit our experimental results,
 as shown by the brown dotted curve in Fig.~\ref{fig:TimeCorr2}(a).
Instead, we find it sufficient to introduce
 an arbitrary ($t_0$-dependent) factor $c$
 to Eq.~\eqref{eq:Singha1}, i.e.,
\begin{equation}
 \frac{C_{\rm t}(t,t_0)}{C_{\rm t}(t_0,t_0)} \approx cF_{\rm Singha}(t/t_0; b) \qquad (t \neq t_0) \label{eq:Singha2}
\end{equation}
 in order to fit all the experimental curves
 as exemplified by the black dashed curves in Fig.~\ref{fig:TimeCorr2}(a),
 except for $t=t_0$ at which the left and right hand sides are
 equal to $1$ and $c$, respectively, by construction.
This suggests that the experimentally obtained correlation function
 includes a fast decaying term as follows:
\begin{equation}
 C_{\rm t}(t,t_0) \sim t_0^{2/3} \[ (1-c) F_{\rm fast}(t-t_0; t_0) + cF_{\rm Singha}(t/t_0; b)\] ,  \label{eq:Singha3}
\end{equation}
 with a function $F_{\rm fast}(t-t_0; t_0)$
 that satisfies $F_{\rm fast}(0;t_0) = 1$
 and decays much faster than the data interval in Fig.~\ref{fig:TimeCorr2}(a),
 namely $0.5\unit{s}$.
The best nonlinear fits of Eq.~\eqref{eq:Singha2} to our experimental data
 provide $t_0$-dependent values for the two fitting parameters $b$ and $c$
 as shown in Fig.~\ref{fig:TimeCorr2}(b).
From those obtained at late times,
 we roughly estimate
 $b \approx 0.8(1)$,
 while the factor $c$ increases with time toward one,
 perhaps by a power law, $1-c \sim t_0^{-\delta'}$ with $\delta' \in [2/3,1]$,
 as shown in the inset of Fig.~\ref{fig:TimeCorr2}(b)%
\footnote{
Note that for larger $t_0$ we have less data points for $C_{\rm t}(t,t_0)$
 and thus the estimates of $b$ and $c$ have larger uncertainties.}.
If $\delta' = 2/3$, we have
 $C_{\rm t}(t,t_0) \sim c' F_{\rm fast}(t-t_0; t_0) + t_0^{2/3} c(t_0) F_{\rm Singha}(t/t_0; b)$ with a constant $c'$, which supports an expectation
 that $F_{\rm fast}(t-t_0; t_0)$ is a microscopic contribution
 decoupled from the macroscopic evolution of the interfaces.
In any case, it is this nontrivial $t_0$-dependence of the factor $c$
 that has apparently hindered a successful rescaling
 by the ansatz \eqref{eq:TemporalCorrFuncAnsatz}
 in Fig.~\ref{fig:TimeCorr1}(b).
In fact, Eqs.~\eqref{eq:FSinghaDef} and \eqref{eq:Singha2}
 with $c \to 1$ and a constant $b$
 imply that the ansatz \eqref{eq:TemporalCorrFuncAnsatz}
 asymptotically holds.
Moreover, it also follows that a non-zero correlation remains forever,
 or
 $\displaystyle{\lim_{t\to\infty} C_{\rm t}(t,t_0) \sim t_0^{2/3} > 0}$,
 and thus $\bar\lambda = 1/3$ for the circular interfaces.
We finally note that a similar result is obtained with the correlation function
 defined with the samplewise average,
 $C_{\rm t}(t,t_0) \equiv \expct{(h(x,t) - \expct{h(x,t)}_{\rm s})(h(x,t_0) - \expct{h(x,t_0)}_{\rm s})}$,
 except that the value of $b$ changes to $1.2(2)$ in this case.

\begin{figure}[t]
 \begin{center}
  \includegraphics[clip]{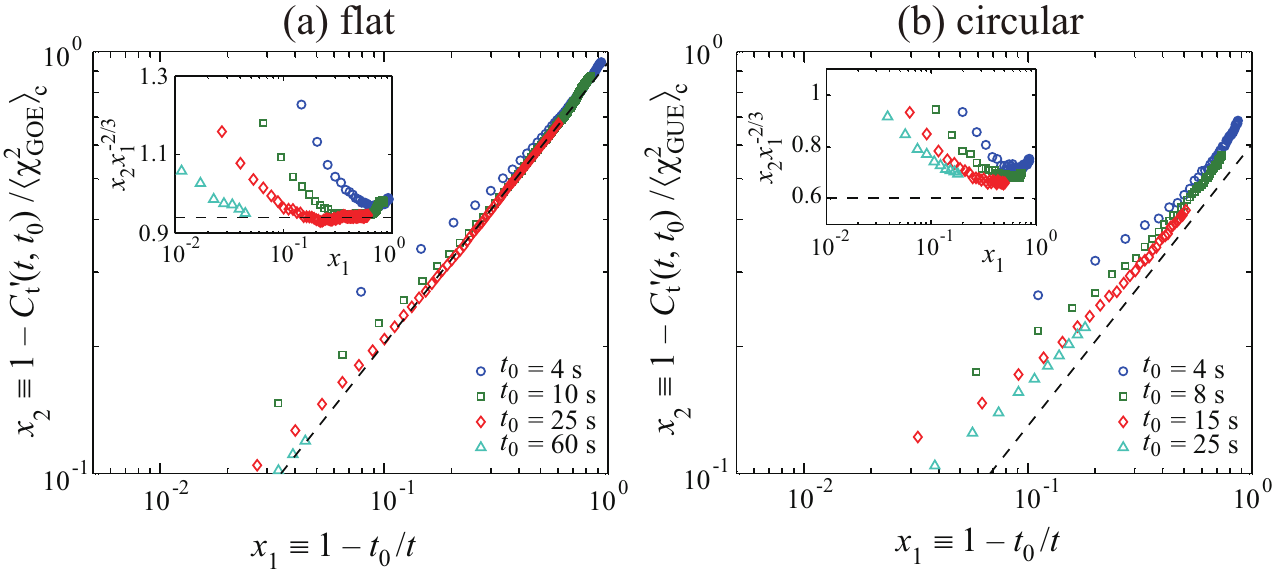}
  \caption{Short-time behavior of the temporal correlation function $C_{\rm t}(t,t_0)$ for the flat (a) and circular (b) interfaces. The main panels show $x_2 \equiv 1-C'_{\rm t}(t,t_0) / \cum{\chi^2}$ against $x_1 \equiv 1-t_0/t$ with $C'_{\rm t}(t,t_0) \equiv C_{\rm t}(t,t_0) / (\Gamma^2 t_0t)^{1/3}$ and the corresponding $\chi$ for each case. The insets show the same data in the rescaled axes, $x_2 x_1^{-2/3}$ against $x_1$. The dashed lines indicate the right hand side of Eq.~\eqref{eq:TemporalCorrFuncShort} with the value of $R$ estimated from the shown data (a) or from the value of $b$ (b).}
  \label{fig:TimeCorrShort}
 \end{center}
\end{figure}%

Short-time behavior of the temporal correlation function is also of interest.
It is known that
 the leading terms are given, in our notation, by
\begin{equation}
 C_{\rm t}(t,t_0) \simeq (\Gamma^2 t_0 t)^{1/3} \cum{\chi^2} \[ 1-\frac{R}{2}\(1-\frac{t_0}{t}\)^{2/3}\],  \label{eq:TemporalCorrFuncShort}
\end{equation}
 for $t-t_0 \ll t_0$, with a
 coefficient $R$ which is universal at least for the flat interfaces
 \cite{Krug.etal-PRA1992,Kallabis.Krug-EL1999}.
Our data for the flat case indeed confirm this
 [Fig.~\ref{fig:TimeCorrShort}(a)],
 providing an estimate $R/2 \approx 0.94$ in agreement
 with the value found in past numerical studies
 $R = 1.8(1)$ \cite{Krug.etal-PRA1992,Kallabis.Krug-EL1999}.
This short-time behavior is less clearly seen for the circular interfaces
 [Fig.~\ref{fig:TimeCorrShort}(b)]
 because of the $t_0$-dependence of the factor $c$,
 but assuming Eq.~\eqref{eq:Singha2} and $c \to 1$ for $t_0\to\infty$,
 we have $R/2 \approx 3b^{2/3}/2^{5/3}\Gamma(2/3)$ \cite{Singha-JSM2005}.
Our estimate $b \approx 0.8(1)$ then yields $R/2 \approx 0.6$,
 which is indicated by the dashed lines in Fig.~\ref{fig:TimeCorrShort}(b)
 and is significantly different from the value for the flat case.
It is not clear if the coefficient $R$ and the parameter $b$
 are universal in the circular case,
 because the former is given as an amplitude ratio
 of the second-order cumulants of the growing and stationary interfaces,
 the latter of which is never attained in the circular case.
Simulations of an off-lattice Eden model by one of the authors
 \cite{Takeuchi-JSM2012} yield $b=1.22(8)$
 and $R/2 \approx 0.68$ (estimated independently of $b$),
 to be compared with $b \approx 0.8(1)$
 and $R/2 \approx 0.6$ in our experiment.

\subsection{Temporal persistence}

Dynamic aspects of the stochastic processes are not fully characterized
 by the two-point correlation function.
Another interesting and useful property in this context
 constitutes first-passage problems \cite{Majumdar-CS1999},
 which concern stochastic time
 at which a given event occurs for the first time.
A central quantity of interest is the persistence probability,
 which is the probability that the local fluctuations
 do not change sign up to time $t$.
It is known to exhibit simple power-law decay
 $P_\pm(t,t_0) \sim t^{-\theta_\pm}$ in various general situations
 such as critical behavior and phase separation,
 though it is rarely solved by analytic means since it involves
 infinite-point correlation \cite{Majumdar-CS1999}.
It has also been studied for fluctuating interfaces,
 first for linear processes \cite{Krug.etal-PRE1997}
 and then for KPZ-class interfaces
 \cite{Kallabis.Krug-EL1999,Singha-JSM2005,Merikoski.etal-PRL2003},
 without analytic results in the latter case.
The present section is devoted to showing experimental results
 on this nontrivial quantity.

\begin{figure}[t]
 \begin{center}
  \includegraphics[clip]{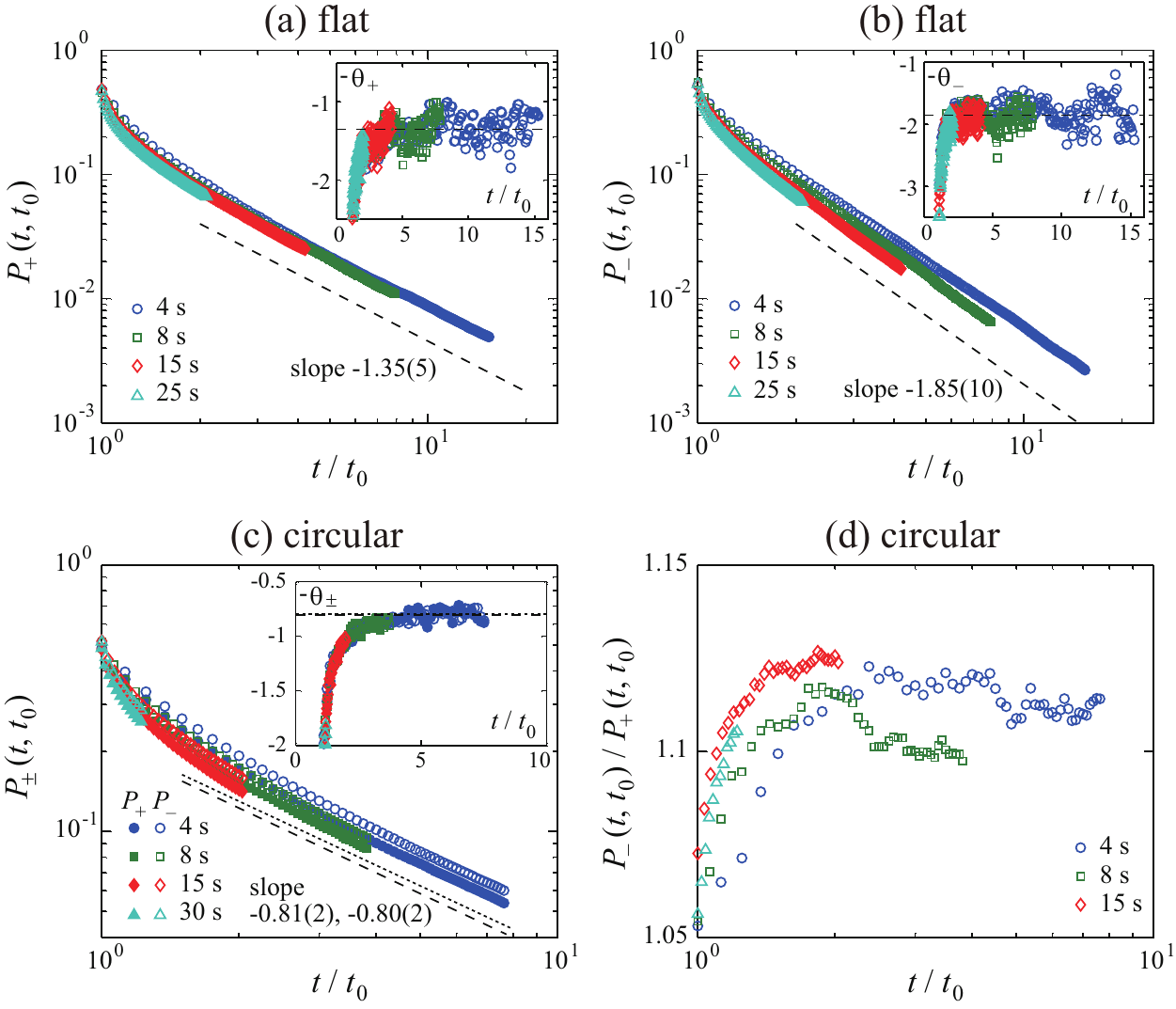}
  \caption{Temporal persistence probabilities $P_\pm(t,t_0)$ for the flat (a,b) and circular (c,d) interfaces, with different $t_0$ as shown in the legends. The panels (a-c) shows the positive and negative persistence probabilities as functions of $t/t_0$, while the panel (d) indicates the ratio of the positive and negative persistence probabilities shown in (c), which confirms $\theta_+ = \theta_-$ for the circular interfaces. The insets show the running exponents $-\theta_\pm(t,t_0) \equiv \rd[\log P_\pm(t,t_0)]/\rd[\log(t/t_0)]$ averaged over $1.0\unit{s}$ in the unit of $t$. Note that, in the panel (c), including its inset, the data for both positive and negative fluctuations are shown (solid and open symbols, respectively). The dashed lines show the indicated values of the persistence exponents $\theta_\pm$. In the panel (c), the dashed and dotted lines correspond to $\theta_+ = 0.81(2)$ and $\theta_- = 0.80(2)$, respectively.}
  \label{fig:TimePersistence}
 \end{center}
\end{figure}%

We define the persistence probability $P_\pm(t,t_0)$
 as the joint probability that the interface fluctuation
 $\delta h(x,t) \equiv h(x,t) - \expct{h}$
 at a fixed location is positive (negative) at time $t_0$
 and maintains the same sign until time $t$.
Figure \ref{fig:TimePersistence}(a-c) displays
 the results for the positive and negative fluctuations,
 $P_+(t,t_0)$ and $P_-(t,t_0)$, respectively,
 with varying $t$ and fixed values of $t_0$.
We find, for both flat and circular interfaces, that
 the persistence probabilities indeed decay algebraically for large $t$:
\begin{equation}
 P_\pm(t,t_0) \sim (t/t_0)^{-\theta_\pm}  \label{eq:PersistenceProb}
\end{equation}
 with the persistence exponents $\theta_\pm$,
 which can be different
 for the positive and negative fluctuations.
In order to check the quality of the power-law decays
 and to measure the persistence exponents, we plot in the insets
 the running exponents
 $\theta_\pm(t,t_0) \equiv -\rd[\log P_\pm(t,t_0)]/\rd[\log(t/t_0)]$
 as functions of $t/t_0$.
The values of $\theta_\pm(t,t_0)$ with different $t_0$
 turn out to overlap in this representation, and, in particular,
 converge to constants, which substantiate
 the power laws \eqref{eq:PersistenceProb}
 with well-defined time-independent exponents $\theta_\pm$.
We estimate them at
\begin{equation}
 \begin{cases} \theta_+ = 1.35(5) \\ \theta_- = 1.85(10) \end{cases} \text{(flat)} \qquad \text{and} \qquad \begin{cases} \theta_+ = 0.81(2) \\ \theta_- = 0.80(2) \end{cases} \text{(circular)},   \label{eq:EstimateTimePersistence}
\end{equation}
 which are remarkably different between the two cases.
In passing, if we use the samplewise average to define the sign, i.e.,
 $\delta h(x,t) \equiv h(x,t) - \expct{h}_{\rm s}$,
 we find $\theta_+ = 0.88(2)$ and $\theta_- = 0.91(2)$
 for the circular case.
We believe however that the physically relevant figures are
 the preceding ones \eqref{eq:EstimateTimePersistence}
 obtained with the ensemble average.

The persistence exponents have also been measured in a few past studies.
For the flat interfaces,
 numerical estimates of $\theta_+ = 1.21(6)$ and $\theta_- = 1.61(8)$
 were reported for a model in the KPZ class with $\lambda>0$,
 and $\theta_+ = 1.67(7)$ and $\theta_- = 1.15(8)$ for $\lambda<0$
 \cite{Kallabis.Krug-EL1999}.
We find small discrepancies from our values \eqref{eq:EstimateTimePersistence}
 for both larger and smaller exponents,
 but these are probably due to the different choice
 of the reference time $t_0$:
 in the numerical study \cite{Kallabis.Krug-EL1999},
 $t_0$ was fixed to be a single Monte Carlo step
 after the flat initial condition and therefore not yet in the KPZ regime.
The persistence was also measured in the slow combustion experiment
 \cite{Merikoski.etal-PRL2003};
 although for the stationary interfaces
 the authors found a power-law decay in agreement with theory,
 for the growing ones,
 they found no asymmetry between $P_+(t,t_0)$ and $P_-(t,t_0)$
 and could not identify power-law decays within their time window.
For the circular interfaces, past simulations of an on-lattice Eden model
 reported $\theta_+ = 0.88(2)$ and $\theta_- = 0.80(2)$ \cite{Singha-JSM2005},
 which are reasonably close to our estimates \eqref{eq:EstimateTimePersistence}.
After the comparison to these earlier studies, however,
 the most important finding we reach in the present work is that,
 while the positive and negative persistence probabilities are asymmetric
 for the flat interfaces, this broken symmetry is somehow recovered
 for the circular case [compare $\theta_+$ and $\theta_-$
 in Eq.~\eqref{eq:EstimateTimePersistence}].
This is clearly confirmed by plotting the ratio $P_-(t,t_0)/P_+(t,t_0)$
 in Fig.~\ref{fig:TimePersistence}(d),
 which asymptotically shows a plateau indicating $\theta_+ = \theta_-$.
A similar result is also obtained
 by one of the authors' simulations of an off-lattice Eden model,
 namely $\theta_+ = 0.81(3)$ and $\theta_- = 0.77(4)$
 \cite{Takeuchi-JSM2012}.
The asymmetry in the flat case is naturally attributed
 to the nonlinear term of the KPZ equation \eqref{eq:KPZEqDef},
 which pulls back negative fluctuations and pushes forward positive humps,
 leading to $\theta_+ < \theta_-$ if $\lambda > 0$
 (and contrary if $\lambda < 0$) \cite{Kallabis.Krug-EL1999}.
We have no explanation why and how this asymmetry is cancelled
 for the circular interfaces.

\subsection{Spatial persistence}  \label{sec:SpatialPersistence}

Similarly to the temporal persistence studied in the preceding section,
 we can also argue a persistence property in space,
 which is known to be nontrivial as well
\cite{Majumdar.Bray-PRL2001,Constantin.etal-PRE2004,Majumdar.Dasgupta-PRE2006}.
It is quantified by the spatial persistence probability%
\footnote{
The persistence probability argued in the present paper is
 sometimes called the survival probability in the literature.
In this case, the persistence probability indicates
 a related but different quantity; it is defined in terms of
 fluctuations from the leftmost (or rightmost) height
 of each segment of the interfaces,
 instead of the global average $\expct{h}$
 \cite{Constantin.etal-PRE2004,Majumdar.Dasgupta-PRE2006}.
We, however, define the persistence probability
 with the global average,
 in order to be consistent with the definition
 of the temporal persistence probability for the growing interfaces
 \cite{Kallabis.Krug-EL1999,Singha-JSM2005,Merikoski.etal-PRL2003},
 for which one \textit{needs} to use the global average
 since the height grows.
Our definition is also more common, as far as we follow,
 in other subjects such as critical phenomena
 \cite{Majumdar-CS1999,Takeuchi.etal-PRE2009}.}
 $P_\pm^{\rm (s)}(l;t)$,
 defined as the probability that a positive (negative) fluctuation
 continues over length $l$ in a spatial profile of the interfaces at time $t$.
For the stationary interfaces in the KPZ class,
 since their spatial profile is equivalent to
 the one-dimensional Brownian motion,
 their spatial persistence is mapped
 to the temporal persistence of the Brownian motion
 \cite{Majumdar.Bray-PRL2001,Majumdar.Dasgupta-PRE2006}
\footnote{
It is interesting to note that,
 for general cases of the \textit{linear} growth equation
 (with an arbitrary value of the dynamic exponent $z$),
 stationary interfaces are equivalent
 to the \textit{fractional} Brownian motion,
 which allows us to compute the value of the persistence exponent exactly
 \cite{Majumdar.Bray-PRL2001}.}.
Its analogue for the growing interfaces has been, however,
 studied only in the slow combustion experiment to our knowledge,
 which reported a power law
 $P_\pm^{\rm (s)}(l;t) \sim l^{-1/2}$ \cite{Merikoski.etal-PRL2003}.

\begin{figure}[t]
 \begin{center}
  \includegraphics[clip]{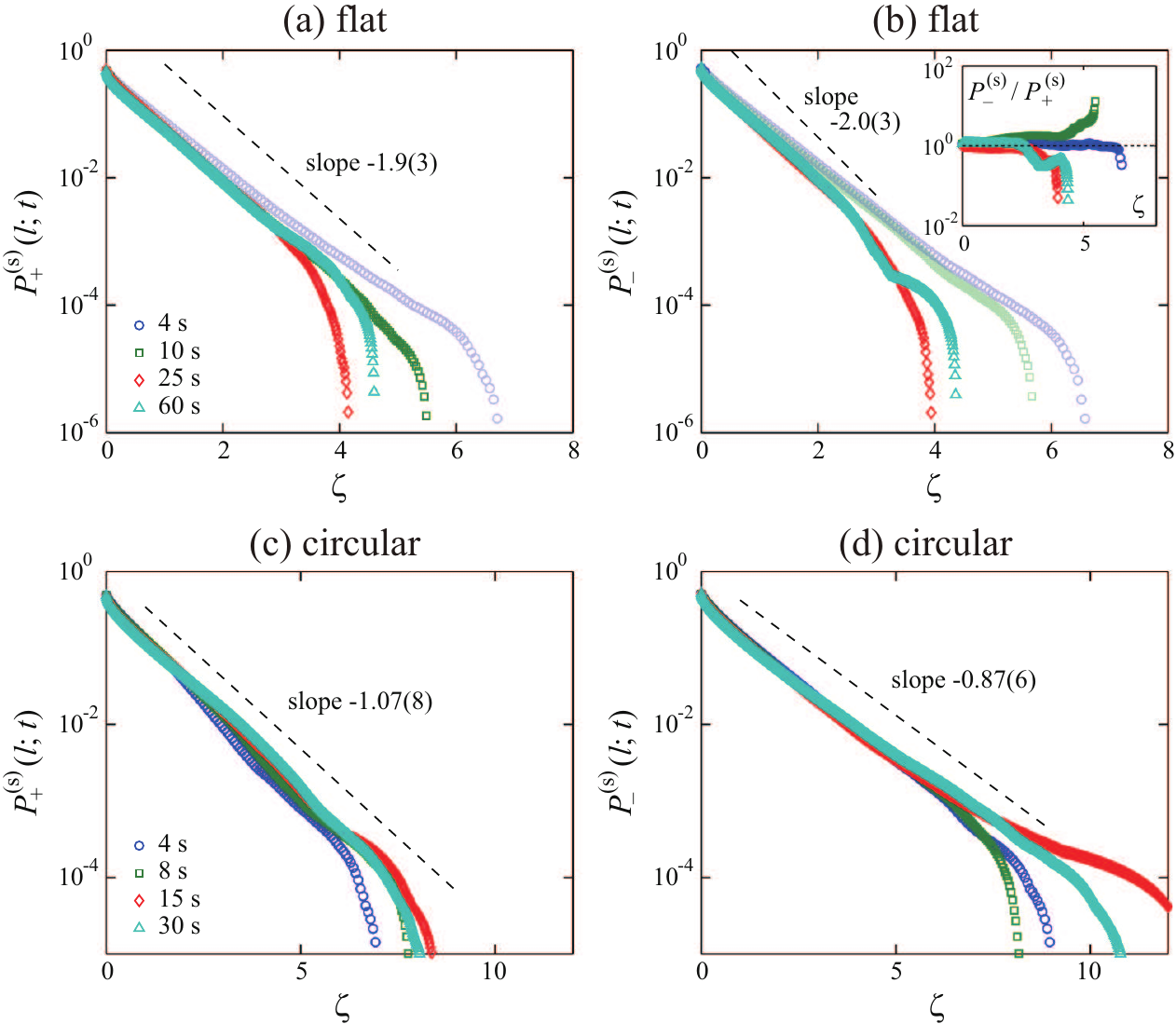}
  \caption{Spatial persistence probabilities $P^{\rm (s)}_\pm(l;t)$ for the flat (a,b) and circular (c,d) interfaces, with different $t$ as shown in the legends. The length $l$ is shown in the rescaled unit $\zeta \equiv (Al/2)(\Gamma t)^{-2/3}$ [also notice the different scales between (a,b) and (c,d)]. Data for early times that deviate from the asymptotic exponential decay are shown by light-color symbols for the sake of visibility. The dashed lines are guides for the eyes indicating the estimated values of the decay coefficients $\kappa^{\rm (s)}_\pm$. The inset of the panel (b) displays the ratio $P^{\rm (s)}_-(l;t) / P^{\rm (s)}_+(l;t)$ for the flat interfaces, which suggests $\kappa^{\rm (s)}_+ = \kappa^{\rm (s)}_-$ in this case.}
  \label{fig:SpacePersistence}
 \end{center}
\end{figure}%

We measure this quantity for our growing interfaces
 and plot it in Fig.~\ref{fig:SpacePersistence},
 for both positive [panels (a,c)] and negative (b,d) fluctuations
 as well as for the flat (a,b) and circular (c,d) interfaces.
In all these cases, we identify exponential decays
 within our statistical accuracy, instead of any power laws,
 as opposed to the slow combustion experiment \cite{Merikoski.etal-PRL2003}.
This is simply written as
\begin{equation}
 P_\pm^{\rm (s)}(l;t) \sim \e^{-\kappa^{\rm (s)}_\pm \zeta},  \label{eq:SpatialPersistence}
\end{equation}
 with the dimensionless length scale $\zeta \equiv (Al/2)(\Gamma t)^{-2/3}$.
We confirm that the coefficients $\kappa^{\rm (s)}_\pm$ defined thereby
 do not depend on time for large $t$ (Fig.~\ref{fig:SpacePersistence}).
Their values, however, do depend on the global shape of the interfaces,
 like in other statistical properties we have studied so far.
Using the data at five different times near the end of the time series,
 namely $60\unit{s} \leq t \leq 63\unit{s}$
 and $28.5\unit{s} \leq t \leq 30.5\unit{s}$
 for the flat and circular interfaces, respectively, we find
\begin{equation}
 \begin{cases} \kappa^{\rm (s)}_+ = 1.9(3) \\ \kappa^{\rm (s)}_- = 2.0(3) \end{cases} \text{(flat)} \qquad \text{and} \qquad \begin{cases} \kappa^{\rm (s)}_+ = 1.07(8) \\ \kappa^{\rm (s)}_- = 0.87(6) \end{cases} \text{(circular)}.   \label{eq:EstimateSpacePersistence}
\end{equation}
We notice here, besides the clear geometry dependence,
 the equality $\kappa^{\rm (s)}_+ = \kappa^{\rm (s)}_-$
 for the flat interfaces.
It is also supported by plotting the ratio
 $P_-^{\rm (s)}(l;t)/P_+^{\rm (s)}(l;t)$
 in the inset of Fig.~\ref{fig:SpacePersistence}(b),
 which shows no significant increase or decrease in a systematic manner.
In contrast, for the circular case, we recognize a slight asymmetry
 between the positive and negative fluctuations,
 though we do not reach a definitive conclusion on it as mentioned below.
We also remark that apparent deviations from the exponential decay
 in Fig.~\ref{fig:SpacePersistence} do not seem statistically significant,
 because we find that the plots can deviate both upward and downward
 without any systematic variation in time,
 even at latest consecutive times.
We however stress that the exponential decay itself is convincing
 in all our experimental data.

Having used the same rescaled length scale $\zeta$
 as for the analysis of the spatial correlation function
 (Sect.~\ref{sec:SpaceCorrFunc}),
 we expect that the values of $\kappa^{\rm (s)}_\pm$
 in Eq.~\eqref{eq:EstimateSpacePersistence} are universal
 in the $(1+1)$-dimensional KPZ class.
Moreover, given the expected equivalence of the asymptotic spatial profile
 of the flat and circular interfaces to the Airy$_1$ and Airy$_2$ processes
 \cite{Prahofer.Spohn-JSP2002,Prolhac.Spohn-JSM2011a,Sasamoto-JPA2005,Borodin.etal-CMP2008,Bornemann.etal-JSP2008}, respectively, 
 our results may also shed light on the \textit{temporal} persistence
 of the Airy processes, as well as, for the circular case,
 that of the largest-eigenvalue dynamics in Dyson's Brownian motion
 for GUE random matrices.
The latter is studied in the appendix
 by direct simulations of GUE Dyson's Brownian motion;
 we then find exponential decay of the persistence probability
 with the rates $\kappa_+ = 0.90(8)$ and $\kappa_- = 0.90(6)$,
 which are indeed close to the values of $\kappa^{\rm (s)}_\pm$
 found for the circular interfaces.
We however recognize a small discrepancy
 between $\kappa^{\rm (s)}_+$ and $\kappa_+$
 and in particular notice that there is no asymmetry
 between the positive and negative fluctuations
 for the results of Dyson's Brownian motion.
On the one hand, this suggests that the true asymptotic values
 of $\kappa^{\rm (s)}_\pm$
 for the curved KPZ class are also identical,
 but somewhat obscured in our data because of statistical error
 and finite-time effect.
We indeed notice in Fig.~\ref{fig:SpacePersistence}(c)
 that the apparent value of $\kappa^{\rm (s)}_+$
 seems to decrease with increasing time.
On the other hand, to our knowledge,
 no theoretical study has computed the persistence probability
 for the Airy processes or for Dyson's Brownian motion;
 correspondence to the spatial persistence in the KPZ class
 should therefore be explicitly checked.
Although one of the authors numerically finds $\kappa^{\rm (s)}_+ = 0.90(2)$
 and $\kappa^{\rm (s)}_- = 0.89(4)$ for an off-lattice Eden model
 \cite{Takeuchi-JSM2012}, in quantitative agreement with the results
 for GUE Dyson's Brownian motion,
 theoretical estimation for $\kappa^{\rm (s)}_\pm$ would be necessary
 to give a firm conclusion on this issue.
In any case, it is interesting that the exponential decay
 is identified in the spatial persistence,
 especially for the circular interfaces whose two-point correlation function
 decays no faster than $g_2(\zeta) \sim \zeta^{-2}$.
This result may remind us of Newell and Rosenblatt's theorem
 \cite{Newell.Rosenblatt-AMS1962}
 for \textit{Gaussian} stationary processes,
 which states that the persistence probability decays exponentially
 if the two-point correlation function decays by a power law
 $g(\zeta) \sim \zeta^{-\mu}$ with $\mu > 1$.
It would be intriguing to investigate the possibility
 of extending Newell and Rosenblatt's theorem
 to non-Gaussian stationary processes
 such as the Airy$_1$ and Airy$_2$ processes.
Note however that, in our experiment,
 because of the limited precision of the data for the spatial persistence,
 we cannot exclude the possibility
 of superexponential decay for large length scales $\zeta$
 (Fig.~\ref{fig:SpacePersistence}),
 in particular for the flat case corresponding to the Airy$_1$ process.

\subsection{Extreme-value statistics}  \label{sec:ExtremeValueStat}

\begin{figure}[t]
 \begin{center}
  \includegraphics[clip]{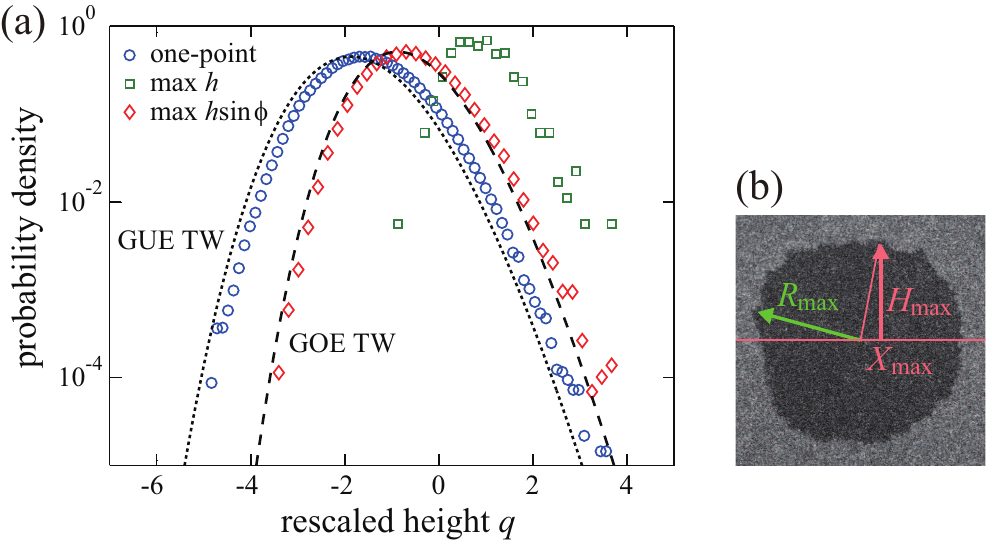}
  \caption{Maximal height distributions for the circular interfaces at $t=30\unit{s}$. (a) The distributions of the local radius $h$ (blue circles), the maximal radius $R_{\rm max} \equiv \max h$ (green squares) and the maximal height on fictitious substrates (see text), $H_{\rm max} \equiv \max (h \sin\phi)$ (red diamonds), are shown in the rescaled axis $q \equiv (h - v_\infty t)/(\Gamma t)^{1/3}$. The dashed and dotted lines indicate the GOE and GUE TW distributions, respectively (with the factor $2^{-2/3}$ for the former). (b) Sketch of the definitions of the maximal radius $R_{\rm max}$ and the maximal height $H_{\rm max}$. $X_{\rm max}$ indicates the position on the fictitious substrate that gives the maximal height $H_{\rm max}$.}
  \label{fig:MaxHeight1}
 \end{center}
\end{figure}

In this section, we turn our focus to extreme-value statistics
 \cite{Gumbel-Book1958,Clusel.Bertin-IJMPB2008}
 for the interface fluctuations, studying in particular
 the distribution of their maximal values.
First of all, let us note that the TW distributions are unbounded.
The maximal height is therefore not an intrinsic quantity
 for the flat interfaces, for which the system size (or the lateral size) $L$
 can be taken arbitrarily large at any times.
For this reason we focus here only on the circular interfaces,
 for which $L$, or the circumference, is finite at finite times
 and therefore the maximal height is a well-defined, intrinsic quantity.
The asymptotic one-point distribution in this case
 is given by the GUE TW distribution
 [Fig.~\ref{fig:Dist};
 reproduced by blue circles in Fig.~\ref{fig:MaxHeight1}(a)].

\begin{figure}[t]
 \begin{center}
  \includegraphics[clip]{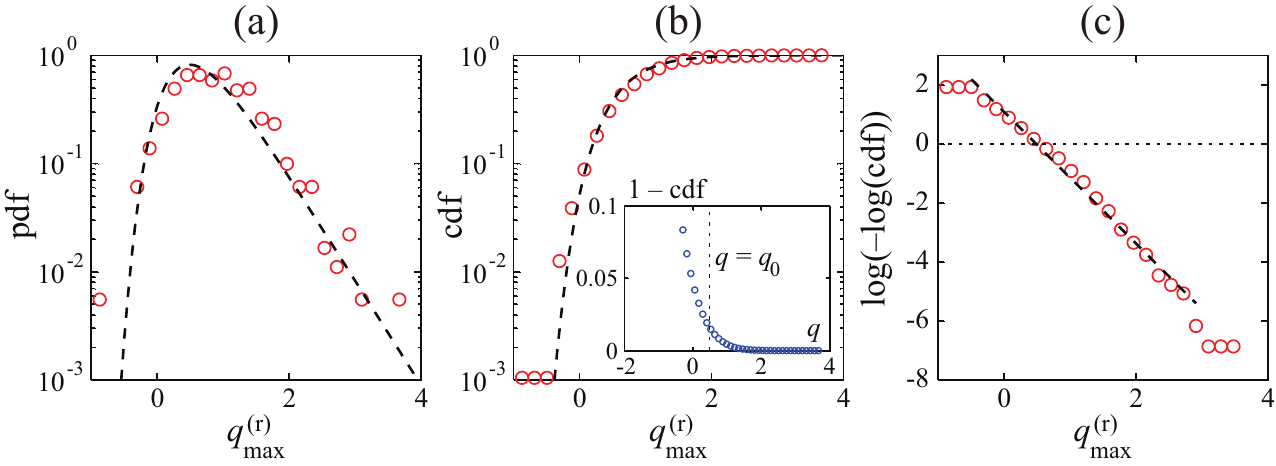}
  \caption{Probability density function (a) and cumulative distribution function (b,c) for the rescaled maximal radius $q_{\rm max}^{\rm (r)} \equiv (R_{\rm max} - v_\infty t)/(\Gamma t)^{1/3}$ at $t=30\unit{s}$. The dashed lines show the results of the fitting of Eq.~\eqref{eq:GumbelDistDef} to the data in the panel (c). The inset of the panel (b) displays the cumulative distribution function for the rescaled local height $q$, with the dotted line indicating the value of $q_0$ from the fitting of the Gumbel distribution.}
  \label{fig:MaxHeight2}
 \end{center}
\end{figure}%

First we measure $R_{\rm max}(t) \equiv \max_x h(x,t)$
 for each interface at a fixed time $t = 30\unit{s}$,
 which we call the maximal radius in this section
 [Fig.~\ref{fig:MaxHeight1}(b)].
Its distribution, in the rescaled axis
 $q_{\rm max}^{\rm (r)} \equiv (R_{\rm max} - v_\infty t)/(\Gamma t)^{1/3}$,
 is by definition on the right of the one-point distribution
 [green squares in Fig.~\ref{fig:MaxHeight1}(a)].
It turns out to be the Gumbel distribution
 \cite{Gumbel-Book1958,Clusel.Bertin-IJMPB2008},
 characterized by the double exponential function
 of the cumulative distribution function (cdf)
\begin{equation}
 \cdf(q_{\rm max}^{\rm (r)}) = \exp[-\e^{-g(q_{\rm max}^{\rm (r)} - q_0)}],  \label{eq:GumbelDistDef}
\end{equation}
 as identified in Fig.~\ref{fig:MaxHeight2},
 with evidence of the double exponential regime
 in Fig.~\ref{fig:MaxHeight2}(c).
Fitting Eq.~\eqref{eq:GumbelDistDef} to the experimental data therein,
 we obtain $q_0 = 0.49(6)$ and $g = 2.23(11)$.
The parameter $q_0$ is called the characteristic largest value
 in extreme-value statistics and is connected to the one-point distribution
 by $\cdf(q = q_0) = 1-1/n$
 with the effective number $n$ of independent samples \cite{Gumbel-Book1958}.
From our experimental data for the one-point distribution
 [inset of Fig.~\ref{fig:MaxHeight2}(b)],
 we have $n=62(9)$, which is indeed in the same order as
 the dimensionless circumference $2\pi\expct{h}(A/2)(\Gamma t)^{-2/3} = 20.6$ 
 at $t=30\unit{s}$.
The realization of the Gumbel distribution is not very surprising
 if one is aware that it arises generically from one-point distributions
 that decay faster than any power law \cite{Clusel.Bertin-IJMPB2008},
 like the TW distributions.

\begin{figure}[t]
 \begin{center}
  \includegraphics[clip]{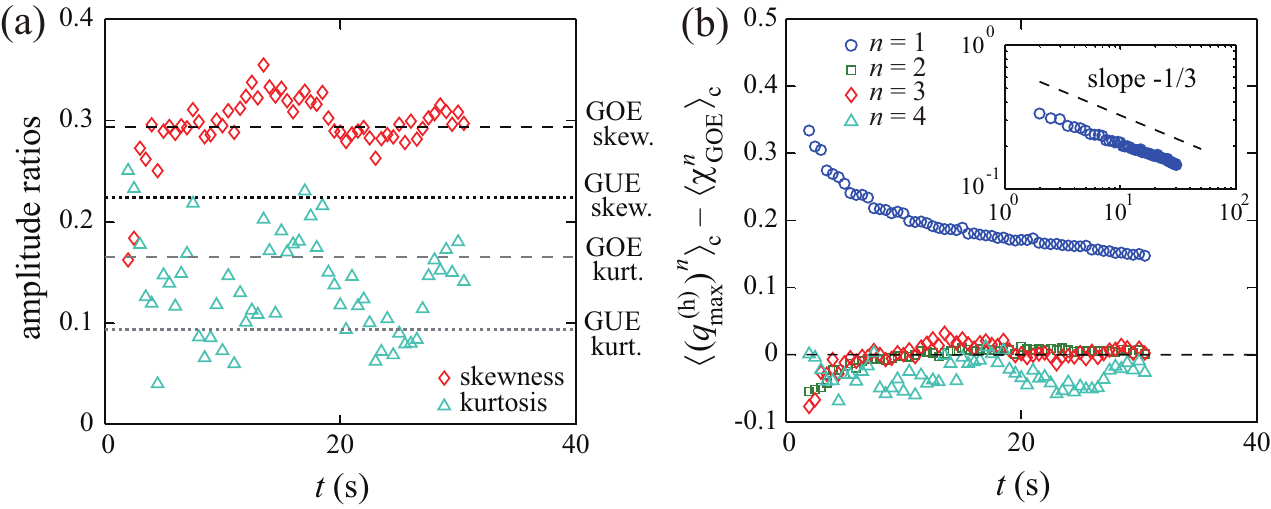}
  \caption{Distribution of the maximal height $H_{\rm max}$ with respect to the fictitious substrates. (a) Skewness $\cum{H_{\rm max}^3}/\cum{H_{\rm max}^2}^{3/2}$ (diamonds) and kurtosis $\cum{H_{\rm max}^4}/\cum{H_{\rm max}^2}^{2}$ (triangles) against time $t$. The horizontal lines indicate the values of the skewness (black) and the kurtosis (gray) for the GUE (dotted) and GOE (dashed) TW distributions \cite{Prahofer.Spohn-PRL2000}. (b) Finite-time corrections in the $n$th-order cumulants, $\cum{(q_{\rm max}^{\rm (h)})^n} - \cum{\chi_{\rm GOE}^n}$. The inset shows the data for $n=1$ in the logarithmic scales, with the dashed line indicating the exponent $-1/3$.}
  \label{fig:MaxHeight3}
 \end{center}
\end{figure}

The circular shape of the growing interfaces allows us to argue
 another interesting extremal of the interface position,
 namely, the maximal height $H_{\rm max}$ measured
 with respect to a fictitious flat substrate passing through the origin
 [Fig.~\ref{fig:MaxHeight1}(b)].
It is defined by $H_{\rm max} \equiv \max_x h(x,t)\sin\phi(x,t)$
 with the angle $\phi$ between the substrate and the vector connecting
 the origin and the point $h(x,t)$ on the interface.
Rotating the substrate arbitrarily, or varying the direction of $\phi=0$,
 we can accumulate as good statistics for $H_{\rm max}$
 as for the local height $h$.
The histogram of the maximal height $H_{\rm max}$ obtained thereby
 is plotted by red diamonds in Fig.~\ref{fig:MaxHeight1}(a)
 in the rescaled axis 
$q_{\rm max}^{\rm (h)} \equiv (H_{\rm max} - v_\infty t)/(\Gamma t)^{1/3}$.
Interestingly, in contrast to the GUE TW distribution for the local height $h$,
 the maximal height $H_{\rm max}$ obeys the GOE TW distribution
 like the local height of the flat interfaces.
This is also confirmed from the values of the skewness and the kurtosis
 as shown in Fig.~\ref{fig:MaxHeight3}(a).
We also observe due finite-time corrections
 from the GOE TW distribution [Fig.~\ref{fig:MaxHeight3}(b)].
The first-order cumulant shows again a pronounced correction
 decreasing as $t^{-1/3}$
 (inset; though the exponent may look slightly less,
 we consider this is not significant within our accuracy).
For the higher-order cumulants,
 we could not single out reliable functional forms
 within our experimental accuracy.
Theoretically, the GOE TW distribution was indeed identified analytically
 in the maximal height of the curved PNG interfaces at infinite time,
 or, equivalently, in the maximal position of the infinitely many
 non-intersecting Brownian particles with fixed end points
 (called the vicious walkers)
 \cite{Johansson-CMP2003,Forrester.etal-NPB2011,MorenoFlores.etal-a2011,Corwin.etal-a2011,Liechty-a2011,Schehr-a2012}.
This remarkable result is attributed, intuitively,
 to the same time-reversal symmetry in its space-time representation
 as for the local height of the flat interfaces.
In our experiment we have demonstrated that this nontrivial property
 on the extrema is also universal
 and robust enough to control the growth of the real turbulent interfaces.

\begin{figure}[t]
 \begin{center}
  \includegraphics[clip]{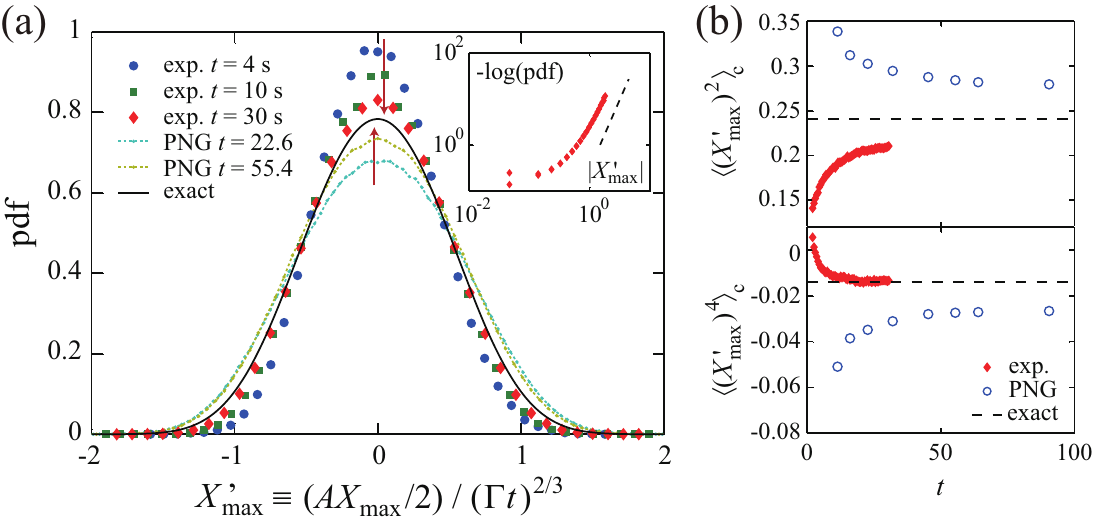}
  \caption{Distribution of the position $X_{\rm max}$ associated with the maximal height $H_{\rm max}$. (a) Probability density function (pdf) in the rescaled unit $X'_{\rm max} \equiv (AX_{\rm max}/2)/(\Gamma t)^{2/3}$ from our experiment (symbols), as well as that from simulations of the PNG model (dotted color lines) and the exact asymptotic solution (solid black line). The numerical data for the PNG model and the numerical evaluation of the exact asymptotic solution were generously offered by Rambeau and Schehr \cite{Rambeau.Schehr-EL2010,Rambeau.Schehr-PRE2011} and by Quastel and Remenik \cite{Quastel.Remenik-a2012a}, respectively. The arrows indicate how the distributions shift with time $t$. The inset shows the experimental data at $t=30\unit{s}$ with the different axes to check the functional form of the tail. The dashed line is a guide for the eyes showing the slope expected from the theoretical prediction ${\rm pdf}(X'_{\rm max}) \sim \exp(-c'_0 |X'_{\rm max}|^3)$ \cite{MorenoFlores.etal-a2011,Schehr-a2012,Quastel.Remenik-a2012a}. (b) Cumulants $\cum{(X'_{\rm max})^n}$ with $n=2$ and $4$ (top and bottom panels, respectively) against $t$. The dashed lines indicate the values estimated from the exact solution, $0.2409$ and $-0.0138$ for $n=2$ and $4$ \cite{Quastel.Remenik-a2012a}, respectively.}
  \label{fig:MaxHeight4}
 \end{center}
\end{figure}%

The horizontal position $X_{\rm max}$
 associated with the maximal height $H_{\rm max}$
 [Fig.~\ref{fig:MaxHeight1}(b)] is also a quantity of interest.
In Fig.~\ref{fig:MaxHeight4}(a), we show histograms of $X_{\rm max}$
 at different times $t$ from our experimental data (symbols),
 together with those from simulations of the PNG model at finite times
 offered by Rambeau and Schehr
 \cite{Rambeau.Schehr-EL2010,Rambeau.Schehr-PRE2011}
 (dotted color lines)
 and the exact asymptotic solution for the vicious walkers
 \cite{Schehr-a2012}
 and for the last passage percolation
 \cite{MorenoFlores.etal-a2011,Quastel.Remenik-a2012a}
 (solid black line; a numerical evaluation by Quastel and Remenik
 \cite{Quastel.Remenik-a2012a} is shown),
 both of which are equivalent to the PNG model.
We find similar curves
 for all the presented distributions
 when plotted in the appropriate dimensionless unit,
 with noticeable finite-time
 effects in the experimental and numerical data 
 toward
 the exact asymptotic solution.
Interestingly, concerning these finite-time effects,
 the experimental data indicate the opposite sign
 from the PNG model at finite times,
 which is more clearly visible in time series of the cumulants
 [Fig.~\ref{fig:MaxHeight4}(b)].
From the theoretical side,
 analytic expressions for the distribution of $X_{\rm max}$ have been
 obtained, first for finite numbers of vicious walkers
 \cite{Rambeau.Schehr-EL2010,Rambeau.Schehr-PRE2011},
 and, very recently, even in the infinite limit
 \cite{MorenoFlores.etal-a2011,Schehr-a2012,Quastel.Remenik-a2012a},
 as indicated by the solid line in Fig.~\ref{fig:MaxHeight4}(a).
The tail of the asymptotic distribution was shown
 to decay as $\exp(-c_0 |X_{\rm max}|^3)$ with a constant $c_0$
 \cite{MorenoFlores.etal-a2011,Schehr-a2012,Quastel.Remenik-a2012a},
 which is indeed identified in our experimental data
 [inset of Fig.~\ref{fig:MaxHeight4}(a)].
Further quantitative comparison of the experimental and numerical data
 to the exact solution would be helpful
 to elucidate the interesting finite-time behavior
 reported in Fig.~\ref{fig:MaxHeight4}.

To end this section, we briefly discuss difference between the two maxima,
 $R_{\rm max}$ and $H_{\rm max}$.
Given that both are the maximum of weakly correlated random variables,
 $h$ and $h\sin\phi$, respectively,
 and that the arc length grows faster than the correlation length
 ($t$ vs $t^{2/3}$),
 it is noteworthy that $R_{\rm max}$ and $H_{\rm max}$ have
 the different limiting distributions as evidenced in the present section.
This is an interesting example
 where collections of identical variables and of non-identical ones
 result in a remarkable difference in extreme-value statistics
 \cite{Schehr-PC2012}.

\subsection{What if applied voltage is varied?}

All the experimental results presented so far
 were obtained at the fixed applied voltage $V = 26\unit{V}$
 applied to the convection cell of the liquid crystal.
Finally we briefly mention how these results change
 if the applied voltage is varied.

\begin{figure}[t]
 \begin{center}
  \includegraphics[clip]{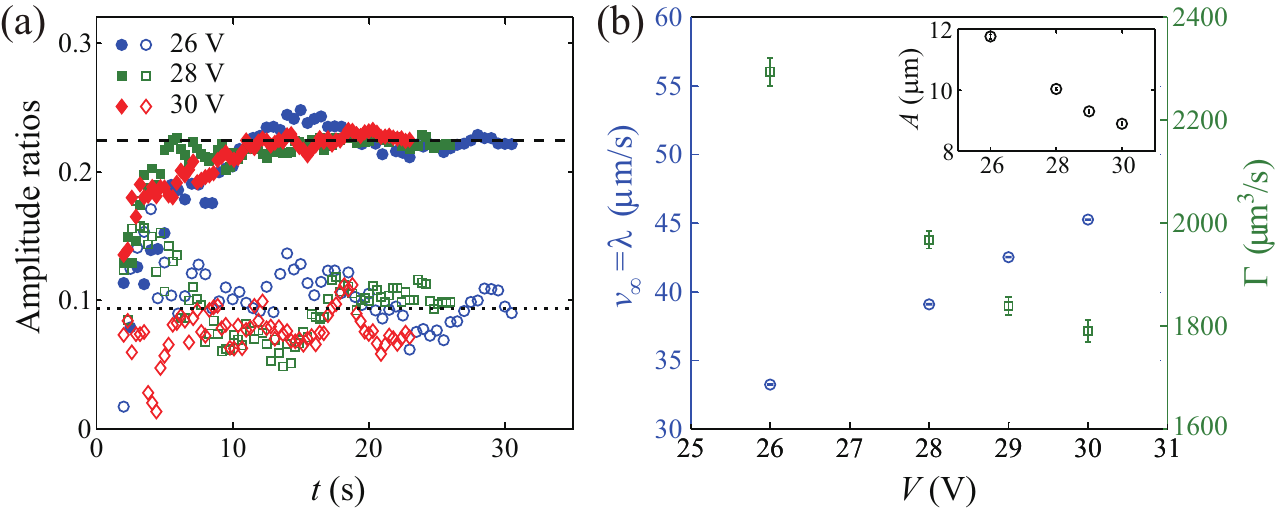}
  \caption{Experimental results from the circular interfaces at different applied voltages $26\unit{V} \leq V \leq 30\unit{V}$. (a) Time series of the skewness $\expct{h^3}_{\rm c} / \expct{h^2}_{\rm c}^{3/2}$ (filled symbols) and the kurtosis $\expct{h^4}_{\rm c} / \expct{h^2}_{\rm c}^2$ (empty symbols) at different voltages $V$ (as shown in the legend). The horizontal lines indicate the values of the skewness (dashed line) and the kurtosis (dotted line) for the GUE TW distribution. (b) Estimated values of the parameters $v_\infty = \lambda$ (blue circles), $\Gamma$ (green squares) and $A$ (inset) as functions of $V$.}
  \label{fig:VariousV}
 \end{center}
\end{figure}

First, for higher voltages,
 we confirmed that all the results are reproduced,
 as far as we explicitly checked statistical properties
 that are expected to be universal,
 such as the scaling exponents and the asymptotic distributions.
In particular, for all the voltages we studied
 in the range $26\unit{V} \leq V \leq 30\unit{V}$,
 the asymptotic one-point distribution is always given
 by the GUE and GOE TW distributions for the circular and flat interfaces,
 respectively, as demonstrated for example
 from the values of the skewness and the kurtosis
 [Fig.~\ref{fig:VariousV}(a)].
The change in the applied voltage is reflected
 in the values of the non-universal parameters $v_\infty$ and $\Gamma$,
 which are connected to the three parameters in the KPZ equation
 \eqref{eq:KPZEqDef}
 via Eqs.~\eqref{eq:GammaDef} and \eqref{eq:VinfEqLambda}.
We estimate them in the same manner as for $V=26\unit{V}$
 and find that, with increasing applied voltage $V$,
 $v_\infty$ increases and $\Gamma$ decreases
 [Fig.~\ref{fig:VariousV}(b)].
This indicates that the interface grows faster as expected,
 but with smaller amplitudes of the fluctuations.
We expect that these claims hold for even higher voltages,
 though it becomes difficult to confirm
 because of more frequent spontaneous nucleation of DSM2 nuclei
 \cite{Kai.etal-PRL1990}.

In contrast, when one lowers the voltage by a few volts from $26\unit{V}$,
 the system enters the spatiotemporal intermittency regime,
 in which the DSM2 clusters are not compact any more
 and replaced by small patches of DSM2, moving around amidst the DSM1 region
 \cite{Takeuchi.etal-PRL2007,Takeuchi.etal-PRE2009}.
The dynamics in this regime is described in coarse-grained scales
 by local spreading and vanishing of DSM2 patches,
 which correspond in microscopic scales
 to proliferation, turbulent diffusion and annihilation of disclinations.
This coarse-grained picture is similar to the model called the contact process,
 albeit oversimplified,
 defined by nearest-neighbor contaminations (spreading) and local recessions
 that take place at constant rates on a lattice
 \cite{Harris-AP1974,Hinrichsen-AP2000}.
Indeed, the DSM1-DSM2 turbulence and the contact process were shown to share
 the same critical behavior at the onset of the spatiotemporal intermittency
 \cite{Takeuchi.etal-PRL2007,Takeuchi.etal-PRE2009,Takeuchi-PRE2008},
 known as the directed percolation universality class \cite{Hinrichsen-AP2000},
 at least when the planar alignment of liquid-crystal molecules is chosen.
We can then interpret that the DSM2 growth for higher voltages
 is realized by rapidly increasing ratio of the contamination rate
 to the recession rate with increasing voltage.
In this line of thoughts
 the realization of the KPZ-class dynamics would be a reasonable consequence,
 because in the limit of the infinitely rapid contamination rate
 the contact process reduces to a variant of the Eden model,
 a representative numerical model which belongs to the KPZ class
 \cite{Barabasi.Stanley-Book1995}.
One of the authors indeed shows by simulations that
 this type of the Eden model, when placed on continuous space, exhibits
 growing interfaces very similar to the ones observed in our experiment
 \cite{Takeuchi-JSM2012}.
We stress, however, that the actual dynamics of the DSM1-DSM2 turbulence
 is far more complicated \cite{deGennes.Prost-Book1995}
 and neither the correspondence to the contact process nor
 the realization of the KPZ-class interfaces is obvious.

\section{Summary}


Throughout this paper, we have studied growing interfaces
 of the DSM2-turbulent domains in the electrically driven liquid crystal.
We have experimentally realized both circular and flat interfaces
 and carried out detailed statistical analyses
 of their scale-invariant fluctuations.
These confirm, first of all, that the DSM2 growing interfaces
 clearly belong to the $(1+1)$-dimensional KPZ class.
We have
 confirmed not only the universal scaling exponents
 but also shown the validity of the far more detailed universality,
 which controls even the precise form of the distribution function
 and the spatial correlation function in the asymptotic limit.
At this level, the KPZ universality class splits into
 at least two distinct sub-classes corresponding
 to different global geometries,
 namely to the flat and circular (or, more generally, curved) interfaces.
It was previously argued by analytic studies for the few solvable models
\cite{Prahofer.Spohn-PRL2000,Kriecherbauer.Krug-JPA2010,Sasamoto.Spohn-JSM2010,Corwin-RMTA2012}
 and is now evidenced in our experimental system.

\begin{table}[t]
\caption{Summary of the main results for the flat and circular interfaces.$^{\rm a}$}
\label{tbl:Summary}
\begin{tabular}{lllll}
\hline\hline\noalign{\smallskip}
& \multicolumn{1}{c}{flat interfaces} && \multicolumn{2}{c}{circular interfaces} \\
\noalign{\smallskip}\cline{2-2}\cline{4-5}\noalign{\smallskip}
& \multicolumn{1}{c}{our experiment} && \multicolumn{1}{c}{our experiment} & \multicolumn{1}{c}{off-lattice Eden \cite{Takeuchi-JSM2012}} \\
\noalign{\smallskip}\hline\noalign{\smallskip}
Scaling exponents & \multicolumn{4}{c}{$\alpha = 1/2, \quad \beta = 1/3, \quad z = 3/2$} \\
\noalign{\smallskip}
One-point distribution for $q$ & GOE TW distribution && \multicolumn{2}{c}{GUE TW distribution} \\
\noalign{\smallskip}
Finite-time corrections\\
\quad in the $n$th-order cumulants $\cum{q^n}$ & $t^{-n/3}$ ($n = 1,2,3,4$) && $t^{-n/3}$ ($n=1,3$) & $t^{-2/3}$ ($n=1,2$) \\
\noalign{\smallskip}
Spatial correlation function $C'_{\rm s}(\zeta;t)$ & Airy$_1$ covariance && \multicolumn{2}{c}{Airy$_2$ covariance} \\
\noalign{\smallskip}
Finite-time correction\\
\quad in $C_{\rm s}^{\rm int}(t) \equiv \int_0^\infty C'_{\rm s}(\zeta;t) \rd\zeta$ &---&& $t^{-1/3}$ & $t^{-1/3}$ \\
\noalign{\smallskip}
Spatial persistence probability$^{\rm b}$ & $\kappa^{\rm (s)}_+ = 1.9(3)$ && $\kappa^{\rm (s)}_+ = 1.07(8)$ & $\kappa^{\rm (s)}_+ = 0.90(2)$ \\
\quad $P_\pm^{\rm (s)}(l) \sim \e^{-\kappa^{\rm (s)}_\pm \zeta}$ & $\kappa^{\rm (s)}_- = 2.0(3)$ && $\kappa^{\rm (s)}_- = 0.87(6)$ & $\kappa^{\rm (s)}_- = 0.89(4)$ \\
\noalign{\smallskip}
Temporal correlation function\\
\quad $C'_{\rm t}(t,t_0) \sim (t/t_0)^{-\bar\lambda}$ & $\bar\lambda = 1$ && $\bar\lambda = 1/3$ & $\bar\lambda = 1/3$ \\
\noalign{\smallskip}
Fitting by the modified form&&&$b \approx 0.8(1)$ & $b \approx 1.22(8)$\\
\quad of Singha's correlation \eqref{eq:Singha2}$^{\rm c,d}$ & --- && $1-c \sim t^{-\delta'}$ & $c=1$\\
\noalign{\smallskip}
Short-time behavior\\
\quad of $C_{\rm t}(t,t_0)$ [Eq.~\eqref{eq:TemporalCorrFuncShort}]$^{\rm e}$ & $R/2 \approx 0.94$ && $R/2 \approx 0.6$ & $R/2 \approx 0.68$\\
\noalign{\smallskip}
Temporal persistence probability & $\theta_+ = 1.35(5)$ && $\theta_+ = 0.81(2)$ & $\theta_+ = 0.81(3)$\\
\quad $P_\pm(t,t_0) \sim (t/t_0)^{-\theta_\pm}$ & $\theta_- = 1.85(10)$ && $\theta_- = 0.80(2)$ & $\theta_- = 0.77(4)$\\
\noalign{\smallskip}
Extreme-value statistics\\
\quad maximal radius $q_{\rm max}^{\rm (r)}$ & --- && \multicolumn{2}{c}{Gumbel distribution}\\
\quad maximal height $q_{\rm max}^{\rm (h)}$ & --- && \multicolumn{2}{c}{GOE TW distribution}\\
\quad finite-time correction in $\expct{q_{\rm max}^{\rm (h)}}$ & --- && $t^{-1/3}$ & $t^{-2/3}$ \\
\quad position $X'_{\rm max}$ of $q_{\rm max}^{\rm (h)}$ & --- && \multicolumn{2}{c}{see Fig.~\ref{fig:MaxHeight4}}\\
\noalign{\smallskip}
\hline\hline
\end{tabular}\\
$^{\rm a}$$q \equiv (h-v_\infty t)/(\Gamma t)^{1/3}$ is the rescaled height; $\zeta \equiv (Al/2)(\Gamma t)^{-2/3}$ is the rescaled length. $q_{\rm max}^{\rm (r)}$ and $q_{\rm max}^{\rm (h)}$ are rescaled as the former and $X'_{\rm max}$ as the latter. The other rescaled variables are defined by $C'_{\rm s}(\zeta;t) \equiv C_{\rm s}(l;t)/(\Gamma t)^{2/3}$, $C'_{\rm t}(t,t_0) \equiv C_{\rm t}(t,t_0) / (\Gamma^2 t_0t)^{1/3}$.\\
$^{\rm b}$Simulations of Dyson's Brownian motion for GUE random matrices give $\kappa_+ = 0.90(8)$ and $\kappa_- = 0.90(6)$ (see Appendix).\\
$^{\rm c}$The estimate of $b$ for our experiment is obtained by averaging the values at late times. This might be underestimated if the value of $b$ at each $t_0$ increases slowly but indefinitely with $t_0$, such as by a power law $b_\infty - b(t_0) \sim t_0^{-\delta''}$. This power law is indeed suggested for the off-lattice Eden model \cite{Takeuchi-JSM2012}, though not definitively.\\
$^{\rm d}$The value of the exponent $\delta'$ is roughly estimated to be in the range $\delta' \in [2/3, 1]$.\\
$^{\rm e}$The values of $R/2$ are estimated directly from the short-time behavior of $C_{\rm t}(t,t_0)$, except for the experimental circular interfaces showing rather strong finite-time effects in this respect. For this case, we give in the table the value of $R/2$ obtained from $b$ (see Sect.~\ref{sec:TimeCorrFunc}).
\end{table}

We have then extended our analyses to the statistical properties
 that remain out of reach of rigorous theoretical treatment,
 especially those related to the temporal correlation.
Our experimental results are summarized in Table \ref{tbl:Summary},
 together with the numerical results for an off-lattice Eden model
 obtained by one of the authors \cite{Takeuchi-JSM2012}.
We notice here that, among the properties we have studied,
 it is only the values of the scaling exponents that are shared
 by both circular and flat interfaces.
For all the rest, the different geometries lead to different results,
 sometimes even with \textit{qualitative} differences,
 such as the superexponential decay of the spatial correlation
 in the flat interfaces,
 as well as the lasting temporal correlation
 and the symmetry between the positive and negative temporal persistence
 in the circular interfaces.

In view of the recent theoretical developments
 on the universal fluctuations of the KPZ class
 \cite{Kriecherbauer.Krug-JPA2010,Sasamoto.Spohn-JSM2010,Corwin-RMTA2012},
 our experimental results lead to a good number of remarks and conclusions.
While we refer the readers to the corresponding sections of the present paper
 for details, we consider that the following are particularly worth stressing:

\begin{itemize}

\item \textit{Distribution function and spatial correlation function.}
We have found the GOE and GUE TW distributions
 and the Airy$_1$ and Airy$_2$ covariance for the flat and circular interfaces,
 respectively, in agreement with the rigorous results
 for the TASEP and PASEP, the PNG model and the KPZ equation
 \cite{Kriecherbauer.Krug-JPA2010,Sasamoto.Spohn-JSM2010,Corwin-RMTA2012}.
Our experimental results obviously do not rely
 on any elaborate mathematical mappings used more or less in common
 in these analytical studies,
 and hence underpin the robustness of the universality in these quantities.

\item \textit{Finite-time corrections in the distribution function.}
We have found that the finite-time corrections for the $n$th-order cumulants
 $\cum{q^n}$ are in the order of $\mathcal{O}(t^{-n/3})$ for $n \leq 4$,
 except that we could not extract any finite-time corrections for $n=2$ and $4$
 from our data of the circular interfaces.
As detailed in Sect.~\ref{sec:DistFunc}, although these do not contradict
 any analytical results for the solvable models
 \cite{Sasamoto.Spohn-PRL2010,Sasamoto.Spohn-NPB2010,Ferrari.Frings-JSP2011,Baik.Jenkins-a2011},
 only some of the properties are confirmed to be shared.
One clearly needs further study to distinguish universal and non-universal
 aspects of this finite-time effect.
In particular, explaining the vanishing corrections in the even-order cumulants
 of the circular interfaces, as well as the absence
 of the $\mathcal{O}(t^{-1/3})$ correction
 in the first-order cumulant of the off-lattice Eden model
 \cite{Takeuchi-JSM2012}, is an interesting issue left for future studies.

\item \textit{Temporal correlation.}
The temporal correlation function and the temporal persistence probability
 are important statistical quantities
 that remain out of reach of the rigorous theoretical developments,
 and at the same time exhibit clear differences
 between the flat and circular interfaces
 (see Table \ref{tbl:Summary}).
We believe that the evidenced difference
 in the temporal correlation function of the circular interfaces
 is essentially captured by Singha's approximative theory
 \cite{Singha-JSM2005},
 though a modification was needed to
 fit our experimental results at finite times.
More importantly, Singha explicitly assumed
 the circular growth of the interfaces.
It therefore remains to be clarified to what extent our results
 on the temporal correlation function
 and the temporal persistence hold for the general situation
 of the curved interfaces.

\item \textit{Spatial persistence.}
We have found exponential decay of the spatial persistence probability,
 $P_\pm^{\rm (s)}(l) \sim \e^{-\kappa^{\rm (s)}_\pm \zeta}$,
 for both flat and circular interfaces,
 albeit with different values of the coefficients $\kappa^{\rm (s)}_\pm$
 in the appropriately rescaled unit.
Given the correspondence to the Airy processes and, for the circular case,
 to Dyson's Brownian motion,
 this result may also shed light on the temporal persistence properties
 of these stochastic processes.
From another viewpoint, one may check the expected equivalence
 up to the persistence probability,
 which formally concerns infinite-point correlation functions.
In this respect we present numerical studies on Dyson's Brownian motion
 in the appendix and obtain roughly consistent results
 for the GUE/circular case, though better precision is required
 to draw a conclusion on it
 (see Appendix and Sect.~\ref{sec:SpatialPersistence}).
Mathematical approach and refined numerical evaluation
 for the persistence of Dyson's Brownian motion
 and that of the Airy processes are indispensable to put these observations
 on firmer ground.

\item \textit{Extreme-value statistics.}
We have also obtained quantitative results on it for our circular interfaces,
 in particular on the distribution of the maximal height $H_{\rm max}$
 and its position $X_{\rm max}$ on fictitious substrates
 (see Sect.~\ref{sec:ExtremeValueStat}).
Our experimental results are in good agreement
 with the asymptotic analytical predictions
 for both $H_{\rm max}$ and $X_{\rm max}$,
 confirming the GOE TW distribution for the former.
The finite-time corrections in their cumulants have also been measured
 and found to have, for $X_{\rm max}$,
 the opposite sign from those in the PNG model.
Further quantitative analyses and theoretical accounts
 for these interesting finite-time effects
 are left for future studies.
\end{itemize}

Despite these new experimental results,
 one should be aware that the range of the experimentally investigated
 quantities or situations still remains quite narrow,
 compared with the vast theoretical explorations marked in recent years.
For example, stationary interfaces are not realized
 in the liquid-crystal turbulence yet,
 which constitute an as important situation as the growing interfaces
 and are theoretically predicted to have another nontrivial distribution
 called the $F_0$ distribution \cite{Kriecherbauer.Krug-JPA2010,Corwin-RMTA2012,Baik.Rains-JSP2000,Prahofer.Spohn-PRL2000,Ferrari.Spohn-CMP2006,Imamura.Sasamoto-a2011}.
To add, many other situations have been considered in analytical work
 and shown to have a variety of intriguing distributions,
 such as the TW distribution for Gaussian symplectic ensemble
 \cite{Prahofer.Spohn-PRL2000,Sasamoto.Imamura-JSP2004}
 and that for (GOE)$^2$
 \cite{Baik.Rains-JSP2000,Prahofer.Spohn-PRL2000}%
\footnote{
It describes the distribution of the largest eigenvalue
 among all eigenvalues of two random matrices independently drawn from GOE.},
 expected under certain conditions with boundary or external sources.
In the recent review \cite{Corwin-RMTA2012},
 six such sub-universality classes are argued.
It is a challenging open problem
 to test these predictions in a real experiment,
 though it is statistically much more demanding
 because one cannot take advantage
 of spatial homogeneity as in the present study.
As a complementary approach, it is also essential
 to find other experimental examples
 for the universal fluctuations of the KPZ class.
Systems showing the KPZ scaling exponents, whether directly or indirectly,
 are of course primary candidates.
Above all the paper combustion experiment
 \cite{Maunuksela.etal-PRL1997,Myllys.etal-PRE2001,Myllys.etal-PRL2000}
 is a quite promising and interesting case, for which the KPZ scaling exponents
 were clearly found \cite{Maunuksela.etal-PRL1997,Myllys.etal-PRE2001}
 but different results from ours were reported
 on the temporal and spatial persistence \cite{Merikoski.etal-PRL2003}.
Although the same group performed a test for the KPZ universal distribution
 in this system \cite{Miettinen.etal-EPJB2005},
 we consider that it is not statistically significant
 to draw a conclusion on it \cite{Takeuchi-c2012};
 we hope that the data will be analyzed again
 along the same lines as the present study or in any equivalent way.

On the theoretical side, in addition to further expanding the realm
 of exact solutions and analytic expressions for particular models,
 it would be fundamentally important to have a general theoretical framework
 that can explain the evidenced detailed yet
 geometry-dependent (or, equivalently, initial-condition-dependent)
 universality of the $(1+1)$-dimensional KPZ class.
A promising approach in this direction has recently been undertaken
 by studies of non-perturbative renormalization group
 \cite{Canet.etal-PRL2010,Canet.etal-PRE2011} and
 renormalization group combined with variational formulation
 \cite{Corwin.Quastel-a2011}.
We believe that such cooperative progress
 in experimental and theoretical investigations will afford
 a further understanding toward this remarkable universality,
 which underlies the general phenomenon of the growing interfaces
 with deep connection to apparently unrelated areas of physics and mathematics.

\begin{acknowledgements}
The authors acknowledge enlightening discussions
 with many theoreticians:
 T.~Sasamoto, H.~Spohn, M.~Pr\"ahofer, G.~Schehr,
 J.~Rambeau, H.~Chat\'e, P.~Ferrari, to name but a few.
We are grateful to T.~Sasamoto for his continuing and scrupulous support
 on the theoretical side of the subject,
 to G.~Schehr
 for drawing our attention to extreme-value statistics
 and to P.~Ferrari with respect to the finite-time corrections
 in the second- and higher-order cumulants of the local height.
We also wish to thank our colleagues who kindly sent us
 theoretical curves and numerical data used in this paper:
 M.~Pr\"ahofer for the theoretical curves of the TW distributions,
 F.~Bornemann for those of the Airy$_1$ and Airy$_2$ covariance
 obtained by his accurate algorithm \cite{Bornemann-MC2010},
 J.~Rambeau and G.~Schehr for their numerical data on the PNG model
 partly presented in their work
 \cite{Rambeau.Schehr-EL2010,Rambeau.Schehr-PRE2011},
 and J.~Quastel and D.~Remenik for the theoretical curve
 of the asymptotic distribution of $X_{\rm max}$,
 numerically evaluated very recently by them \cite{Quastel.Remenik-a2012a}.
Critical reading of the manuscript and useful comments
 by J.~Krug, J.~Rambeau, T.~Sasamoto,
 G.~Schehr and H.~Spohn are also much appreciated.
This work is supported in part
 by Grant for Basic Science Research Projects from The Sumitomo Foundation
 and by the JSPS Core-to-Core Program ``International research network
 for non-equilibrium dynamics of soft matter.''
\end{acknowledgements}

\appendix
\def\thesection{Appendix:}
\section[\hspace{25pt}Simulations of Dyson's Brownian motion]{Simulations of Dyson's Brownian motion}  \label{sec:DysonBM}

%

In Sect.~\ref{sec:SpaceCorrFunc} we have experimentally shown
 that the spatial two-point correlation of the circular and flat interfaces
 is indeed given, asymptotically, by that of the temporal correlation
 of the Airy$_2$ and Airy$_1$ processes, respectively.
Theoretically, however, this correspondence is expected to be far beyond;
 the spatial profile of the interfaces itself is
 considered to be statistically equivalent
 to the locus of the Airy processes, or, for the curved interfaces,
 to that of the largest-eigenvalue dynamics
 in Dyson's Brownian motion of GUE random matrices.
Our analysis on spatial correlation of the interfaces
 may therefore shed light also on the temporal correlation
 of these stochastic processes.
In this context, particularly interesting,
 and not investigated yet to our knowledge,
 is their temporal persistence property,
 which may be realized as the spatial persistence of the interfaces
 studied in Sect.~\ref{sec:SpatialPersistence}.
This correspondence might be mathematically not obvious,
 because the asymptotic equivalence in the \textit{moments}
 of the Airy$_2$ process and those of the largest eigenvalue
 in Dyson's Brownian motion for GUE
 has not been rigorously proved yet \cite{Bornemann.etal-JSP2008}.
Here, performing direct simulations of Dyson's Brownian motion,
 we shall indeed probe this bridge to the interface problem
 at the level of the persistence property,
 which should also help interpret our experimental results
 shown in Sect.~\ref{sec:SpatialPersistence}.

Dyson's Brownian motion is defined as the time evolution
 of the eigenvalues of a random matrix, taken from GUE or GOE here,
 whose independent elements exhibit uncorrelated Ornstein-Uhlenbeck processes
 \cite{Mehta-Book2004}.
Specifically, for an $N \times N$ Hermitian or symmetric matrix $M$,
 we consider the process
\begin{equation}
\diff{M(t)}{t} = -\gamma M(t) + \Xi(t),  \label{eq:DysonBMDef}
\end{equation}
 where $\gamma$ is a constant scalar and $\Xi(t)$ is a matrix
 with independent white-noise elements $\Xi_{ij}(t)$
 preserving the same symmetry as $M$;
 these matrix elements satisfy $\expct{\Xi_{ij}(t)} = 0$,
 $\expct{\Xi_{ii}(t)\Xi_{i'i'}(t')} = \delta_{ii'}\delta(t-t')$
 for the diagonal elements and 
 $\expct{[\re\Xi_{ij}(t)][\re \Xi_{i'j'}(t')]}$
 ($=\expct{[\im\Xi_{ij}(t)][\im \Xi_{i'j'}(t')]}$ for GUE)
 $= (1/2)\delta_{ii'}\delta_{jj'}\delta(t-t')$
 for the nondiagonal elements $i>j$.
The one-point distribution for $M(t)$ defined thereby remains to be
 that for GUE or GOE at any time $t$.
We then focus on the largest eigenvalue $\lambda_1(t)$ of the matrix $M(t)$,
 which is rescaled as
\begin{align}
 &\lambda_{\rm GUE}(t') \equiv \sqrt{2\gamma}N^{1/6}\[ \lambda_1(t'/\gamma N^{1/3}) - \sqrt{2N/\gamma}\],  \label{eq:LambdaGUEDef} \\
 &\lambda_{\rm GOE}(t') \equiv 2^{1/3}\gamma^{1/2}N^{1/6}\[ \lambda_1(2^{1/3}t'/\gamma N^{1/3}) - \sqrt{N/\gamma}\],  \label{eq:LambdaGOEDef}
\end{align}
 for GUE and GOE, respectively.
The factors for this rescaling are determined,
 after Bornemann \textit{et al.} \cite{Bornemann.etal-JSP2008},
 in such a way that
 $\lambda_{\rm GUE}(t')$ and $\lambda_{\rm GOE}(t')$ have
 the same values of the covariance
 and its derivative at zero as the Airy$_2$ and Airy$_1$ processes,
 respectively.
Note that the factors for the GOE case are different from those
 used by Bornemann \textit{et al.} \cite{Bornemann.etal-JSP2008}
 because of our somewhat unconventional definition
 for the Airy$_1$ process, which is however useful in the context
 of the growing interfaces
 (see footnote \ref{ft:Airy1Def} on page \pageref{ft:Airy1Def}).

We numerically integrate Eq.~\eqref{eq:DysonBMDef}
 using the standard numerical scheme for the Ornstein-Uhlenbeck process
 \cite{Gillespie-PRE1996}, which is exact for any finite time step $\Delta t$.
We then compute the temporal persistence probability
 $P_\pm(\zeta)$
 for the fluctuations of $\lambda_{\rm GUE}(t')$ and $\lambda_{\rm GOE}(t')$
 as functions of a period $\zeta$
 over which the sign of positive or negative fluctuations remains unchanged.
Since the measurement of the persistence probability is influenced
 by the choice of the discrete time step $\Delta t$,
 we use in the following $\Delta t = 10^{-2} N^{-1/3}$ and $10^{-3} N^{-1/3}$
 and check that the results are not affected in a significant way.
Concerning the other parameters, we fix $\gamma=1$ without loss of generality,
 and compare $N=64$ and $256$ to confirm that
 our results reflect the property in the asymptotic limit $N\to\infty$.

\begin{figure}[t]
 \begin{center}
  \includegraphics[clip]{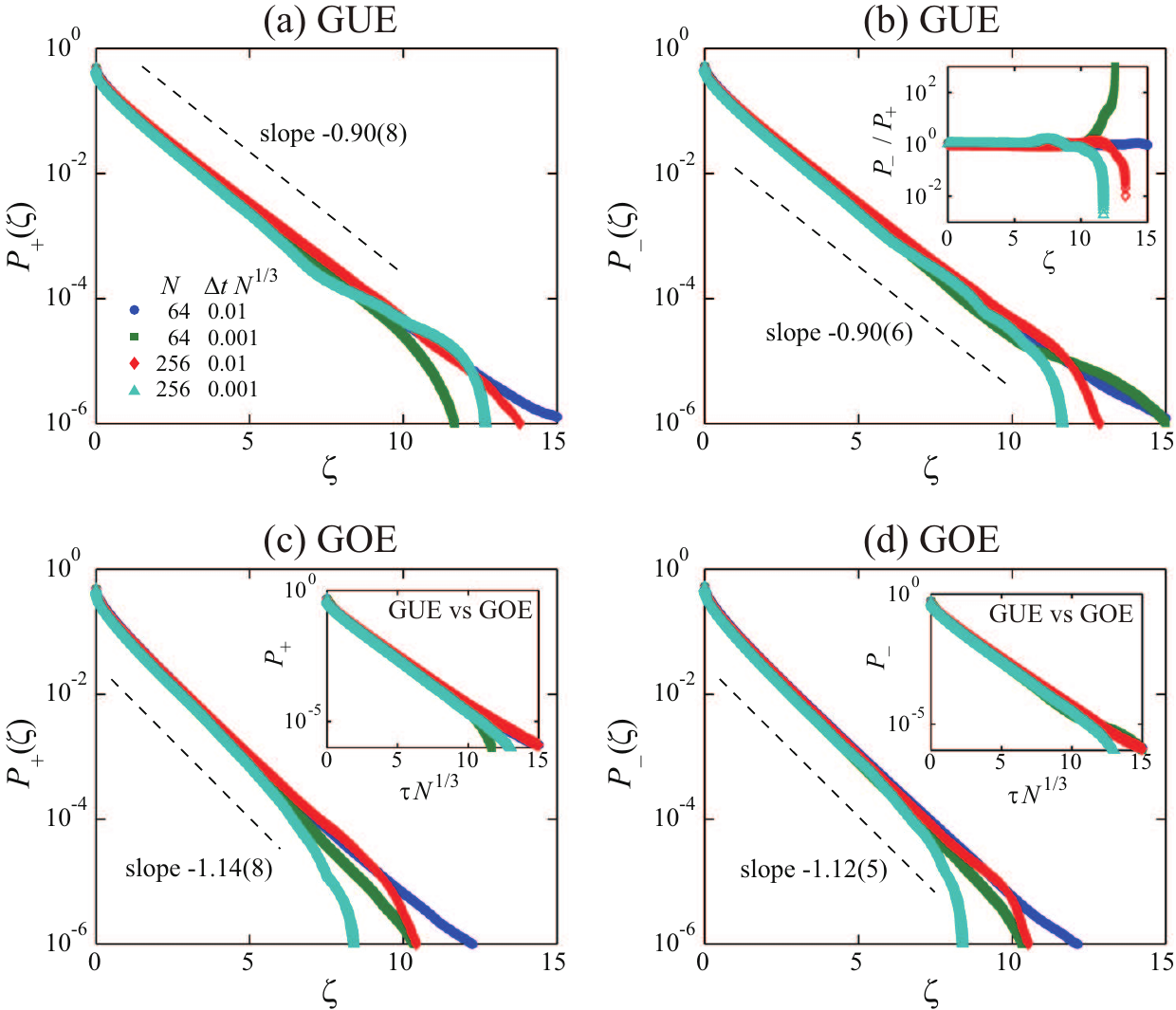}
  \caption{Temporal persistence probability $P_\pm(\zeta)$ for Dyson's Brownian motion of GUE (a,b) and GOE (c,d) random matrices. The four series of colored symbols in the main panels correspond to different $N$ and $\Delta t$ as shown in the legend of the panel (a). The dashed lines are guides for the eyes indicating the estimated values of $\kappa_\pm$ for each case. The inset of the panel (b) shows the ratio $P_-(\zeta) / P_+(\zeta)$ for the four data series shown in the panels (a,b). The insets of the panels (c,d) display the data for both GUE (blue and green) and GOE (red and turquoise) with $N=64$ in the original time unit of Dyson's Brownian motion, which overlap reasonably well within statistical accuracy.}
  \label{fig:DysonBM}
 \end{center}
\end{figure}%

Figure \ref{fig:DysonBM} shows the results obtained
 from more than 1000 and 50 independent simulations for $N=64$ and 256,
 respectively, both of length $10^6$ time steps.
It clearly shows that, for both GUE and GOE and
 for both positive and negative fluctuations, the persistence probability
 decays exponentially within our statistical accuracy,
 $P_\pm(\zeta) \sim \exp(-\kappa_\pm \zeta)$,
 as we have found for the spatial persistence of the interface fluctuations
 (Fig.~\ref{fig:SpacePersistence}).
Concerning the coefficient $\kappa_\pm$, we find here
\begin{equation}
 \begin{cases} \kappa^{\rm (s)}_+ = 0.90(8) \\ \kappa^{\rm (s)}_- = 0.90(6) \end{cases} \text{(GUE)} \qquad \text{and} \qquad \begin{cases} \kappa^{\rm (s)}_+ = 1.14(8) \\ \kappa^{\rm (s)}_- = 1.12(5) \end{cases} \text{(GOE)}.   \label{eq:EstimateDysonBMPersistence}
\end{equation}
 in the unit defined
 by Eqs.~\eqref{eq:LambdaGUEDef} and \eqref{eq:LambdaGOEDef}.
This shows no significant difference
 between the positive and negative fluctuations for both cases,
 which is also confirmed from the ratio $P_-(\zeta) / P_+(\zeta)$
 [inset of Fig.~\ref{fig:DysonBM}(b) for GUE].
In contrast, the values of $\kappa_\pm$ seem to be slightly different
 between GUE and GOE,
 but this turns out to result from the different normalizations of the time
 in Eqs.~\eqref{eq:LambdaGUEDef} and \eqref{eq:LambdaGOEDef}.
Measuring the duration $\tau$
 in the original time unit of Dyson's Brownian motion,
 we find that the persistence probabilities $P_\pm(\tau)$
 for GUE and GOE overlap reasonably well
 for both positive and negative fluctuations
 [insets of Fig.~\ref{fig:DysonBM}(c,d)].

The appropriate rescaled time units used to define $\kappa_\pm$
 allow us to make a direct comparison
 to the experimental values $\kappa^{\rm (s)}_\pm$
 for the spatial persistence of the growing interfaces.
For the circular case, we have found
 $\kappa^{\rm (s)}_+ = 1.07(8)$ and $\kappa^{\rm (s)}_- = 0.87(6)$,
 to be compared with the GUE values
 $\kappa_+ = 0.90(8)$ and $\kappa_- = 0.90(6)$.
While the values for $\kappa^{\rm (s)}_-$ and $\kappa_-$
 are in good agreement, we notice that
 those for $\kappa^{\rm (s)}_+$ and $\kappa_+$
 seem to be slightly different.
In particular, the apparent asymmetry
 between the positive and negative spatial persistence
 in the experiment is not reproduced in the temporal persistence
 of GUE Dyson's Brownian motion.
Two possibilities can be considered;
 our estimates for $\kappa^{\rm (s)}_+$ and $\kappa_+$
 are not sufficiently precise and/or affected by finite-time effects
 and they actually take the same value,
 or the spatial profile of the circular interfaces
 and the locus of the largest eigenvalue in GUE Dyson's Brownian motion
 are not equivalent at the level of the persistence property.
One of the authors' simulations of an off-lattice Eden model give
 $\kappa^{\rm (s)}_+ = 0.90(2)$ and $\kappa^{\rm (s)}_- = 0.89(4)$
 \cite{Takeuchi-JSM2012} and thus support the former possibility,
 but we do not single out either of them at present.
This should be clarified by further study with better precision,
 preferably with numerical estimation
 of the persistence probability for the Airy processes,
 for which Bornemann's method for evaluating Fredholm determinants
 \cite{Bornemann-MC2010,Bornemann.etal-JSP2008} could be utilized.

\bibliographystyle{spmpsci}      
\bibliography{KPZfullpaper,otherrefs}  

\begin{thebibliography}{100}
\providecommand{\url}[1]{{#1}}
\providecommand{\urlprefix}{URL }
\expandafter\ifx\csname urlstyle\endcsname\relax
  \providecommand{\doi}[1]{DOI~\discretionary{}{}{}#1}\else
  \providecommand{\doi}{DOI~\discretionary{}{}{}\begingroup
  \urlstyle{rm}\Url}\fi

\bibitem{Amar.Family-PRA1992}
Amar, J.G., Family, F.: Universality in surface growth: Scaling functions and
  amplitude ratios.
\newblock Phys. Rev. A \textbf{45}, 5378--5393 (1992)

\bibitem{Amir.etal-CPAM2011}
Amir, G., Corwin, I., Quastel, J.: Probability distribution of the free energy
  of the continuum directed random polymer in 1 + 1 dimensions.
\newblock Commun. Pure Appl. Math. \textbf{64}, 466--537 (2011)

\bibitem{Baik.etal-JAMS1999}
Baik, J., Deift, P., Johansson, K.: On the distribution of the length of the
  longest increasing subsequence of random permutations.
\newblock J. Am. Math. Soc. \textbf{12}, 1119-1178 (1999)

\bibitem{Baik.Jenkins-a2011}
Baik, J., Jenkins, R.: Limiting distribution of maximal crossing and nesting of
  poissonized random matchings.
\newblock arXiv 1111.0269 (2011)

\bibitem{Baik.Rains-JSP2000}
Baik, J., Rains, E.M.: Limiting distributions for a polynuclear growth model
  with external sources.
\newblock J. Stat. Phys. \textbf{100}, 523--541 (2000)

\bibitem{Baik.Rains-DMJ2001}
Baik, J., Rains, E.M.: The asymptotics of monotone subsequences of involutions.
\newblock Duke Math. J. \textbf{109}, 205-281 (2001)

\bibitem{Baik.Rains-MSRIP2001}
Baik, J., Rains, E.M.: Symmetrized random permutations.
\newblock In: P.~Bleher, A.~Its (eds.) Random Matrix Models and Their
  Applications, \emph{MSRI Publications}, vol.~40, pp. 1--19. Cambridge Univ.
  Press, Cambridge (2001)

\bibitem{Barabasi.Stanley-Book1995}
Barabasi, A.L., Stanley, H.E.: Fractal Concepts in Surface Growth.
\newblock Cambridge Univ. Press, Cambridge (1995)

\bibitem{Bornemann-MC2010}
Bornemann, F.: On the numerical evaluation of {Fredholm} determinants.
\newblock Math. Comput. \textbf{79}, 871--915 (2010)

\bibitem{Bornemann.etal-JSP2008}
Bornemann, F., Ferrari, P., Pr\"ahofer, M.: The {Airy}$_1$ process is not the
  limit of the largest eigenvalue in {GOE} matrix diffusion.
\newblock J. Stat. Phys. \textbf{133}, 405--415 (2008)

\bibitem{Borodin.etal-JSP2007}
Borodin, A., Ferrari, P., Pr\"ahofer, M., Sasamoto, T.: Fluctuation properties
  of the {TASEP} with periodic initial configuration.
\newblock J. Stat. Phys. \textbf{129}, 1055--1080 (2007)

\bibitem{Borodin.etal-CMP2008}
Borodin, A., Ferrari, P., Sasamoto, T.: Large time asymptotics of growth models
  on space-like paths ii: {PNG} and parallel {TASEP}.
\newblock Commun. Math. Phys. \textbf{283}, 417--449 (2008)

\bibitem{Calabrese.LeDoussal-PRL2011}
Calabrese, P., Le~Doussal, P.: Exact solution for the {Kardar-Parisi-Zhang}
  equation with flat initial conditions.
\newblock Phys. Rev. Lett. \textbf{106}, 250603 (2011)

\bibitem{Calabrese.etal-EL2010}
Calabrese, P., Le~Doussal, P., Rosso, A.: Free-energy distribution of the
  directed polymer at high temperature.
\newblock Europhys. Lett. \textbf{90}, 20002 (2010)

\bibitem{Canet.etal-PRL2010}
Canet, L., Chat\'e, H., Delamotte, B., Wschebor, N.: Nonperturbative
  renormalization group for the {Kardar-Parisi-Zhang} equation.
\newblock Phys. Rev. Lett. \textbf{104}, 150601 (2010)

\bibitem{Canet.etal-PRE2011}
Canet, L., Chat\'e, H., Delamotte, B., Wschebor, N.: Nonperturbative
  renormalization group for the {Kardar-Parisi-Zhang} equation: General
  framework and first applications.
\newblock Phys. Rev. E \textbf{84}, 061128 (2011)

\bibitem{Clusel.Bertin-IJMPB2008}
Clusel, M., Bertin, E.: Global fluctuations in physical systems: a subtle
  interplay between sum and extreme value statistics.
\newblock Int. J. Mod. Phys. B \textbf{22}, 3311--3368 (2008)

\bibitem{Constantin.etal-PRE2004}
Constantin, M., Das~Sarma, S., Dasgupta, C.: Spatial persistence and survival
  probabilities for fluctuating interfaces.
\newblock Phys. Rev. E \textbf{69}, 051603 (2004)

\bibitem{Corwin-RMTA2012}
Corwin, I.: The {Kardar-Parisi-Zhang} equation and universality class.
\newblock Random Matrices: Theory and Applications \textbf{1}, 1130001 (2012)

\bibitem{Corwin.etal-AIHPBPS2012}
Corwin, I., Ferrari, P.L., P\'ech\'e, S.: Universality of slow decorrelation in
  {KPZ} growth.
\newblock Ann. Inst. H. Poincar\'e B Probab. Statist. \textbf{48}, 134-150
  (2012)

\bibitem{Corwin.Quastel-a2011}
Corwin, I., Quastel, J.: Renormalization fixed point of the {KPZ} universality
  class.
\newblock arXiv 1103.3422 (2011)

\bibitem{Corwin.etal-a2011}
Corwin, I., Quastel, J., Remenik, D.: Continuum statistics of the {Airy}$_2$
  process.
\newblock arXiv 1106.2717 (2011)

\bibitem{Dotsenko-EL2010}
Dotsenko, V.: Bethe ansatz derivation of the {Tracy-Widom} distribution for
  one-dimensional directed polymers.
\newblock Europhys. Lett. \textbf{90}, 20003 (2010)

\bibitem{Family.Vicsek-JPA1985}
Family, F., Vicsek, T.: Scaling of the active zone in the {Eden} process on
  percolation networks and the ballistic deposition model.
\newblock J. Phys. A \textbf{18}, L75--L81 (1985)

\bibitem{Ferrari-JSM2008}
Ferrari, P.L.: Slow decorrelations in {Kardar-Parisi-Zhang} growth.
\newblock J. Stat. Mech. \textbf{2008}, P07022 (2008)

\bibitem{Ferrari.Frings-JSP2011}
Ferrari, P.L., Frings, R.: Finite time corrections in {KPZ} growth models.
\newblock J. Stat. Phys. \textbf{144}, 1123--1150 (2011)

\bibitem{Ferrari.Spohn-CMP2006}
Ferrari, P.L., Spohn, H.: Scaling limit for the space-time covariance of the
  stationary totally asymmetric simple exclusion process.
\newblock Commun. Math. Phys. \textbf{265}, 1--44 (2006)

\bibitem{FerreiraJr.Alves-JSM2006}
Ferreira~Jr, S.C., Alves, S.G.: Pitfalls in the determination of the
  universality class of radial clusters.
\newblock J. Stat. Mech. \textbf{2006}, P11007 (2006)

\bibitem{Forrester.etal-NPB2011}
Forrester, P.J., Majumdar, S.N., Schehr, G.: Non-intersecting {Brownian}
  walkers and {Yang-Mills} theory on the sphere.
\newblock Nucl. Phys. B \textbf{844}, 500--526 (2011)

\bibitem{Forster.etal-PRA1977}
Forster, D., Nelson, D.R., Stephen, M.J.: Large-distance and long-time
  properties of a randomly stirred fluid.
\newblock Phys. Rev. A \textbf{16}, 732--749 (1977)

\bibitem{Frisch-Book1995}
Frisch, U.: Turbulence: The Legacy of A. N. Kolmogorov.
\newblock Cambridge Univ. Press, Cambridge (1995)

\bibitem{deGennes.Prost-Book1995}
de~Gennes, P.G., Prost, J.: The Physics of Liquid Crystals, \emph{International
  Series of Monographs on Physics}, vol.~83, 2 edn.
\newblock Oxford Univ. Press, New York (1995)

\bibitem{Gillespie-PRE1996}
Gillespie, D.T.: Exact numerical simulation of the {Ornstein-Uhlenbeck} process
  and its integral.
\newblock Phys. Rev. E \textbf{54}, 2084--2091 (1996)

\bibitem{Gumbel-Book1958}
Gumbel, E.J.: Statistics of Extremes.
\newblock Columbia Univ. Press, republished by Dover Publications, New York,
  2004, New York (1958)

\bibitem{HalpinHealy.Zhang-PR1995}
Halpin-Healy, T., Zhang, Y.C.: Kinetic roughening phenomena, stochastic growth,
  directed polymers and all that. aspects of multidisciplinary statistical
  mechanics.
\newblock Phys. Rep. \textbf{254}, 215--414 (1995)

\bibitem{Harris-AP1974}
Harris, T.E.: Contact interactions on a lattice.
\newblock Ann. Probab. \textbf{2}, 969-988 (1974)

\bibitem{Henkel-Book1999}
Henkel, M.: Conformal Invariance and Critical Phenomena.
\newblock Springer-Verlag, Berlin, Heidelberg, New York (1999)

\bibitem{Henkel.etal-PRE2012}
Henkel, M., Noh, J.D., Pleimling, M.: Phenomenology of aging in the
  kardar-parisi-zhang equation.
\newblock Phys. Rev. E \textbf{85}, 030102 (2012)

\bibitem{Hinrichsen-AP2000}
Hinrichsen, H.: Non-equilibrium critical phenomena and phase transitions into
  absorbing states.
\newblock Adv. Phys. \textbf{49}, 815--958 (2000)

\bibitem{Huergo.etal-PRE2010}
Huergo, M.A.C., Pasquale, M.A., Bolz\'an, A.E., Arvia, A.J., Gonz\'alez, P.H.:
  Morphology and dynamic scaling analysis of cell colonies with linear growth
  fronts.
\newblock Phys. Rev. E \textbf{82}, 031903 (2010)

\bibitem{Huergo.etal-PRE2011}
Huergo, M.A.C., Pasquale, M.A., Gonz\'alez, P.H., Bolz\'an, A.E., Arvia, A.J.:
  Dynamics and morphology characteristics of cell colonies with radially
  spreading growth fronts.
\newblock Phys. Rev. E \textbf{84}, 021917 (2011)

\bibitem{Imamura.Sasamoto-a2011}
Imamura, T., Sasamoto, T.: Exact solution for the stationary {KPZ} equation.
\newblock arXiv 1111.4634 (2011)

\bibitem{Johansson-CMP2000}
Johansson, K.: Shape fluctuations and random matrices.
\newblock Commun. Math. Phys. \textbf{209}, 437--476 (2000)

\bibitem{Johansson-CMP2003}
Johansson, K.: Discrete polynuclear growth and determinantal processes.
\newblock Commun. Math. Phys. \textbf{242}, 277--329 (2003)

\bibitem{Kai.Zimmermann-PTPS1989}
Kai, S., Zimmermann, W.: Pattern dynamics in the electrohydrodynamics of
  nematic liquid crystals.
\newblock Prog. Theor. Phys. Suppl. \textbf{99}, 458-492 (1989)

\bibitem{Kai.etal-PRL1990}
Kai, S., Zimmermann, W., Andoh, M., Chizumi, N.: Local transition to turbulence
  in electrohydrodynamic convection.
\newblock Phys. Rev. Lett. \textbf{64}, 1111--1114 (1990)

\bibitem{Kallabis.Krug-EL1999}
Kallabis, H., Krug, J.: Persistence of {Kardar-Parisi-Zhang} interfaces.
\newblock Europhys. Lett. \textbf{45}, 20--25 (1999)

\bibitem{Kardar.etal-PRL1986}
Kardar, M., Parisi, G., Zhang, Y.C.: Dynamic scaling of growing interfaces.
\newblock Phys. Rev. Lett. \textbf{56}, 889--892 (1986)

\bibitem{Kriecherbauer.Krug-JPA2010}
Kriecherbauer, T., Krug, J.: A pedestrian's view on interacting particle
  systems, {KPZ} universality and random matrices.
\newblock J. Phys. A \textbf{43}, 403001 (2010)

\bibitem{Krug-JPA1989}
Krug, J.: Classification of some deposition and growth processes.
\newblock J. Phys. A \textbf{22}, L769--L773 (1989)

\bibitem{Krug-AP1997}
Krug, J.: Origins of scale invariance in growth processes.
\newblock Adv. Phys. \textbf{46}, 139--282 (1997)

\bibitem{Krug.etal-PRE1997}
Krug, J., Kallabis, H., Majumdar, S.N., Cornell, S.J., Bray, A.J., Sire, C.:
  Persistence exponents for fluctuating interfaces.
\newblock Phys. Rev. E \textbf{56}, 2702--2712 (1997)

\bibitem{Krug.etal-PRA1992}
Krug, J., Meakin, P., Halpin-Healy, T.: Amplitude universality for driven
  interfaces and directed polymers in random media.
\newblock Phys. Rev. A \textbf{45}, 638--653 (1992)

\bibitem{Kuennen.Wang-JSM2008}
Kuennen, E.W., Wang, C.Y.: Off-lattice radial {Eden} cluster growth in two and
  three dimensions.
\newblock J. Stat. Mech. \textbf{2008}, P05014 (2008)

\bibitem{Liechty-a2011}
Liechty, K.: The limiting distribution of the maximal height of the outermost
  path of nonintersecting {Brownian} excursions and discrete {Gaussian}
  orthogonal polynomials.
\newblock arXiv 1111.4239 (2011)

\bibitem{Majumdar-CS1999}
Majumdar, S.N.: Persistence in nonequilibrium systems.
\newblock Curr. Sci. \textbf{77}, 370--375 (1999)

\bibitem{Majumdar.Bray-PRL2001}
Majumdar, S.N., Bray, A.J.: Spatial persistence of fluctuating interfaces.
\newblock Phys. Rev. Lett. \textbf{86}, 3700--3703 (2001)

\bibitem{Majumdar.Dasgupta-PRE2006}
Majumdar, S.N., Dasgupta, C.: Spatial survival probability for one-dimensional
  fluctuating interfaces in the steady state.
\newblock Phys. Rev. E \textbf{73}, 011602 (2006)

\bibitem{Maunuksela.etal-PRL1997}
Maunuksela, J., Myllys, M., K\"ahk\"onen, O.P., Timonen, J., Provatas, N.,
  Alava, M.J., Ala-Nissila, T.: Kinetic roughening in slow combustion of paper.
\newblock Phys. Rev. Lett. \textbf{79}, 1515--1518 (1997)

\bibitem{Meakin-PR1993}
Meakin, P.: The growth of rough surfaces and interfaces.
\newblock Phys. Rep. \textbf{235}, 189--289 (1993)

\bibitem{Mehta-Book2004}
Mehta, M.L.: Random Matrices, \emph{Pure and Applied Mathematics}, vol. 142, 3
  edn.
\newblock Elsevier, San Diego (2004)

\bibitem{Merikoski.etal-PRL2003}
Merikoski, J., Maunuksela, J., Myllys, M., Timonen, J., Alava, M.J.: Temporal
  and spatial persistence of combustion fronts in paper.
\newblock Phys. Rev. Lett. \textbf{90}, 024501 (2003)

\bibitem{Mezard.etal-Book1987}
M\'ezard, M., Parisi, G., Virasoro, M.: Spin Glass Theory and Beyond: An
  Introduction to the Replica Method and Its Applications, \emph{World
  Scientific Lecture Notes in Physics}, vol.~9.
\newblock World Scientific, Singapore (1987)

\bibitem{Miettinen.etal-EPJB2005}
Miettinen, L., Myllys, M., Merikoski, J., Timonen, J.: Experimental
  determination of {KPZ} height-fluctuation distributions.
\newblock Eur. Phys. J. B \textbf{46}, 55--60 (2005)

\bibitem{MorenoFlores.etal-a2011}
Moreno~Flores, G., Quastel, J., Remenik, D.: Endpoint distribution of directed
  polymers in 1+1 dimensions.
\newblock arXiv 1106.2716 (2011)

\bibitem{Myllys.etal-PRE2001}
Myllys, M., Maunuksela, J., Alava, M., Ala-Nissila, T., Merikoski, J., Timonen,
  J.: Kinetic roughening in slow combustion of paper.
\newblock Phys. Rev. E \textbf{64}, 036101 (2001)

\bibitem{Myllys.etal-PRL2000}
Myllys, M., Maunuksela, J., Alava, M.J., Ala-Nissila, T., Timonen, J.: Scaling
  and noise in slow combustion of paper.
\newblock Phys. Rev. Lett. \textbf{84}, 1946--1949 (2000)

\bibitem{Newell.Rosenblatt-AMS1962}
Newell, G.F., Rosenblatt, M.: Zero crossing probabilities for {Gaussian}
  stationary processes.
\newblock Ann. Math. Stat. \textbf{33}, 1306-1313 (1962)

\bibitem{Oliveira.etal-PRE2012}
Oliveira, T.J., Ferreira, S.C., Alves, S.G.: Universal fluctuations in
  {Kardar-Parisi-Zhang} growth on one-dimensional flat substrates.
\newblock Phys. Rev. E \textbf{85}, 010601 (2012)

\bibitem{Paiva.FerreiraJr-JPA2007}
Paiva, L.R., Ferreira~Jr, S.C.: Universality class of isotropic on-lattice
  {Eden} clusters.
\newblock J. Phys. A \textbf{40}, F43--F49 (2007)

\bibitem{Prahofer.Spohn-PA2000}
Pr\"ahofer, M., Spohn, H.: Statistical self-similarity of one-dimensional
  growth processes.
\newblock Physica A \textbf{279}, 342--352 (2000)

\bibitem{Prahofer.Spohn-PRL2000}
Pr\"ahofer, M., Spohn, H.: Universal distributions for growth processes in
  $1+1$ dimensions and random matrices.
\newblock Phys. Rev. Lett. \textbf{84}, 4882--4885 (2000)

\bibitem{Prahofer.Spohn-JSP2002}
Pr\"ahofer, M., Spohn, H.: Scale invariance of the {PNG} droplet and the {Airy}
  process.
\newblock J. Stat. Phys. \textbf{108}, 1071--1106 (2002)

\bibitem{Prolhac.Spohn-PRE2011}
Prolhac, S., Spohn, H.: Height distribution of the {Kardar-Parisi-Zhang}
  equation with sharp-wedge initial condition: Numerical evaluations.
\newblock Phys. Rev. E \textbf{84}, 011119 (2011)

\bibitem{Prolhac.Spohn-JSM2011a}
Prolhac, S., Spohn, H.: Two-point generating function of the free energy for a
  directed polymer in a random medium.
\newblock J. Stat. Mech. \textbf{2011}, P01031 (2011)

\bibitem{Quastel.Remenik-a2012a}
Quastel, J., Remenik, D.: Tails of the endpoint distribution of directed
  polymers.
\newblock arXiv 1203.2907 (2012)

\bibitem{Rambeau.Schehr-EL2010}
Rambeau, J., Schehr, G.: Extremal statistics of curved growing interfaces in
  1+1 dimensions.
\newblock Europhys. Lett. \textbf{91}, 60006 (2010)

\bibitem{Rambeau.Schehr-PRE2011}
Rambeau, J., Schehr, G.: Distribution of the time at which $n$ vicious walkers
  reach their maximal height.
\newblock Phys. Rev. E \textbf{83}, 061146 (2011)

\bibitem{Rodriguez-Laguna.etal-JSM2011}
Rodr\'iguez-Laguna, J., Santalla, S.N., Cuerno, R.: Intrinsic geometry approach
  to surface kinetic roughening.
\newblock J. Stat. Mech. \textbf{2011}, P05032 (2011)

\bibitem{Sasamoto-JPA2005}
Sasamoto, T.: Spatial correlations of the {1D KPZ} surface on a flat substrate.
\newblock J. Phys. A \textbf{38}, L549--L556 (2005)

\bibitem{Sasamoto-PC2012}
Sasamoto, T.: private communication (2012)

\bibitem{Sasamoto.Imamura-JSP2004}
Sasamoto, T., Imamura, T.: Fluctuations of the one-dimensional polynuclear
  growth model in half-space.
\newblock J. Stat. Phys. \textbf{115}, 749--803 (2004)

\bibitem{Sasamoto.Spohn-JSM2010}
Sasamoto, T., Spohn, H.: The 1 + 1-dimensional {Kardar-Parisi-Zhang} equation
  and its universality class.
\newblock J. Stat. Mech. \textbf{2010}, P11013 (2010)

\bibitem{Sasamoto.Spohn-NPB2010}
Sasamoto, T., Spohn, H.: Exact height distributions for the {KPZ} equation with
  narrow wedge initial condition.
\newblock Nucl. Phys. B \textbf{834}, 523--542 (2010)

\bibitem{Sasamoto.Spohn-PRL2010}
Sasamoto, T., Spohn, H.: One-dimensional {Kardar-Parisi-Zhang} equation: An
  exact solution and its universality.
\newblock Phys. Rev. Lett. \textbf{104}, 230602 (2010)

\bibitem{Schehr-a2012}
Schehr, G.: Extremes of {$N$} vicious walkers for large {$N$}: application to
  the directed polymer and {KPZ} interfaces.
\newblock arXiv 1203.1658 (2012)

\bibitem{Schehr-PC2012}
Schehr, G.: private communication (2012)

\bibitem{Singha-JSM2005}
Singha, S.B.: Persistence of surface fluctuations in radially growing surfaces.
\newblock J. Stat. Mech. \textbf{2005}, P08006 (2005)

\bibitem{Stanley-Book1987}
Stanley, H.E.: Introduction to Phase Transitions and Critical Phenomena,
  \emph{International Series of Monographs on Physics}, vol.~46.
\newblock Oxford Univ. Press, Oxford (1987)

\bibitem{Takeuchi-PRE2008}
Takeuchi, K.A.: Scaling of hysteresis loops at phase transitions into a
  quasiabsorbing state.
\newblock Phys. Rev. E \textbf{77}, 030103(R) (2008)

\bibitem{Takeuchi-c2012}
Takeuchi, K.A.: {Comment on ``Experimental determination of KPZ
  height-fluctuation distributions'' by L. Miettinen \textit{et al.}} (2012).
\newblock \urlprefix\url{http://publ.kaztake.org/miet-com.pdf}

\bibitem{Takeuchi-JSM2012}
Takeuchi, K.A.: Statistics of circular interface fluctuations in an off-lattice
  {Eden} model.
\newblock J. Stat. Mech. \textbf{2012}, P05007 (2012)

\bibitem{Takeuchi.etal-PRL2007}
Takeuchi, K.A., Kuroda, M., Chat\'e, H., Sano, M.: Directed percolation
  criticality in turbulent liquid crystals.
\newblock Phys. Rev. Lett. \textbf{99}, 234503 (2007)

\bibitem{Takeuchi.etal-PRE2009}
Takeuchi, K.A., Kuroda, M., Chat\'e, H., Sano, M.: Experimental realization of
  directed percolation criticality in turbulent liquid crystals.
\newblock Phys. Rev. E \textbf{80}, 051116 (2009)

\bibitem{Takeuchi.Sano-PRL2010}
Takeuchi, K.A., Sano, M.: Universal fluctuations of growing interfaces:
  Evidence in turbulent liquid crystals.
\newblock Phys. Rev. Lett. \textbf{104}, 230601 (2010)

\bibitem{Takeuchi.etal-SR2011}
Takeuchi, K.A., Sano, M., Sasamoto, T., Spohn, H.: Growing interfaces uncover
  universal fluctuations behind scale invariance.
\newblock Sci. Rep. \textbf{1}, 34 (2011)

\bibitem{Tracy.Widom-CMP2009}
Tracy, C., Widom, H.: Asymptotics in {ASEP} with step initial condition.
\newblock Commun. Math. Phys. \textbf{290}, 129--154 (2009)

\bibitem{Tracy.Widom-CMP1994}
Tracy, C.A., Widom, H.: Level-spacing distributions and the {Airy} kernel.
\newblock Commun. Math. Phys. \textbf{159}, 151--174 (1994)

\bibitem{Tracy.Widom-CMP1996}
Tracy, C.A., Widom, H.: On orthogonal and symplectic matrix ensembles.
\newblock Commun. Math. Phys. \textbf{177}, 727--754 (1996)

\bibitem{Vicsek.etal-PA1990}
Vicsek, T., Cserz\H{o}, M., Horv\'ath, V.K.: Self-affine growth of bacterial
  colonies.
\newblock Physica A \textbf{167}, 315--321 (1990)

\bibitem{Wakita.etal-JPSJ1997}
Wakita, J.i., Itoh, H., Matsuyama, T., Matsushita, M.: Self-affinity for the
  growing interface of bacterial colonies.
\newblock J. Phys. Soc. Jpn. \textbf{66}, 67-72 (1997)

\end{thebibliography}

\end{document}